%% file: aanda.tex
\pdfoutput=1
\documentclass{aa}  

\usepackage{graphicx}
\usepackage{natbib}
\usepackage{txfonts}

\usepackage[dvipsnames]{xcolor}
\usepackage[colorlinks=true, linkcolor={blue!70!black}, citecolor={blue!70!black}, urlcolor={blue!70!black}]{hyperref}

\usepackage{CJKutf8}

\begin{document}

   \title{Mapping the "invisible" circumgalactic medium around a z $\sim$ 4.5 radio galaxy with MUSE}


   \author{Wuji Wang (\begin{CJK*}{UTF8}{gbsn}王无忌\end{CJK*}) \inst{\ref{inst1},\ref{inst2}}
        \and
        Dominika Wylezalek\inst{\ref{inst1},\ref{inst2}}
        \and 
        Carlos De Breuck \inst{\ref{inst2}}
        \and
        Joël Vernet\inst{\ref{inst2}}
        \and
        Andrew Humphrey
         \inst{\ref{inst3}}
        \and Montserrat Villar Mart\'in \inst{\ref{inst4}}
        \and  Matthew D. Lehnert \inst{\ref{inst5}}
        \and Sthabile Kolwa \inst{\ref{inst6},\ref{inst7}}
          }

   \institute{   Astronomisches Rechen-Institut, Zentrum f\"{u}r Astronomie der Universit\"{a}t Heidelberg, M\"{o}nchhofstr. 12-14, D-69120 Heidelberg, Germany\label{inst1}\\
                \email{wuji.wang@uni-heidelberg.de}
   \and
   European Southern Observatory, Karl-Schwarzchild-Str. 2, D-85748 Garching, Germany\label{inst2}
   \and
    Instituto de Astrof\'{\i}sica e Ci\^{e}ncias do Espa\c{c}o, Universidade do Porto, CAUP, Rua das Estrelas, PT4150-762 Porto, Portugal\label{inst3}
   \and
   Centro de Astrobiolog\'{\i}a, CSIC-INTA, Ctra. de Torrej\'{o}n a Ajalvir, km 4, 28850 Torrej\'{o}n de Ardoz, Madrid, Spain\label{inst4}
   \and
   Univ Lyon, Univ Lyon1, Ens de Lyon, CNRS, Centre de Recherche Astrophysique de Lyon UMR5574, F-69230, Saint-Genis-Laval, France\label{inst5}
  \and 
  Inter-University Institute for Data Intensive Astronomy, Department of Astronomy, University of Cape Town, Rondebosch 7701, South Africa\label{inst6}
  \and
  Physics Department, University of Johannesburg, 5 Kingsway Ave, Rossmore, Johannesburg, 2092, South Africa\label{inst7}
             }

   \date{Received 16 June 2021; accepted 19 July 2021}

 
  \abstract{In this paper we present Multi Unit Spectroscopic Explorer (MUSE) integral field unit spectroscopic observations of the $\sim70\times30$ kpc$^2$ Ly$\alpha$ halo around the radio galaxy \object{4C04.11} at $z = 4.5077$. High-redshift radio galaxies (HzRGs) are hosted by some of the most massive galaxies known at any redshift and are unique markers of concomitant powerful active galactic nucleus (AGN) activity and star formation episodes. We map the emission and kinematics of the Ly$\alpha$  across the halo as well as the kinematics and column densities of eight \ion{H}{i} absorbing systems at $-3500 < \Delta v < 0$ km s$^{-1}$. We find that the strong absorber at $\Delta v \sim 0\,\rm km\,s^{-1}$ has a high areal coverage ($30\times30$ kpc$^2$), being detected across a large extent of the Ly$\alpha$ halo, a significant column density gradient along the southwest to northeast direction, and a velocity gradient along the radio jet axis. We propose that the absorbing structure, which is also seen in \ion{C}{iv} and \ion{N}{v} absorption, represents an outflowing metal-enriched shell driven by a previous AGN or star formation episode within the galaxy and is now caught up by the radio jet, leading to jet-gas interactions. These observations provide evidence that feedback from AGN in some of the most massive galaxies in the early Universe may play an important role in redistributing material and metals in their environments.

  }
  


   \maketitle
%

\section{Introduction}\label{sec:introduction}

There is significant observational and theoretical evidence that supermassive black holes (SMBHs) in the centers of galaxies play a crucial role in the evolution of their hosts \citep[e.g.,][]{2008ARA&A..46..475H}. The powerful nuclear activities caused by actively accreting SMBHs -- active galactic nuclei (AGN) -- can lead to a substantial release of energy \citep[][]{1998A&A...331L...1S} and impact the evolution of their host galaxies \citep[][]{2013ARA&A..51..511K,2014ARA&A..52..589H}. 
The most powerful AGN (quasars, $L_{\rm bol}>10^{45}\rm \,erg\,s^{-1}$) can easily heat and photo-ionize their surrounding gas, sometimes even on scales of tens of kiloparsecs, well into the circumgalactic medium \citep[CGM; e.g.,][]{2017ARA&A..55..389T}. The detailed mechanisms and timescales relevant to AGN-driven feedback are still not fully understood \citep[][]{2012ARA&A..50..455F}, but large samples of galaxies with high spatial resolution using the modern surveys, primarily at low redshift right now, for example Sloan Digital Sky Survey-IV Mapping Nearby Galaxies at Apache Point Observatory \citep[SDSS-IV MaNGA, e.g.,][]{2018MNRAS.474.1499W}, help in assessing its prevalence and nature. However, it is the epoch at $z \sim 2-3$ that marks the peak of both star formation and quasar activity \citep[cosmic noon, $z\sim 2-3$;][]{2014ARA&A..52..415M,2020ARA&A..58..661F}, and probing the feedback processes in AGN at that epoch is essential. 

The CGM \citep[see][for a detailed review]{2017ARA&A..55..389T} is now understood to be a key component in disentangling the feedback processes in active galaxies. It links the smaller-scale interstellar medium (ISM) of the galaxy to the larger-scale intergalactic medium (IGM), not only in a geometrical way but also by acting as the reservoir fueling star formation and the central black hole, where the feedback interacts with the galactic environment and where the gas recycling during galaxy evolution is controlled. This complex environment is multiphase
and has been observed in numerous surveys \citep[e.g.,][]{2013ApJ...777...59T,2014ApJ...796..136B,2015ApJ...813....7P,2015ApJ...813...46B} at low redshift. A prominent feature of the CGM around active galaxies is the Ly$\alpha$ (Lyman-$\alpha$) emission line, which is also ubiquitously observed at high redshift \citep[e.g.,][]{2001ApJ...556...87H,2003ApJ...592..755R,2006AN....327..175V, 2007NewAR..51..194V,2013ApJ...768L...3H,2014Natur.506...63C,2016A&A...587A..98W,2018Natur.562..229W,2018MNRAS.473.3907A,2019MNRAS.482.3162A,2020ApJ...904..164N}. Ly$\alpha$ is the transition of the hydrogen electron from the $2p$ orbit to its ground state. It can happen primarily through collisional excitation and recombination \citep[see][for a detailed review of Ly$\alpha$ emission mechanisms and radiative transfer]{2014PASA...31...40D,2017arXiv170403416D}. In extragalactic studies, the recombination production of Ly$\alpha$ emission can be generated by photoionization by young stars and/or AGN (fluorescence). This fluorescence emission on larger scales (CGM and IGM) can also be due to UV background radiation. Additionally, collisional excitation can play an important role in the emission seen in outflows and infalling gas \citep[][]{2020ARA&A..58..617O}. The bright Ly$\alpha$ emission line, along with other UV lines excited by the central or background sources, provides a useful tool for studying the galactic environments in the early Universe. Additionally, \ion{H}{i} and metal absorption features observed in the CGM are powerful tracers of feedback signatures as well as tracers of infalling pristine gas \citep[e.g., low metallicity absorption in a $z\sim 2.7$ submillimeter galaxy;][]{2021ApJ...908..188F}. The sensitive integral field spectrographs on the largest ground-based telescopes, such as MUSE \citep[Multi-Unit Spectroscopic Explorer;][]{Bacon2010,2014Msngr.157...13B} and KCWI (Keck Cosmic Web Imager; \citealp{2012SPIE.8446E..13M}; see \citealp{2019ApJS..245...23C} for observation of Ly$\alpha$ halos with KCWI)
, are perfectly suited for mapping these UV features as they move into the optical band for high-redshift sources.


This paper focuses on the population of high-redshift radio galaxies \citep[HzRGs; $L_{\rm 500\, MHz} > 10^{26} \, \rm W \, Hz^{-1}$][]{2008A&ARv..15...67M}, which are some of the most massive galaxies known at any redshift \citep[with a narrow range in stellar masses of $(1-6) \times 10^{11}\,\rm M_{\odot}$ for $1<z<5.2$;][]{2010ApJ...725...36D}. Their energetic radio jets are unique markers of concomitant powerful AGN activity, which place them amongst the most active sources at and near cosmic noon. High-redshift radio galaxies have furthermore been shown to be powerful beacons of dense (proto-)cluster environments in the early Universe \citep[e.g.,][]{1996ApJ...471L..11L,2003AJ....125.2759S,2002ApJ...569L..11V,2003NewAR..47..353V,2004A&A...424L..17V,2005A&A...431..793V,2007A&A...461..823V,2013ApJ...769...79W}. The quasar-level AGN activity \citep[][]{2008A&ARv..15...67M} at the center is blocked by the thick dusty torus acting as the "coronograph" \citep[][]{2001A&A...366....7V}; this makes HzRGs true obscured type-2 quasars, allowing us to probe their host galaxies and CGM without strong AGN contamination (e.g., for unobscured quasars, see \citealp{2019MNRAS.482.3162A}, and for radio-quite type-2 sources, see \citealp{2017ApJ...837...71C}). Comprehensive studies using near-infrared integral field unit (IFU) instruments show that the ionized gas in HzRGs is highly perturbed ($FWHM \sim \rm 1000\,km\,s^{-1}$) at kiloparsec scales and is aligned with the radio jets \citep[e.g.,][]{2006ApJ...650..693N,2007A&A...475..145N,2008A&A...491..407N,2017A&A...599A.123N,2017A&A...600A.121N,2015A&A...579A..89C,2016A&A...586A.152C}. This implies that the energy and momentum transfer between the central quasar and their ISM is likely due to the jets. Radio-mode feedback may therefore play a fundamental role during the evolution of HzRGs. Recently, \citet{2019A&A...621A..27F} combined infrared and millimeter data and deduced a more robust result of a relatively low star formation rate (SFR) for a sample of HzRGs, suggesting evidence of rapid quenching compared to previous studies \citep[][]{2014A&A...566A..53D}. Using a small sample of HzRGs, \citet{2011A&A...525A..43N} shows that they are going through a transition phase from active to passive. These observations indicate that HzRGs are on a different track of evolution compared to radio-quiet objects, assembling most of their stellar mass early \citep[$z\sim3$;][]{2007ApJS..171..353S,2010ApJ...725...36D} , and that radio jets may actively affect their quenching. However, there is also circumstantial evidence showing that the jet can induce star formation. \citet{2006MNRAS.369.1103H} found that HzRGs ($z>2$ in the sample) with smaller radio sources and more perturbed gas (emission line) kinematics show lower UV continuum polarization, which could be due to the presence of more luminous young stellar populations and can possibly be explained by the interaction between radio jets and the ISM that enhances star formation. Besides, there is also an anticorrelation between the rest frame submillimeter flux density and radio size in HzRGs \citep{2011MNRAS.418...74H}, although it is not clear if the physics behind this is feedback-induced star formation, a simultaneous triggering of star formation and the radio-loud AGN activity, or simply environmental effects. Some well-studied HzRGs show evidence of having high SFRs \citep[e.g., 4C41.47 and PKS 0529-549;][]{2020A&A...639L..13N,2019A&A...621A..27F}. In these sources, we may interestingly be witnessing both the jets compressing the gas, leading to enhanced SFRs \citep[e.g.,][]{2017ApJ...850..171F}, and the feedback from the AGN and star formation quenching it \citep[][]{2019A&A...624A..81M}.

One of the most prominent features of HzRGs is their gaseous halos, which often reach out to more than 100 kpc from the nucleus, well into the CGM \citep[e.g.,][]{1996A&A...313...25V,1997A&A...317..358V,2003MNRAS.346..273V}, and which have different dynamical states \citep[from more perturbed inner regions to quieter outer regions; e.g.,][]{2017A&A...602L...6V}. The halos are observed in all strong emission lines \citep[e.g., Ly$\alpha$ to H$\alpha$,][]{1993ARA&A..31..639M,2008A&ARv..15...67M} and are metal-enriched, often detected in \ion{N}{v}$\lambda$1240\AA\ and \ion{C}{iv}$\lambda$1548\AA. 
The CGM is not only the venue of the feedback but also an essential path from which IGM gas can fuel the growth of SMBHs and star formation, as suggested by various cosmological models \citep[e.g.,][]{2005Natur.435..629S,2011MNRAS.418.1796F}. \citet{2019Sci...366...97U} observed (proto-)cluster-scale gas filaments that may be tracing infalling gas. Observations of the CGM around HzRGs \citep[e.g.,][]{2007MNRAS.375..705H,2008MNRAS.390.1505H,2017A&A...602L...6V} provide evidence of inflowing gas in both absorption and emission with the scale of 10s $\times$ 10s kpc$^{2}$. In addition to the neutral and ionized gas, the molecular and dust phases have also been studied using the Actacama Large Milimeter/submilimeter Array (ALMA; or other millimeter telescopes), which traces the environment of stellar components in the galaxies to show a comprehensive view of galaxy evolution in the early Universe \citep[e.g.,][]{2016A&A...586A.124G,2021A&A...645A.120F}. 

Many HzRGs have deep extended absorbers associated with them \citep{1997A&A...317..358V}. These associated absorbers offer a unique opportunity for probing the neutral CGM, without the requirement of direct ionization by the central AGN
\citep[][]{1995MNRAS.277..389R,1997A&A...317..358V,2008MNRAS.390.1505H,2003MNRAS.338..263J,2004MNRAS.351.1109W,2018MNRAS.474.3649S,2019A&A...625A.102K}.
The absorbers are usually blueshifted with respect to the host systemic redshift, which can be understood as a potential signature of outflowing gas. Over the past two decades, a series of works have established the picture and have offered evidence for explaining the observed absorption through the scenario of giant expanding shells of gas: \citet{2000A&A...356...23B} argued that the prototypical \ion{H}{i} and \ion{C}{iv} absorber in \object{MRC 0943-242} is probably a giant shell enveloping the line-emitting halo. \citet{2003MNRAS.338..263J} and \citet{2004MNRAS.351.1109W} obtained additional data and further developed the expanding shell idea. Before \citet{2008MNRAS.390.1505H}, who published the first IFU study of the properties of an extended HzRG absorber, works on the absorbers had only used long slit spectroscopy placed along the radio axis, meaning that there was no proof, only suspicion, that the \ion{H}{i} absorbers are not only extended along the radio jet axis. The result of \citet{2008MNRAS.390.1505H}, therefore, reinforced the giant expanding shell hypothesis. \citet{2018MNRAS.481.1401S} studied the Ly$\alpha$ halos of a sample of HzRGs to examine whether extended \ion{H}{i} absorbers are usually extended perpendicular to the radio axis. With the long slit spectroscopic data together with a handful of previously published MUSE observations containing extended \ion{H}{i}/HzRG absorbers \citep[e.g.,][]{2015MNRAS.449.1298S}, it was possible to draw the conclusion that extended \ion{H}{i} absorbers of HzRGs are commonly extended perpendicular to the radio axis. In \citet{2018MNRAS.474.3649S}, the authors measured the line-of-sight velocity as a function of offset from the AGN for the main \ion{H}{i} absorber in \object{MRC 0943-242} and detected a radially decreasing blueshift, consistent with an expanding shell centered on the nucleus. More interestingly, around 30\% of the detected absorbers are redshifted, and their natures are still unclear. This begs the question of whether they are the cooling inflowing IGM gas that models predict dominates the gas accretion of massive galaxies or are due to the emission line gas in the Ly$\alpha$ halo simply outflowing with a higher line-of-sight velocity than the \ion{H}{i} absorber. Absorption features are not unique around type-2 sources like HzRGs; they are also seen in the spectra of type-1 high-redshift quasars \citep[e.g.,][]{2019MNRAS.482.3162A}. These absorbers may also have an important influence on the inferred intrinsic total flux, which is sometimes neglected \citep[e.g., peak Ly$\alpha$ in ][]{2021MNRAS.502..494M}.

\object{4C04.11} (\object{RC J0311+0507}), at $z\sim4.5$, is the focusing target of this work. Radio emission of the source was first discovered with the Russian RATAN-600 instrument \citep[the 600 m diameter ring antenna of the Russian Academy of Sciences;][]{1992AZh....69..673G}. It was observed subsequently by other telescopes in the radio and optical \citep[][]{2006AstL...32..433K,1996BSAO...40....5P,2000A&AT...19..297P,2013EAS....61..439P,2014MNRAS.439.2314P}. The source (RC J0311) was then found to be the same one (4C04.11) registered in the older Cambridge surveys \citep[][]{1958AuJPh..11..360M,1967MmRAS..71...49G}. We note that \citet{2006AstL...32..433K} first obtained the redshift, $z=4.514$, of this target using the Ly$\alpha$ line spectrum taken from the Russian 6 m optical telescope (BTA). It is classified as an FR II source based on the radio morphology \citep[][]{1974MNRAS.167P..31F}. Previous studies show it has a central SMBH with a mass of $\sim 10^{9}\, \rm M_{\odot}$ \citep{2014MNRAS.439.2314P,2017A&A...599A.123N}. \citet{2017ApJ...841..128K} studied the large-scale environment of \object{4C04.11} by searching for surrounding Ly$\alpha$ emitters (LAEs) using the Subaru Telescope, which found that \object{4C04.11} is residing in a low-density region of LAEs. Its X-ray proprieties have been reported by \citet{2020ApJ...899..127S}, including the spectrum photon index ($\Gamma=0.92^{+0.5}_{-0.51}$) and the optical$-$X-ray power law slope, $\alpha_{\rm OX}=-1.31\pm0.08$. That work also reports the absence of extended X-ray structures despite the large radio jet scale.

In this paper we present the results of the MUSE observation for \object{4C04.11,} focusing on the absorption features in its CGM. This radio galaxy is the highest-redshift source in our sample of eight HzRGs with both MUSE and ALMA data. It also has multiple \ion{H}{i} and associated metal absorbers on which we can test the absorption mapping ability of MUSE. Hence, this is a pilot work, and upcoming studies will focus on the spatial characteristics of the CGM absorbers of the whole sample. In Sect. \ref{sec:observation} we present the observation and the optimized data reduction procedure of the target. The methodology used for analyzing emission line and absorption spectra as well as the spatial mapping is shown in Sect. \ref{sec:spectrumexandlinfit}. The results are presented in Sect. \ref{sec:fitresult} and Sect. \ref{sec:spatialmapresults} for the 1D spectrum and 2D mapping, respectively. We discuss some physical explanations from the analyzed results in Sect. \ref{sec:discussion} and propose several models to the observed spatial column density gradient of \ion{H}{i} absorber \#1 in Sect. \ref{sec:hi1spatial}. Finally, we summarize and conclude in Sect. \ref{sec:conclusions}. For this work, we use a flat Lambda cold dark matter cosmology with $\rm H_{0} = 71\,km\,s^{-1}\,Mpc^{-1}$ and $\rm \Omega_{M} = 0.27$. In this cosmology, 1" corresponds to 6.731 kpc at the redshift of our target, 4.5077.

\section{Observations and data reduction}\label{sec:observation}
\subsection{MUSE Observations}

The target of this work, the radio galaxy \object{4C04.11}, was observed by the European Southern Observatory (ESO) Very Large Telescope (VLT) using the instrument MUSE from December 2 to 15, 2015, under the program run 096.B-0752(F) (PI:J.Vernet). The observations were divided into four observing blocks (OBs), where each OB had two exposures of about 30 minutes each. The total integrated time was 4 hours on target. Observations were carried out in the extended wide-field mode of MUSE without the correction from active adaptive optics (WFM-NOAO-E). The wavelength coverage of MUSE is $4650-9300\,\text{\AA}$ and the field of view (FOV) of $60 \times 60\,\rm arcsec^{2}$ with a spatial resolution of $\rm 0.2 \times 0.2\,arcsec^{2}$ and a $1.25\,\text{\AA}\,\rm pix^{-1}$ wavelength sampling. The spectral resolving power of MUSE is approximately $\lambda / \Delta \lambda = 1700-3400,$ which is $\Delta \lambda =  2.82 - 2.74\,\text{\AA}$  or $\Delta v = 180-90$ $\rm km\,s^{-1}$ (blue to red) in terms of resolution \citep[][]{2014Msngr.157...13B}.  

\subsection{Optimized data reduction}\label{sec:datareduc} 

We are interested particularly in the faint extended line emission in the CGM of \object{4C04.11}. Therefore, we explore different data reduction strategies in order to find an optimized method for further analysis. First, we use MUSE Data Reduction Software (MUSE DRS, version 2.6, the newest version is 2.8.x) pipeline\footnote{\url{https://www.eso.org/sci/facilities/paranal/instruments/muse/doc.html}} \citep[][]{2020A&A...641A..28W} by running \textsc{esorex} (a command-line tool can be used for executing VLT/VLTI instrument pipeline) for calibration creation, observation preprocessing and observation post-processing. These three reduction stages are completed in the same default procedures for each method before adjusting the reductions. We explore the options of combining the individual exposures using the MUSE DRS pipeline and the MPDAF \citep[MUSE Python Data Analysis Framework;][]{2016ascl.soft11003B,2019ASPC..521..545P}. Furthermore, we explore the sky subtraction using the pipeline and Zurich Atmosphere Purge \citep[\textsc{ZAP}, a python package developed for MUSE data based on principal component analysis algorithm;][]{2016MNRAS.458.3210S}. 

We evaluate the performance of each reduction method and choose the one that maximizes the signal-to-noise ratio (S/N) for our target by qualitatively comparing the spectra extracted from different data cubes and quantitatively comparing their S/N. We extract spectra from the same apertures as in Sect. \ref{sec:specex} for each cube, respectively. Then, the S/N is calculated using four wavelength ranges for each spectrum ($5600-5900\,\text{\AA}$, line-free range; $6567-6864\,\text{\AA}$, Ly$\alpha$ emission range; $7400-8000\,\text{\AA}$, line-free range; $8300-9200\,\text{\AA}$, \ion{C}{iv} and \ion{He}{ii} emission range ). The performances of all cubes are similar. In the two line-free ranges, the optimized method (Sect. \ref{sec:datareduc}) is $\sim 2\%$ better. As for the emission line ranges, the optimized method is $\sim 5-10\%$ better. The skyline residuals (e.g., Sect. \ref{sec:civheii}) are less severe in the optimized method compared to the other methods, although we still apply masking when analyzing \ion{C}{iv} (Sect. \ref{sec:civheii}). Through this test, we find the most optimized method for reduction of our observation of \object{4C04.11}: all calibrations are done in the standard way following the pipeline; sky subtraction is done along with the pipeline; each derived exposure data cube goes through \textsc{zap} to remove the sky residuals; all exposures are then combined by MPDAF using the median absolute deviation (MAD) \footnote{For an univariate data set $X_{1},X_{2}...X_{n}$, MAD $= {\rm median}|X_{i}-\widetilde{X}|$, where $\widetilde{X} = {\rm median}(X)$} method.

We then perform the astrometric correction to the derived data cube to improve the accuracy of MUSE astrometry. In this step, the \textit{Gaia} Data Release 2 catalog \citep[][]{2016A&A...595A...1G,2018A&A...616A...1G} is adopted for acquiring the precise coordinate of the only field star in our MUSE FOV. The position offset is calculated based on this star (fitted with a 2D Gaussian model) and applied to our MUSE observation. This uncertainty estimated in this astrometry correction is 0.007 arcsec for which a large fraction (> 98\%) comes from the Gaussian fitting of the field star position.

Before we can obtain the data cube for the following scientific analysis, we perform the variance scaling on the variance extension of the data cube using a source-free region of data. The variance extension of the data cube before correction often underestimates the uncertainties due to the incomplete covering of the variance sources \citep[][]{2020A&A...641A..28W}. The variance scale factor is 1.27. The scaled variance extension can then be used for our scientific analysis. 

Finally, we note that comparing to the data cube of our target derived using MUSE DRS version 1.6 (S. Kolwa, priv. comm.), our new data cube has a more homogeneous background due to the implemented \texttt{auto\_calibration} function \citep[see][]{2020A&A...641A..28W} in version 2.6, which refines the IFU-to-IFU and slice-to-slice flux variations. 

We estimate the seeing PSF for the combined optimized data cube to be $\sim 0.97$ arcsec in the wavelength range $6573-6819\,\text{\AA,}$ which is the range of the observed Ly$\alpha$ emission. This is smaller than the extension of the Ly$\alpha$ halo and the \ion{H}{i} absorber \#1 (Sect. \ref{sec:spatialmapresults}). The central part of the Ly$\alpha$ emission halo is not dominated by any unresolved AGN emission
such that a further PSF subtraction is not necessary and will not improve the results. The $5\sigma$ surface brightness detection limit of our data at $6695.86\,\text{\AA}$ (peak of Ly$\alpha$, Table \ref{tab:lineemissionfit}) is $5\times 10^{-18}\,\rm erg\,s^{-1}\,cm^{-2}\,arcsec^{-2}$ summed over 6 wavelength channels ($7.5\,\text{\AA}$) from a 1 arcsec radius aperture. For the spectrum analyzed in this work (Sects. \ref{sec:specex} and \ref{sec:fitresult}), the noise spectrum is also shown, which is the standard deviation derived from the variance extension of the data cube presenting the quality of the reduction (Fig. \ref{fig:lyafit}, \ref{fig:civheiifit}, \ref{fig:nvfit} and \ref{fig:oiiifit}).

   \begin{figure}
   \centering
   \includegraphics[width=\hsize]{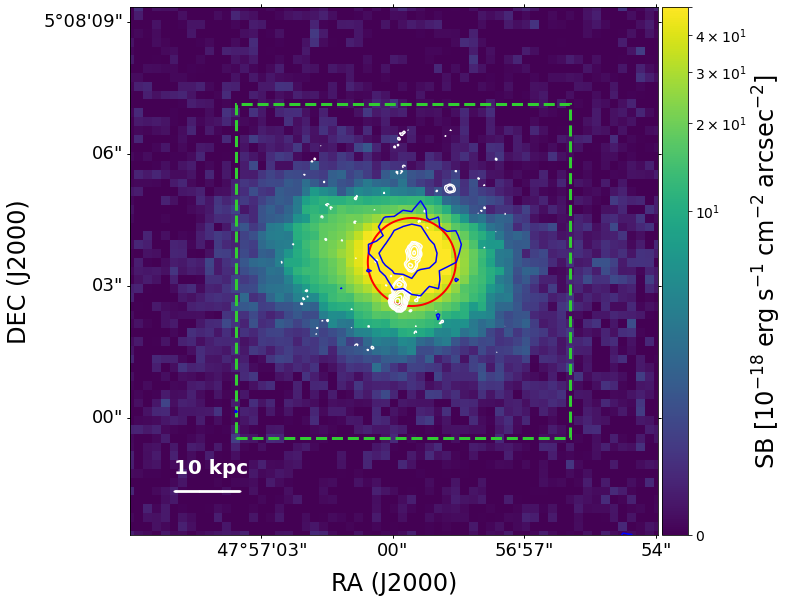}
      \caption{Ly$\alpha$ narrowband surface brightness (SB) image of \object{4C04.11} derived from the data cube using $6704-6710\,\text{\AA}$ in the observed frame. The blue contour indicates the \ion{He}{ii} emission region. The contour is calculated from the pseudo-narrowband image of \ion{He}{ii} using $9028-9044\,\text{\AA}$ in the observed frame with the level of $3\sigma_{\rm \ion{He}{ii}}$ and $2 \times 3\sigma_{\rm \ion{He}{ii}}$, where $\sigma_{\rm \ion{He}{ii}}$ is the standard deviation derived from a source-less region of the \ion{He}{ii} pseudo-narrowband image. The white contour traces the position of the radio jet observed by MERLIN \citep[Multi-Element-Radio-Link-Interferometer-Network;][]{2013EAS....61..439P,2014MNRAS.439.2314P} with the level of 0.45$\times$ (-1, 1, 2, 4, 16, 32, 48) mJy beam$^{-1}$ following \citet{2014MNRAS.439.2314P}. The overlaid red circle with a 1 arcsec radius marks the aperture over which the master spectrum is extracted. The green dashed box shows the FOV of individual panels in Figs. \ref{fig:lyaspatial1} and \ref{fig:lyaspatial2}, which is the region we focus on in the spatial mapping in Sect. \ref{sec:spatialmapresults}. }\label{fig:moment0}
   \end{figure}

\section{Data analysis}\label{sec:spectrumexandlinfit}
\subsection{Single aperture spectrum (master spectrum)}\label{sec:specex}

We first extract a spectrum from a large aperture with the goal of using this high S/N spectrum to optimize our analysis and line fitting procedures. The center of the extraction aperture is at ($\alpha$, $\delta$) = ($47^{\circ}56^{\prime}59^{\prime\prime}.6$, $5^{\circ}08^{\prime}03^{\prime\prime}.5$). This is chosen to be at the pixel with the highest Ly$\alpha$ flux value from the pseudo-narrowband image collapsed between $6704\,\text{\AA}$ and $6710\,\text{\AA}$ \footnote{This is also the position of the flux peak for Ly$\alpha$ narrowband image collapsed between larger wavelength range (e.g., $6573-6819\,\text{\AA}$ covering the entire Ly$\alpha$ line emission) and white light image of \object{4C04.11}.} (Fig. \ref{fig:moment0} red circle). Next, we extracted three spectra using apertures with radii 0.5, 1 and 1.5 arcsec. By comparing the three spectra, we find that: the line flux (Ly$\alpha$) ratio for the 0.5 arcsec to 1 arcsec is proportional to their area ratio, which means that the background does not dominate. But the line flux ratio for the 1 arcsec to 1.5 arcsec spectrum is larger than their area ratio, meaning that the contribution of the background is starting to impact the flux measurement. We therefore choose the spectrum extracted from a 1 arcsec aperture centered on the brightest pixel for the following analysis and refer to this spectrum as the "master spectrum" in the remaining parts of the paper. 

\begin{figure*}
  \centering
        \includegraphics[width=16.4cm,clip]{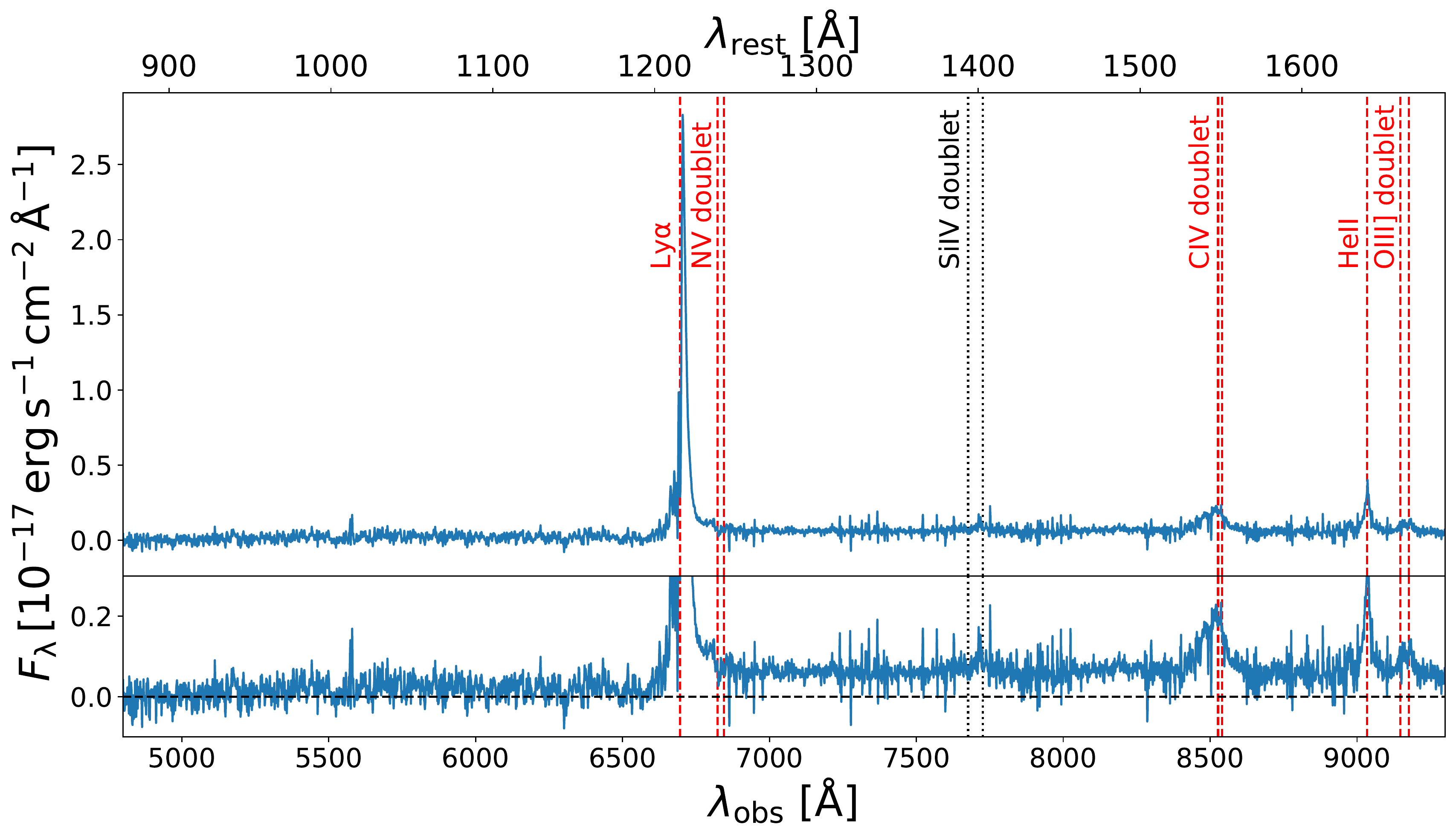}
      \caption{Rest frame UV spectra of \object{4C04.11}. {\it Upper panel}: Full MUSE spectrum extracted from the central 1 arcsec aperture region. We refer to this spectrum as the master spectrum. The detected UV lines (Ly$\alpha$, \ion{C}{iv}, \ion{He}{ii,} and \ion{O}{iii}]) are marked with red dashed lines. We also mark the \ion{N}{v,} which has a low S/N, overlaps with the broad Ly$\alpha$ wing, and is not obvious in this full spectrum (see Sect. \ref{sec:nv}). We use the black dotted line to indicate the position of the undetected \ion{Si}{iv}. {\it Lower panel}: Same plot as the upper panel but zoomed in to show the continuum. We note that the skyline residuals are seen as regions with higher noise. The horizontal black dashed line marks the zero flux level. }
         \label{fig:fullspectrum}
   \end{figure*}

The master spectrum is shown in Fig. \ref{fig:fullspectrum} with the upper panel focusing on emission lines and lower panel focusing on the continuum. In the figure, we mark the emission lines with significant detection that our analysis will focus on: Ly$\alpha$ $\lambda$1216 (hereafter Ly$\alpha$), \ion{C}{iv} $\lambda\lambda$1548, 1551 (hereafter \ion{C}{iv}), \ion{He}{ii} $\lambda$1640 (hereafter \ion{He}{ii}) and \ion{O}{iii}]$\lambda\lambda1660,1666$ (hereafter \ion{O}{iii}]). We also mark the low S/N \ion{N}{v} $\lambda\lambda$1238, 1243 (hereafter \ion{N}{v}). The flux of \ion{N}{v} is indistinctly low and highly absorbed. Additionally, its position is located in the Ly$\alpha$ wing making it hard to detect in the full spectrum (Sect. \ref{sec:nv}). Using the black dotted lines, we indicate the potential positions of \ion{Si}{iv} $\lambda\lambda$1393, 1402 (hereafter \ion{Si}{iv}, the overlapped \ion{O}{iv}] quintuplet is not shown). We perform the fitting of \ion{Si}{iv} following Sect. \ref{sec:fitprocedure}, but the S/N is so low that the line model is poorly constrained. Hence, we consider it as an un-detection.

\subsection{Spectral analysis}\label{sec:fitprocedure}

We fit models to the observed emission lines to study the physical properties of the gaseous halos of \object{4C04.11}, for example the emitted flux and absorber column density. To do this task properly, we use the Gaussian or Lorentzian model to describe the emission and Voigt-Hjerting function \citep[e.g.,][]{2006MNRAS.369.2025T,2007MNRAS.382.1375T} for the absorption. The fitted function (Eq. \ref{eq:fitfunction}) is composed of the emission model(s), $F_{\rm \lambda,G}$ or $F_{\rm \lambda,L}$ (defined as Eq. \ref{eq:lorentzian}), multiplied with the convolved Voigt function(s) (Eq. \ref{eq:cv}). The convolution with the MUSE line-spread function is applied to account for the instrumental resolution. The decision whether Gaussian or Lorentzian function is used for modeling the emission components is made based on several reasons explained in Sect. \ref{sec:fitresult}.  In Appendix \ref{apd:fitequation}, we explain the definitions and equations used in our fitting. The underlying assumption made when fitting the Voigt profile to the absorption is that each of the absorbing cloud gas has a covering factor close to unity (C $\simeq$ 1.0). 

We use different strategies to manage the continua for different lines (see Sect. \ref{sec:fitresult} and Appendix \ref{apd:fitingprodecurenotes} for details of different lines). The basic idea is to fit the continuum around the emission line with the emission part masked. After the continuum fitting, we then apply the nonlinear least-squares (least-squares for short) algorithm, which preforms with $\chi^{2}$ minimization to fit the interested spectra. Because of the number of free parameters used in the fitting or/and insensitivity of the algorithm to one (or some) of the variables, several problems appear when running the least-squares method, for example the covariance matrix from which uncertainties of the fitting are derived cannot be produced. Then we apply a more sophisticated method, Markov chain Monte Carlo \citep[MCMC; using the python package \texttt{emcee};][]{2013PASP..125..306F}, which realizes the fitting through maximizing the likelihood, to better constrain the results and determine the fitting uncertainties. To fulfill this, we perform the MCMC fitting using results from least-squares as initials. We report the results together with the $\chi^{2}_{\nu}$ in Sect. \ref{sec:fitresult}\footnote{Reduced $\chi^{2}$, $\chi^{2}_{\nu}=\frac{{\mathit \chi^2}}{N-N_i}$, which is calculated from the best-fit MCMC model and the data as an indicator of the fitting quality, where $N$ is the number of input data points and $N_{i}$ being the number of free parameters.} .

We note that the reported "1$\sigma$" uncertainties in this paper are either the direct reported value of the 1$\sigma$ confidence level from the algorithm (half the difference between the 15.8 and 84.2 percentiles) or the propagated value from this. Due to the large number of free parameters used in the fitting model and the physically limited parameter ranges, we cannot always explore the entire parameter space, that is to say, the fitting procedure seldom gives us the 3$\sigma$ confidence level. Hence, we take the compromise to report the 1$\sigma$ confidence level for a reference. Some of the reported formal uncertainties are too small compared to the instrumental limitations, for example the uncertainty of the line center and the spectral resolution of MUSE.

\subsection{Spatial mapping method}\label{sec:spatialmethod}
The MUSE observations allow us to spatially and spectrally map the gaseous halo around \object{4C04.11}. We mainly focus on the morphology, kinematics and absorption column density distribution of Ly$\alpha$ emission because of its high surface brightness. The spatial properties of \ion{C}{iv} absorbers are also studied but only in two spatial apertures due to its relatively low S/N. Hence, we describe here the method we use for mapping the Ly$\alpha$ characteristics in this subsection and show the details of \ion{C}{iv} together with its result in Sect. \ref{sec:civspatial}. 

The first step is to spatially bin the data of the Ly$\alpha$ emission region to increase the S/N for the following fitting. We adopt the method from \citet{2015MNRAS.449.1298S}, which starts at the brightest spatial pixel (spaxel) and bins the spaxels around it until the set S/N or the number of spaxels in one bin threshold is reached. The S/N threshold is set to be 13, which is close to the median value of the spaxel-based S/N in the region enclosed by the green box shown in Fig. \ref{fig:moment0} and calculated from the wavelength range $6672-6695\,\text{\AA}$. This wavelength range is slightly bluer than the peak emission wavelength of Ly$\alpha$ because we are interested in the spatial distribution of the absorbers that are located in the blue wing of Ly$\alpha$. The largest length of one tessellation bin is 25 spaxels (5 arcsec), which is $\sim$ 5 times the size of our seeing disk (Sect. \ref{sec:datareduc}) to include any large-scale structures with low S/N. The commonly used Voronoi binning \citep[][]{2003MNRAS.342..345C} method is not suitable for our purposes due to the high S/N gradient across the Ly$\alpha$ nebula ($\sim$150 to $\sim$10 in 20 spaxels). We manually bin some spaxels after running the algorithm to achieve a more homogeneous S/N distribution. There are 64 bins in the final result, which is shown in Fig. \ref{fig:binum_map}. 

Next, we fit the Ly$\alpha$ spectrum extracted from each bin following the description in Sect. \ref{sec:fitprocedure}, namely we first fit with the least-squares method and then used MCMC to refine the fit. We note that only \ion{H}{i} absorbers \#1 and \#2 (see Sect. \ref{sec:lya}) can be identified in all bins. But for consistency we include all eight absorbers in each fit. To minimize the number of free parameters and keep the fitting of absorbers \#3$-$8 less problematic (especially for those bins where they cannot be seen), we fix the positions (velocity shifts) of these 6 absorbers using the values derived from the aperture-extracted Ly$\alpha$ fitting (see Sect. \ref{sec:lya}). We also fix the continuum fitted from each spectrum prior to including the combined Gaussian plus Voigt profiles in the fitting function. The results are presented in Sect. \ref{sec:lyaspatial}.

\subsection{Photometry data and SED fitting}\label{sec:stellarmass}

\object{4C04.11} has multiband photometry available from previous observations, namely, $B$, $V$, $R$, $I$ and $K$ bands reported in \citet{2014MNRAS.439.2314P}, 4 bands of ALLWISE \citep[an extended survey of Wide-field Infrared Survey Explorer;][]{2010AJ....140.1868W,2011ApJ...731...53M} archival data and \textit{Spitzer} IRAC 1 and 2 observation \citep[ID 70135, PI: D. Stern. See][for data reduction and flux measurement]{2013ApJ...769...79W,2014ApJ...786...17W}. In addition, \citet{2020ApJ...899..127S} reported the \textit{Chandra} 0.5-7 keV X-ray continuum detection of our target. Using these data, we preform a spectral energy distribution (SED) fitting with X-CIGALE \citep[X-ray module for Code Investigating GALaxy Emission;][the used photometric data and fitting result are presented in Appendix \ref{apd:sed}]{2019A&A...622A.103B,2020MNRAS.491..740Y} and show the SED fit in the appendix. 

We extract the unattenuated stellar emission flux at rest frame 1.6 $\rm \mu m$ from the fitted SED model from which $M_{\star}$ is estimated using the extrapolated IR mass-to-light ratio and galaxy age relation \citep[e.g., Fig. 2 in][]{2007ApJS..171..353S}. The $ 1.6 \, \rm \mu m$ is a "sweet spot" for deriving $M_{\star}$ of HzRGs. The flux at shorter wavelengths is dominated by young stellar populations (and contaminated by emission lines) and the shape of the SED beyond the stellar emission bump at around $1-2 \rm \,\mu m$ is dominated by AGN-heated dust.  Hence, the flux at $\sim$ $1.6\,\rm \mu m$ is dominated by the bulk of the stellar population.  We consider our stellar mass estimate of $M_{\star} < 6.9 \times 10^{11}\,\rm M_{\odot}$ as an upper limit due to unaccounted for contributions from AGN-heated dust. 

This derived upper limit of the stellar mass is quite high but comparable to other HzRGs \citep[$1<z<5.2$,][]{2010ApJ...725...36D}. Therefore, taking into account the derived upper limit and the stellar masses from a large sample, we set the $M_{\star}$ of 4C04.11 to  $\sim 2 \times 10^{11}\rm \,M_{\odot}$ and use this value for following calculation. Galaxies with $M_{\star}\sim10^{11}\rm \,M_{\odot}$ are extremely rare ($\log{(\Phi\rm /dex^{-1}/Mpc^{-3})} \sim 10^{-6}$) at the redshift ($z=4.5077$) of our object \citep[][]{2017A&A...605A..70D}. This indicates that \object{4C04.11} is a rare galaxy that assembled most of its mass and formed stars when the Universe was very young. It is of great interests to study the different phases of feedback as well as the current environment of such an object. 
  
 To estimate the total (baryonic and dark matter) mass of our object, we assume the $M_{\star}/M_{\rm halo}$ ratio to be 0.02 \citep[see][]{2013ApJ...770...57B}. This ratio has a large uncertainty, especially for objects at $z>4$. For high $M_{\star}$ objects at high redshift with extremely low number density, it is difficult to predict from simulation works \citep[e.g.,][]{2019MNRAS.488.3143B}. The evolutionary trend from \citet{2013ApJ...770...57B} shows that this ratio will be higher in the early Universe for objects that are the progenitors of present day massive galaxies (assumed to be applicable to HzRGs, see Sect. \ref{sec:introduction}). Hence, we adopt a conservative value of 0.02, which is the maximum ratio predicted by \citet{2013ApJ...770...57B}. 
 Then we can calculate the virial radius of the host galaxy, $R_{\rm vir} \simeq 117\,$kpc, using 
\begin{equation}
    R_{{\rm vir}} \simeq 100 \, \mathrm{kpc}\,(M_{{\rm vir}}/10^{12}{\rm M_{\odot}})^{1/3} (1+z)_{3}^{-1}, 
\end{equation}
where $(1+z)_{3} = (1+z)/3$ given by \citet{2013MNRAS.435..999D}. This is accurate to a few percent for a system at $z\gtrsim 1$. 

Using the following equation from the same work, 
\begin{equation}
    V_{\mathrm{vir}} \simeq 200\, \mathrm{km}\,(M_{{\rm vir}}/10^{12}{\rm M_{\odot}})^{1/3}(1+z)_{3}^{1 / 2},
\end{equation}
we also calculate the virial velocity of our target to be $\simeq 583 \, \rm km\,s^{-1}$. The virial temperature is at the order of $10^7$\,K. We note that these should be treated as approximation since we only take the $M_{{\star}}$ derived from the SED fitting as an upper limit and use the maximum predicted $M_{\star}/M_{\rm halo}$ ratio value in calculation.

   \begin{figure}
   \centering
   \includegraphics[width=\hsize]{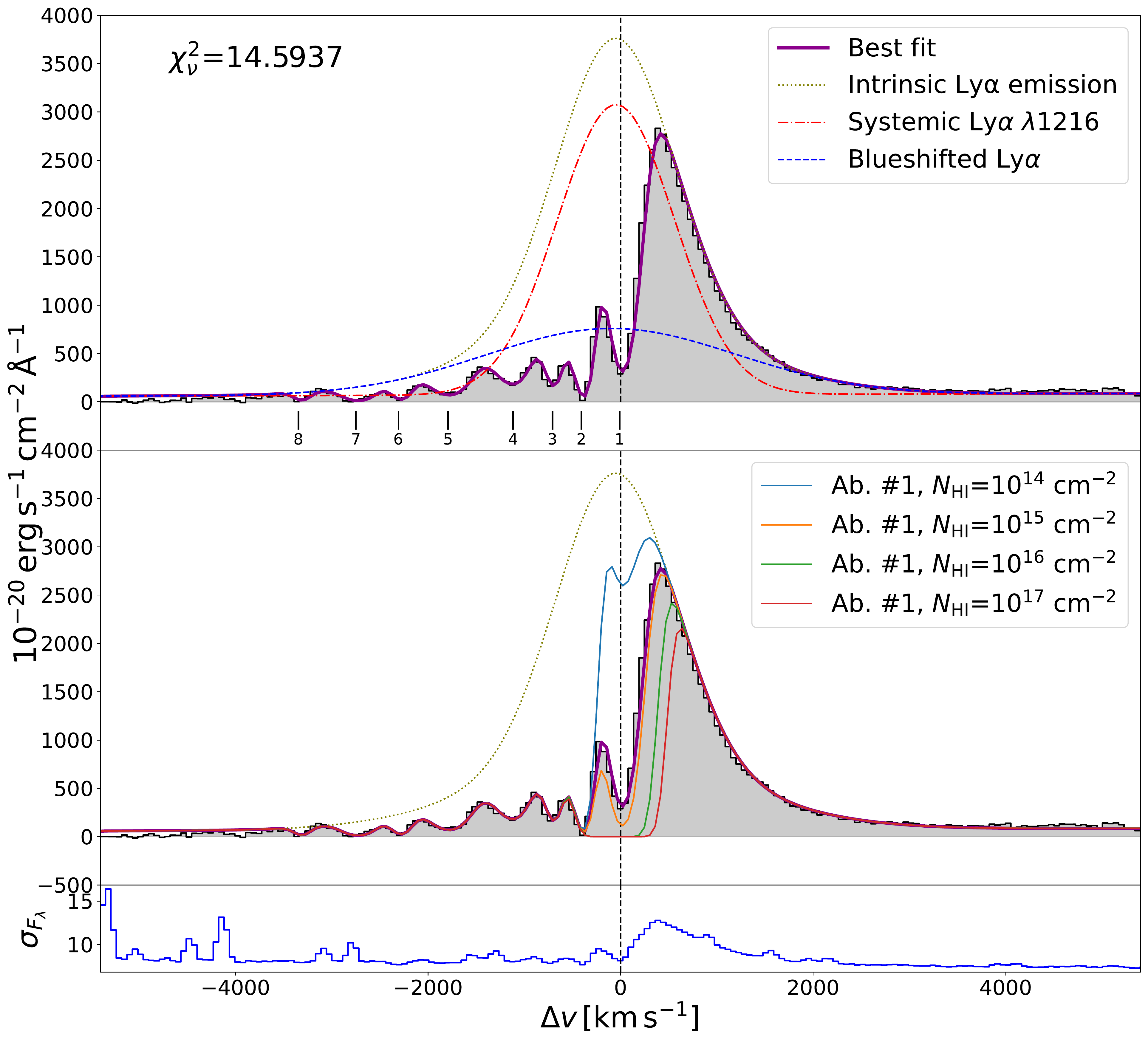}
      \caption{Ly$\alpha$ fitting result and model sensitivity check. {\it Upper panel}: Best fitting result of the master Ly$\alpha$ line using the MCMC method. The dark magenta line represents the best fit, while the dotted olive line traces the overall Gaussian emission. The systemic emission is shown in a red dotted-dashed line, and the blueshifted emission is marked by a blue dashed line. The positions of all eight absorbers are shown in this panel with black vertical bars. The $\chi_{\nu}^{2}$ is reported as an indicator for the quality of the fit. We note that the small flux excess at $\gtrsim 4000\,\rm km\,s^{-1}$ is the contribution from \ion{N}{v} (Sect. \ref{sec:nv}).  {\it Middle panel}: Column density sensitivity check of the Voigt model. The overall intrinsic Gaussian (dotted olive line) and best-fit model (dark magenta line) are the same as in the upper panel. The other lines show how the fitting result will change by only adjusting the column density of absorber \#1 to values at $ \log(N_{\rm \ion{H}{i}}/\rm cm^{-2})$ =14, 15, 16, and 17. These lines demonstrate that this fitting is only sensitive to the $N_{\rm H}$ values around the best fitting result. {\it Lower panel}: Standard deviation (noise) of the spectrum derived from the variance extension of the data cube that is used as the weight (inverse) in the fitting. It is shown in the same units as the spectrum and can be used to trace the skylines. We note that the Ly$\alpha$ line suffers less from strong skylines in $-4\,200 \sim 5\,500\,\rm km \, s^{-1}$ compared to \citet{2003A&A...407.1157H}.}\label{fig:lyafit}
   \end{figure}

\section{Line fitting results}\label{sec:fitresult}

In this section we present the line fitting results of Ly$\alpha$, \ion{N}{v}, \ion{C}{iv} + \ion{He}{ii} and \ion{O}{iii}]. We remind the readers that the metal absorbers are not as robustly detected as \ion{H}{i} absorbers, which have well-defined trough(s) in the spectra in visual check. This is probably due to the depth of the exposure and the spectral resolution. Hence, during the fit, we assume they are at the similar redshift (velocity shift) to the corresponding \ion{H}{i} absorbers. The reasons a subset of the absorbers are considered are presented in corresponding subsections (see Sects. \ref{sec:civheii} and \ref{sec:nv}). We refer them as "detection" if their probability distributions in the corner plots (Appendix \ref{apd:auxmcmc}) are well constrained. To visually distinguish the better and poorly constrained absorbers, we use the short solid bars and dashed bars with lighter colors in the figures showing the fitting results, respectively.

\begin{figure*}
   \centering
   \includegraphics[width=16.4cm,clip]{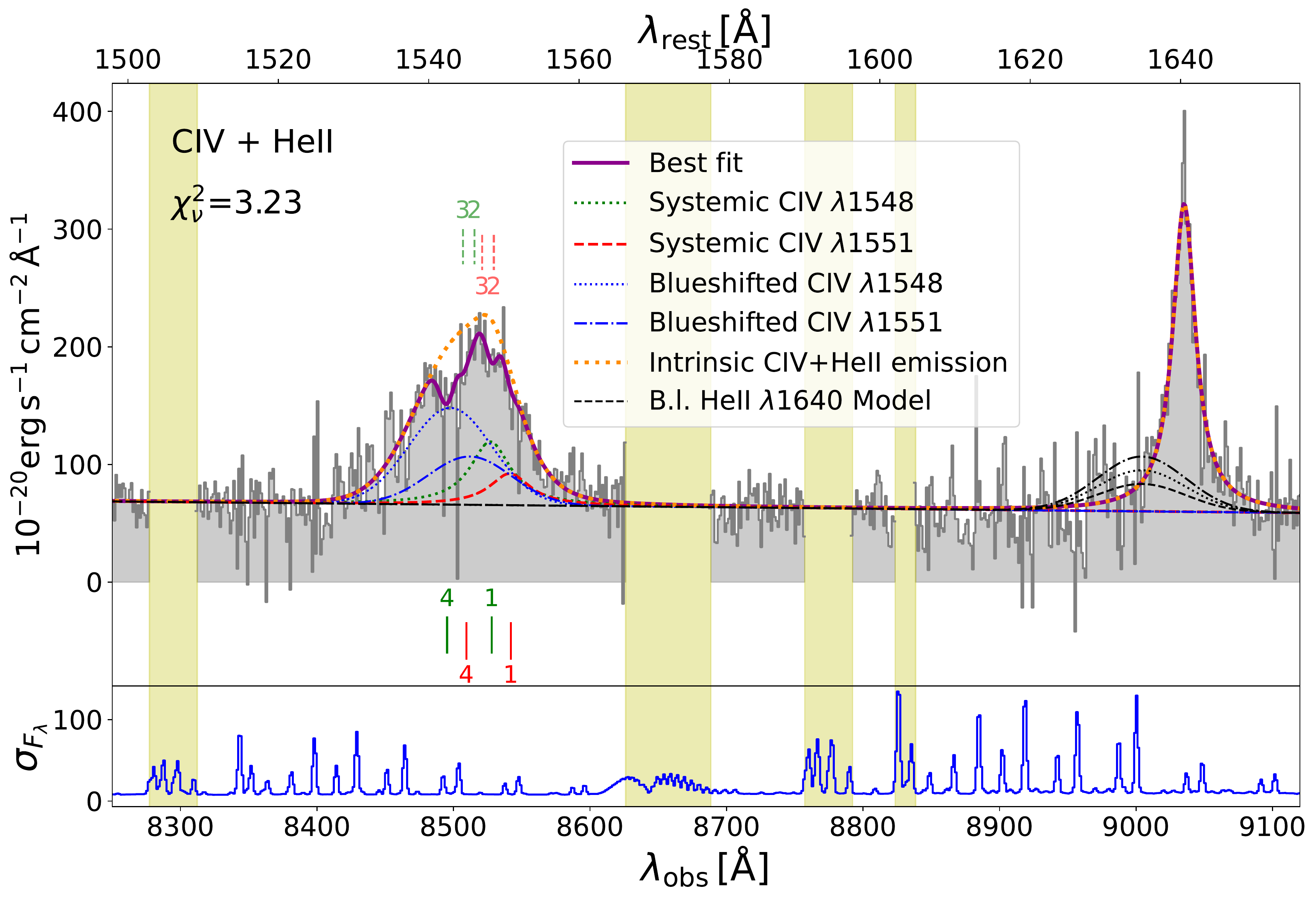}
      \caption{Best fit of the \ion{C}{iv} and \ion{He}{ii} lines from the master spectrum using the MCMC method. We use the dark magenta line to trace the best fit of these two lines. For the intrinsic \ion{C}{iv} and \ion{He}{ii}, the yellow dotted line is used. We mark the systemic emissions of the \ion{C}{iv} doublet in dotted green and dashed red lines and blueshifted emissions in dotted and dot-dash blue lines. The $\chi^{2}_{\nu}$ of the fit is reported to give a hint of the quality of the fit. The yellow shaded regions are excluded during the fit because it is severely affected by skylines (compared with the \citealt{2003A&A...407.1157H} data; see also the standard deviation in the lower panel, which offers an alternative proof of the positions of the skylines). The short green (red) vertical bars with numbers show the positions of the absorbers on top of the \ion{C}{iv}$\lambda$1548 (\ion{C}{iv}$\lambda$1551) line. Absorbers \#2 and \#3 are marginally constrained (see text), and hence their positions are shown in dashed bars with lighter numbers. The black lines are models of blueshifted \ion{He}{ii} lines with \textit{FWHM} and velocity shift fixed to the fitted values from the \ion{C}{iv} blueshifted component. The dashed, dotted, and dash-dotted lines show the line flux of 0.2, 0.3, and 0.4 of the total fitted flux of the blueshifted \ion{C}{iv}. The {\it lower panel} shows the standard deviation (noise) of the spectrum derived from the variance extension of the data cube, which is used as a fitting weight. It is shown in the same units as the spectrum and can be used to trace the skylines.}\label{fig:civheiifit}
   \end{figure*}

We also run a test on fitting the \ion{C}{iv} and \ion{N}{v} without absorption. The overall shape of the \ion{C}{iv} could be fitted without absorbers involved. However, the deep trough around 8500\AA, which is too broad to be influenced by skylines (see Fig. \ref{fig:civheiifit}), cannot be reproduced. As for \ion{N}{v}, the algorithm failed to reproduce the systemic emission component. Hence, we believe the absorbing material is enriched and fit the aforementioned two lines with absorbers.

\begin{table*}
 \caption{Best fitted emission results of the 1D aperture-extracted spectrum using the MCMC method.}\label{tab:lineemissionfit}
 \centering
\begin{tabular}{ l c c c c }
\hline
\hline
Ion & Line center (rest) & Line center (obs.) & Line flux & Line width \\
    & $ \lambda_{0}\,[\text{\AA}]$ & $\lambda\,[\text{\AA}]$ & $F$ [$\rm 10^{-17} \, erg \, s^{-1} \, cm^{-2}$] & $FWHM$ [$\rm km \, s^{-1}$] \\
    
\hline

Ly$\rm \alpha$        & 1215.67 & 6694.47 $\pm$ 0.59 & 101.82 $\pm$ 4.46 & 1426 $\pm$ 23\\
Ly$\rm \alpha$ (b.l.) & 1215.67 & $\sim$ 6693.28  & 49.92 $\pm$ 1.26 & 3055 $\pm$ 38\\

\ion{N}{v}           & 1238.82 & 6823.05  & 1.0 $\pm$ 0.2 & 2087 $\pm$ 250 \\
\ion{N}{v} (b.l.)    & 1238.82 & $\sim$6786.93 & 4.3 $\pm$ 0.2 & 6034 $\pm$ 259  \\

\ion{N}{v}           & 1242.80 & 6844.97  & 0.5 $\pm$ 0.1 & 2087 $\pm$ 250 \\
\ion{N}{v} (b.l.)    & 1242.80 & $\sim$6808.73 & 2.2 $\pm$ 0.1 & 6034 $\pm$ 259  \\

\ion{C}{iv}           & 1548.20 & 8526.98 $\pm$ 0.18 & 3.04 $\pm$ 1.01 & 1264 $\pm$ 245 \\
\ion{C}{iv} (b.l.)    & 1548.20 & 8497.81 $\pm$ 3.20 & 6.27 $\pm$ 0.84 & 2517 $\pm$ 143 \\

\ion{C}{iv}           & 1550.77 & 8541.19 $\pm$ 0.18 & 1.52 $\pm$ 0.50 & 1262 $\pm$ 244 \\
\ion{C}{iv} (b.l.)    & 1550.77 & 8511.97 $\pm$ 3.20 & 3.13 $\pm$ 0.42 & 2513 $\pm$ 142 \\

\ion{He}{ii} &1640.47 & 9035.23 $\pm$ 0.19 & 8.31 $\pm$ 0.15 & 671 $\pm$ 19 \\

\ion{O}{iii}]         & 1660.81 & 9147.24 & 1.00 $\pm$ 0.05 & 907 $\pm$ 68 \\
\ion{O}{iii}]         & 1666.15 & 9176.65 & 1.50 $\pm$ 0.05 & 907 $\pm$ 68 \\

 \hline
\end{tabular}
\tablefoot{The blueshifted component is marked as b.l. The reported errors are either the direct $1\sigma$ error bar output by the MCMC method or calculated through propagation of uncertainty from the MCMC output. We note that the "best" fitting results reported here from the MCMC method are the median values from the MCMC sampling. The line centers of the systemic \ion{N}{v} and \ion{O}{iii}] emissions are fixed to the redshift determined from \ion{He}{ii} in this work during the fitting. }

\end{table*}

\begin{table*}
 \caption{Best absorption fitting results of the 1D aperture-extracted spectrum using the MCMC method.}\label{tab:absorberfit}
 \centering
\begin{tabular}{c c c c c c c}
\hline
\hline
Abs. & Ion & Redshift & Absorber wav. & Velocity & Column density & Doppler \\
\#    & & $z$        & $\lambda$ $[\text{\AA}]$ & $\Delta v$ $[\rm km\,s^{-1}]$ & $\log(N/ \rm cm^{-2})$ & $b$ $[\rm km\,s^{-1}]$ \\
\hline
 1 & Ly$\alpha$ & 4.5075 $\pm$ 0.0001& 6695.34 $\pm$ 0.02 & $-9$ $\pm$ 6 & 14.843$\pm$ 0.004& 187 $\pm$ 1 \\ 
   & \ion{N}{v} & $-$                & 6822.84            & $-$          & 14.99 $\pm$ 0.05  & 387 $\pm$ 12 \\
   & \ion{C}{iv}& 4.5083  $\pm$ 0.0005 & 8527.91 $\pm$ 0.76 & $32$ $\pm$ 38 & 13.9 $\pm$ 0.2  & 198 $\pm$ 43\\ 
  
 2& Ly$\alpha$  & 4.5002 $\pm$ 0.0001 & 6686.44 $\pm$ 0.04 & $-408$ $\pm$ 6 & 15.53 $\pm$ 0.14 & 73 $\pm$ 4 \\  
   & \ion{C}{iv}\tablefootmark{a} & $-$  & 8515.37 & $-$ & $<$12.07   & $-$ \\
   
 3& Ly$\alpha$  & 4.4947 $\pm$ 0.0001 & 6679.76 $\pm$ 0.05 & $-707$ $\pm$ 6  & 14.72$\pm$ 0.01 & 110 $\pm$ 3 \\ 
   & \ion{C}{iv}\tablefootmark{a} & $-$  & 8506.86  & $-$ & $<13.05$    & $-$ \\
   
 4& Ly$\alpha$  & 4.4872 $\pm$ 0.0001 & 6670.61 $\pm$ 0.09 & $-1116$ $\pm$ 8 & 14.85 $\pm$ 0.02 & 265 $\pm$ 7 \\
   & \ion{C}{iv}& 4.4872   $\pm$ 0.0006  & 8495.28 $\pm$ 0.87 & $-1114$ $\pm$ 44& 14.24 $\pm$ 0.08& 271 $\pm$ 42\\
   
 5& Ly$\alpha$ & 4.4748  $\pm$ 0.0001 & 6655.54 $\pm$ 0.12 & $-1791$ $\pm$ 9  & 14.77 $\pm$ 0.01 & 231 $\pm$ 7 \\
 
 6& Ly$\alpha$ & 4.4653 $\pm$ 0.0001 & 6644.04 $\pm$ 0.14 & $-2306$ $\pm$ 10 & 14.70 $\pm$ 0.09 & 88  $\pm$ 10 \\ 
 
 7& Ly$\alpha$ & 4.4572 $\pm$ 0.0002  & 6634.17 $\pm$ 0.19 & $-2748$ $\pm$ 13 & 14.81 $\pm$ 0.04 & 165 $\pm$ 11\\ 
 
 8& Ly$\alpha$ & 4.4462 $\pm$ 0.0002  & 6620.83 $\pm$ 0.22 & $-3345$ $\pm$ 15 & $<$15.46  & $\sim$40 \\ 
 \hline
\end{tabular}
\tablefoot{
The un-reported values and values without uncertainties are fixed parameters during the fit using the values from corresponding \ion{H}{i} absorbers. 
\tablefoottext{a}{The \ion{C}{iv} absorber \#2 and \#3 are poorly constrained (See Appendix \ref{apd:fittingnotescivheii} and \ref{apd:civheiimaster} and Fig. \ref{fig:civheiicorner} for discussion and their probability distributions). We only report the fitted column density results as upper limit.  }}
\end{table*}

\subsection{Ly\texorpdfstring{$\alpha$}{a}}\label{sec:lya}
We use a double-Gaussian model to fit and estimate the un-absorbed emission. We note that Ly$\alpha$ is a resonant line, which makes it difficult to trace the intrinsic velocity range where the photons originated from. The double-Gaussian model used is a simple implementation to fit the high emission peak with a broad wing. This two-component fitting is also applied to the \ion{C}{iv} and \ion{N}{v} but with different velocity shifts (Sects. \ref{sec:civheii} and \ref{sec:nv}). This indicates there are at least two components of gas emission with different physical origins (further discussion in Sect. \ref{sec:emissionstructure}).

We present the best-fit model of Ly$\alpha$ in Fig. \ref{fig:lyafit} upper panel with dark magenta line. In the figure, we mark the positions of eight \ion{H}{i} absorbers.
The best-fit parameters are presented in Table \ref{tab:lineemissionfit} (for emissions) and Table \ref{tab:absorberfit} (for absorption). Figure \ref{fig:lyafit} middle panel shows a $N_{\rm \ion{H}{i}}$ sensitivity test for the model of absorber \#1. We vary the column density of absorber \#1 from $\rm 10^{14}\,cm^{-2}$ to $\rm 10^{17}\,cm^{-2}$ with all other parameters fixed to the best-fit values and find that the profile is only sensitive to the $N_{\rm \ion{H}{i}}$ near the best fit value (dark magenta line shown in the figure). This test shows that the column density variation in one absorber has little influence on the others unless it is saturated.

We include further details on the Ly$\alpha$ fitting procedure in Appendix \ref{apd:lyamasterfitnotes}. The boundary conditions used for the fitting are also presented in Appendix \ref{apd:lyamasterfitnotes}. In Appendix \ref{apd:auxmcmc} we show the corner plot (Fig. \ref{fig:lyacorner}) and acceptance fraction plot (Fig. \ref{fig:lyaaccept}), which traces the correlations between each pair of fitted parameters and quality of the MCMC run, respectively. 

\citet{2019MNRAS.486.2102H} studied the contamination of \ion{O}{v}] $\lambda\lambda$1213.8,1218.3 (hereafter \ion{O}{v}]) and \ion{He}{ii} $\lambda$1215.1 emissions for high-redshift Ly$\alpha$ emitters (Type-2 quasars, HzRGs). In general, the contribution from \ion{He}{ii} $\lambda$1215.1 is insignificant while the \ion{O}{v}] emission can contribute 10\% (or more) to the Ly$\alpha$ + \ion{O}{v}]  + \ion{He}{ii} $\lambda$1215.1 flux if certain ionization parameter and metallicity are given. By using the grid model search, \citet{2019MNRAS.486.2102H} proposed a correlation between \ion{O}{v}] and \ion{N}{v,} which can be used to estimate the significance of the contamination. To test how \ion{O}{v}] will affect the \ion{H}{i} fitting result, which is the primary goal of this work, we run the fit of Ly$\alpha$ including the emission doublet of \ion{O}{v}]. In this test, the total \ion{O}{v}] flux is fixed to $2.5F_{\rm \ion{N}{v}}$ according to \citet{2019MNRAS.486.2102H} with $F_{\rm \ion{N}{v}}$ being the total fitted \ion{N}{v} flux (Sect. \ref{sec:nv}) in this work. We also fix the \textit{FWHM} of \ion{O}{v}], a nonresonant line, to the value derived from \ion{He}{ii} (Sect. \ref{sec:civheii}). The results of the 8 \ion{H}{i} absorbers, especially the $N_{\ion{H}{i}}$, are similar to the fitting results without \ion{O}{v}]. Therefore, we do not include the \ion{O}{v}] into the Ly$\alpha$ fitting in order to avoid introducing more free parameters.

   \begin{figure}
   \centering
   \includegraphics[width=\hsize]{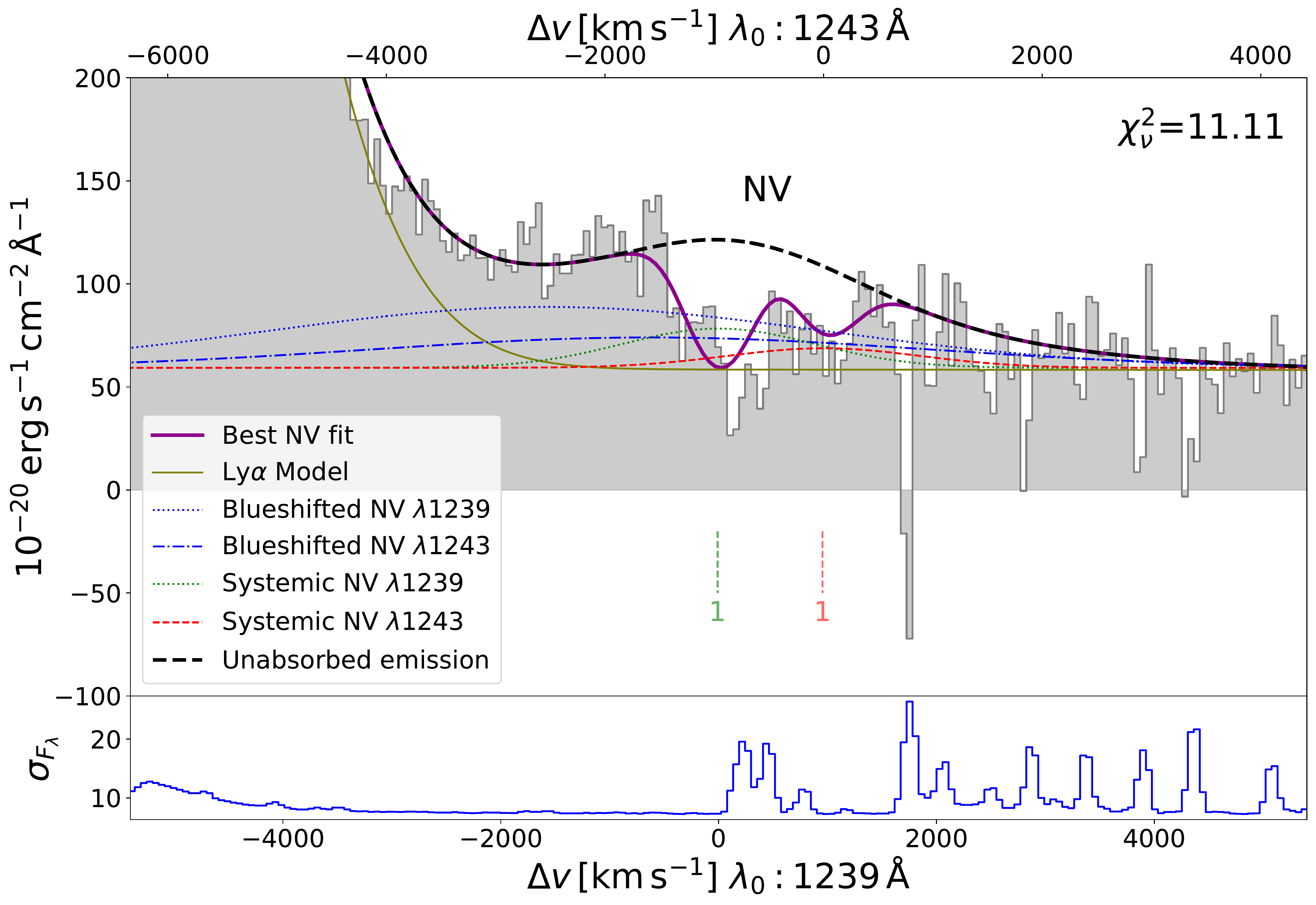}
      \caption{\ion{N}{v} best-fit model from MCMC method. The dark magenta line shows the best \ion{N}{v} fit combined with the Ly$\alpha$ from Sect. \ref{sec:lya}. The black dashed line marks all combined emissions without absorption. The systemic emissions are marked in dotted green, and dashed red curves show the doublet. The zero velocities of the systemic emissions for the doublet components are derived from the \ion{He}{ii} result. The blueshifted components are shown in dotted and dot-dash blue lines for the doublet. The solid olive line shows the Ly$\alpha$ model, which is fixed in the \ion{N}{v} fit. The short green (red) vertical bars with numbers show the positions of the absorbers on top of the \ion{N}{v}$\lambda$1239 (\ion{N}{v}$\lambda$1243) line. The dashed line style and lighter color are used to indicate that \ion{N}{v} absorber \#1 is marginally constrained (see text). The $\chi^{2}_{\nu}$ of the fit is reported to give a hint of the quality of the fit. The {\it lower panel} shows the standard deviation (noise) of the spectrum derived from the variance extension of the data cube that is used as the fitting weight. It is shown in the same units as the spectrum and can be used to trace the skylines.}\label{fig:nvfit}
   \end{figure}


   \begin{figure}
   \centering
   \includegraphics[width=\hsize]{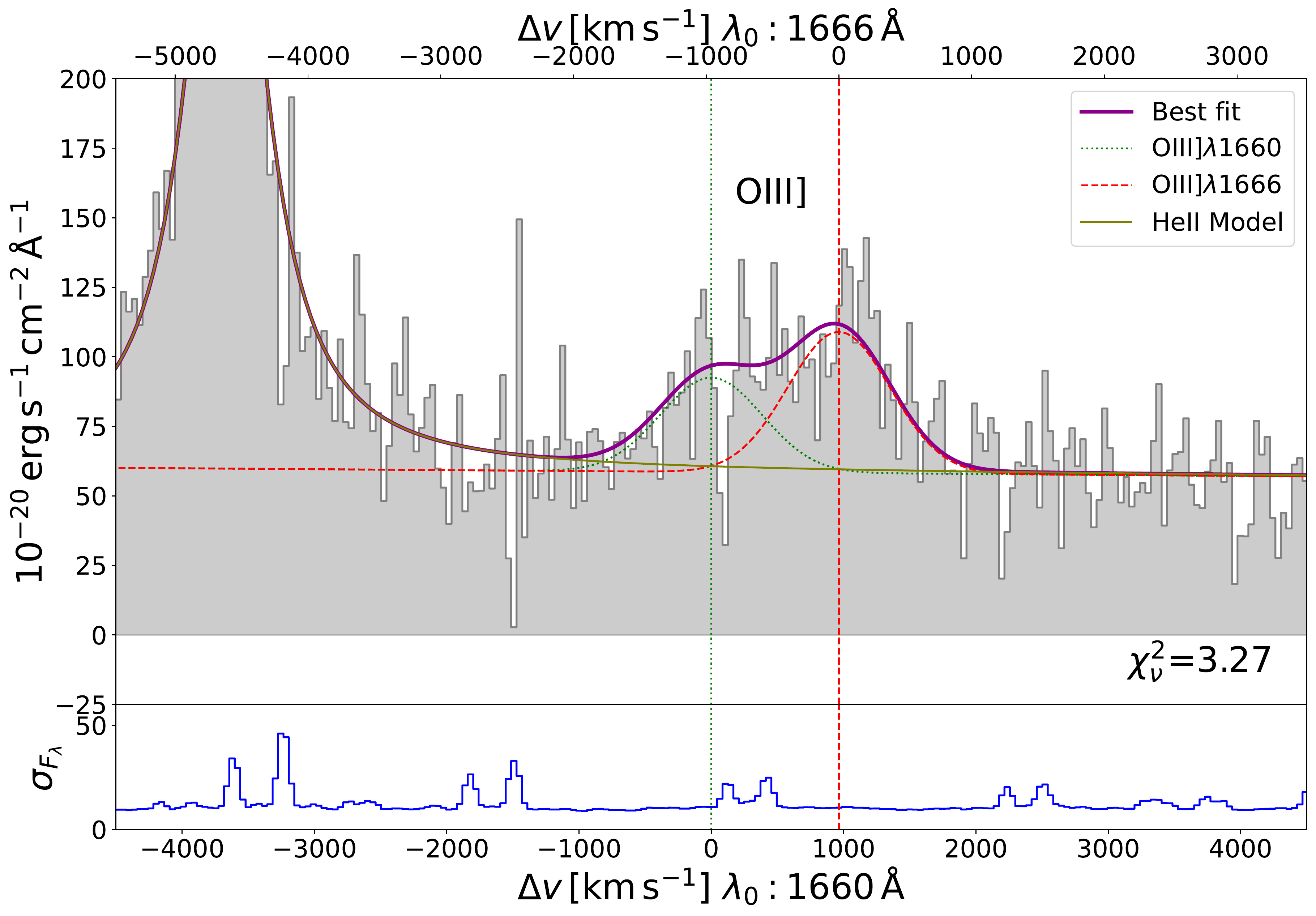}
      \caption{\ion{O}{iii}] best-fit model using the MCMC method. The dark magenta line shows the best \ion{O}{iii}] fit combined with the \ion{He}{ii} from Sect. \ref{sec:civheii}. The emissions of the \ion{O}{iii}] doublet are shown in dotted green and dashed red curves. The line centers of the systemic emissions expected for the doublet components from the \ion{He}{ii} implied redshift are shown in vertical dotted green and red lines. The solid olive line shows the \ion{He}{ii} model that is fixed in the \ion{O}{iii}] fit. The $\chi^{2}_{\nu}$ of the fit is reported to give a hint of the quality of the fit. The {\it lower panel} shows the standard deviation (noise) of the spectrum derived from the variance extension of the data cube that is used as a fitting weight. It is shown in the same units as the spectrum and can be used to trace the skylines.}\label{fig:oiiifit}
   \end{figure}


   \begin{figure*}
   \centering
   \includegraphics[width=16.4cm,clip]{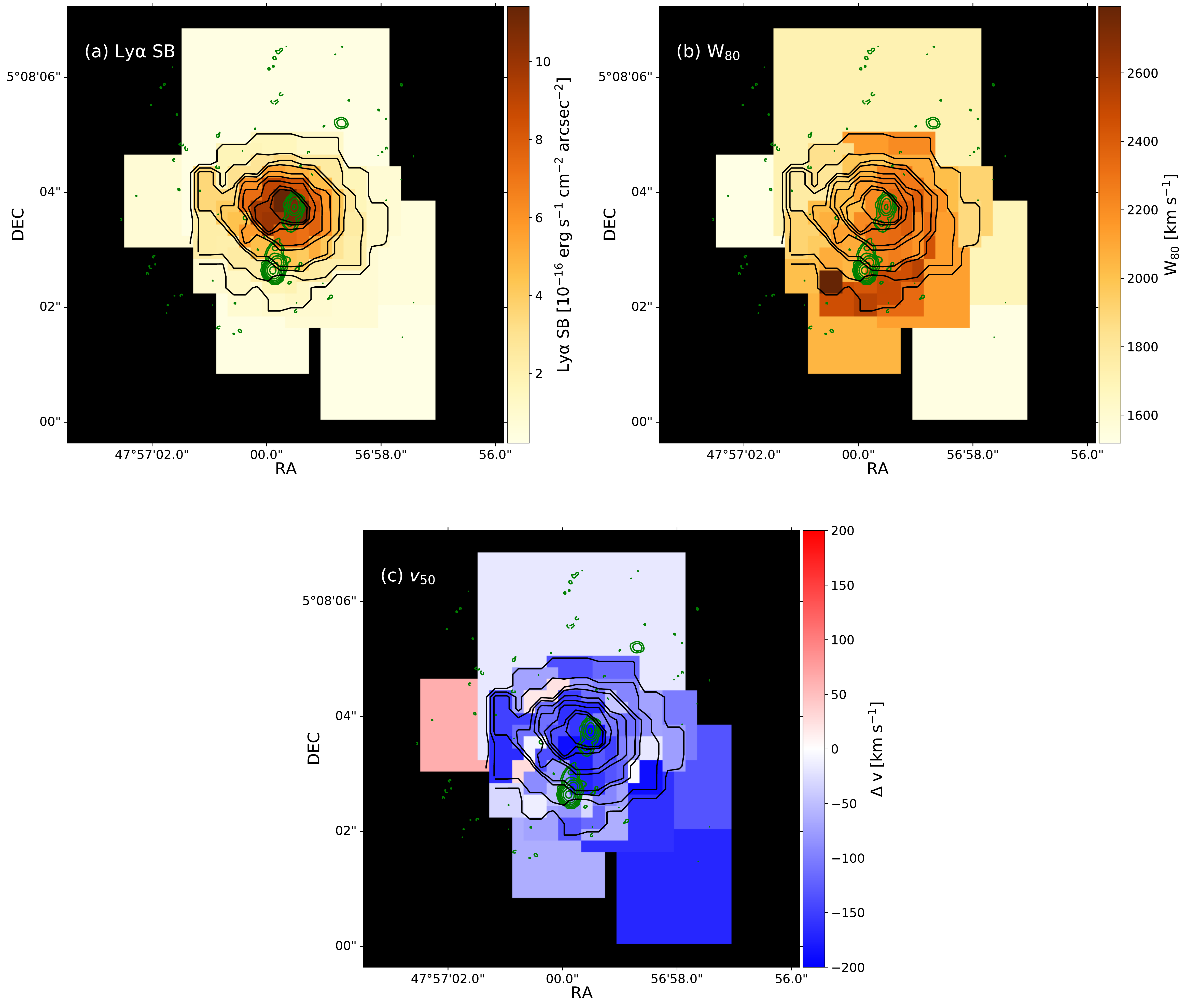}
      \caption{Spatial mapping results of Ly$\alpha$ emission. The contours are the same in all panels, with the green showing the position of the radio source \citep[see the Fig. \ref{fig:moment0} caption for details;][]{2013EAS....61..439P,2014MNRAS.439.2314P} and black tracing the Ly$\alpha$ surface brightness resulting from spatial fitting (the levels are given arbitrarily). All these maps are constructed based on the fitting results, i.e., not directly from the data. (a) Intrinsic surface brightness map  of Ly$\alpha$ emission. (b)  $W_{80}$ map of Ly$\alpha$ emission (nonparametric measurement of the line width; see text). (c)  $v_{50}$ map of the fitted Ly$\alpha$ emission profile (nonparametric measurement of the line-of-sight velocity; see text).}\label{fig:lyaspatial1}
   \end{figure*}

   \begin{figure*}
   \centering
   \includegraphics[width=16.4cm,clip]{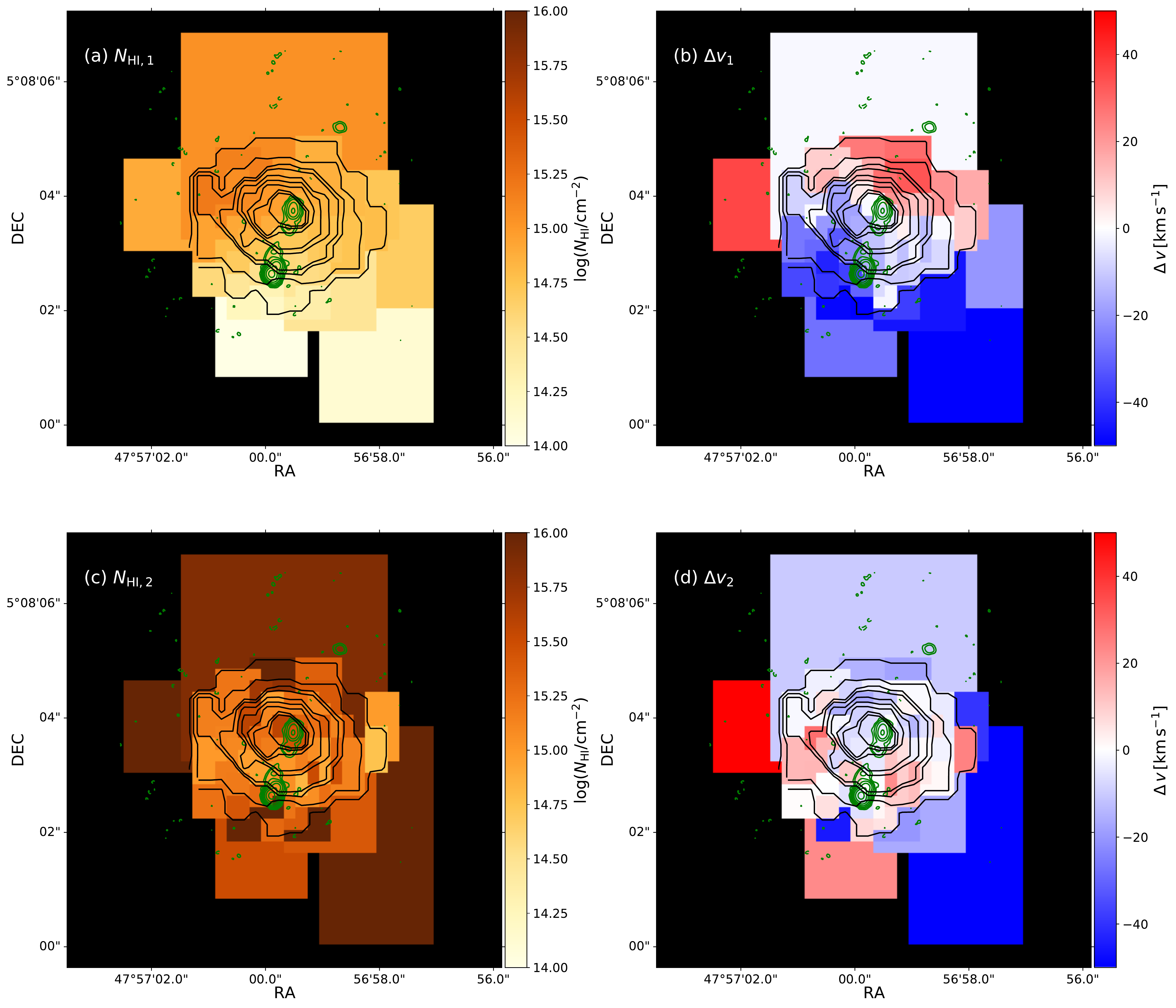}
      \caption{Column density and velocity shift maps of \ion{H}{i} absorber \#1 (panel (a) and (b)) and \#2 (panel (c) and (d)). The black contours are the same as in Fig. \ref{fig:lyaspatial1}. The zero points for the velocity shift maps are chosen individually to be the $\Delta v$ of each absorber as derived from the master spectrum fitting reported in Table \ref{tab:absorberfit}. The maps therefore show the velocity shifts relative to the redshift of the respective absorber.}\label{fig:lyaspatial2}
   \end{figure*}

   \begin{figure}
   \centering
   \includegraphics[width=\hsize]{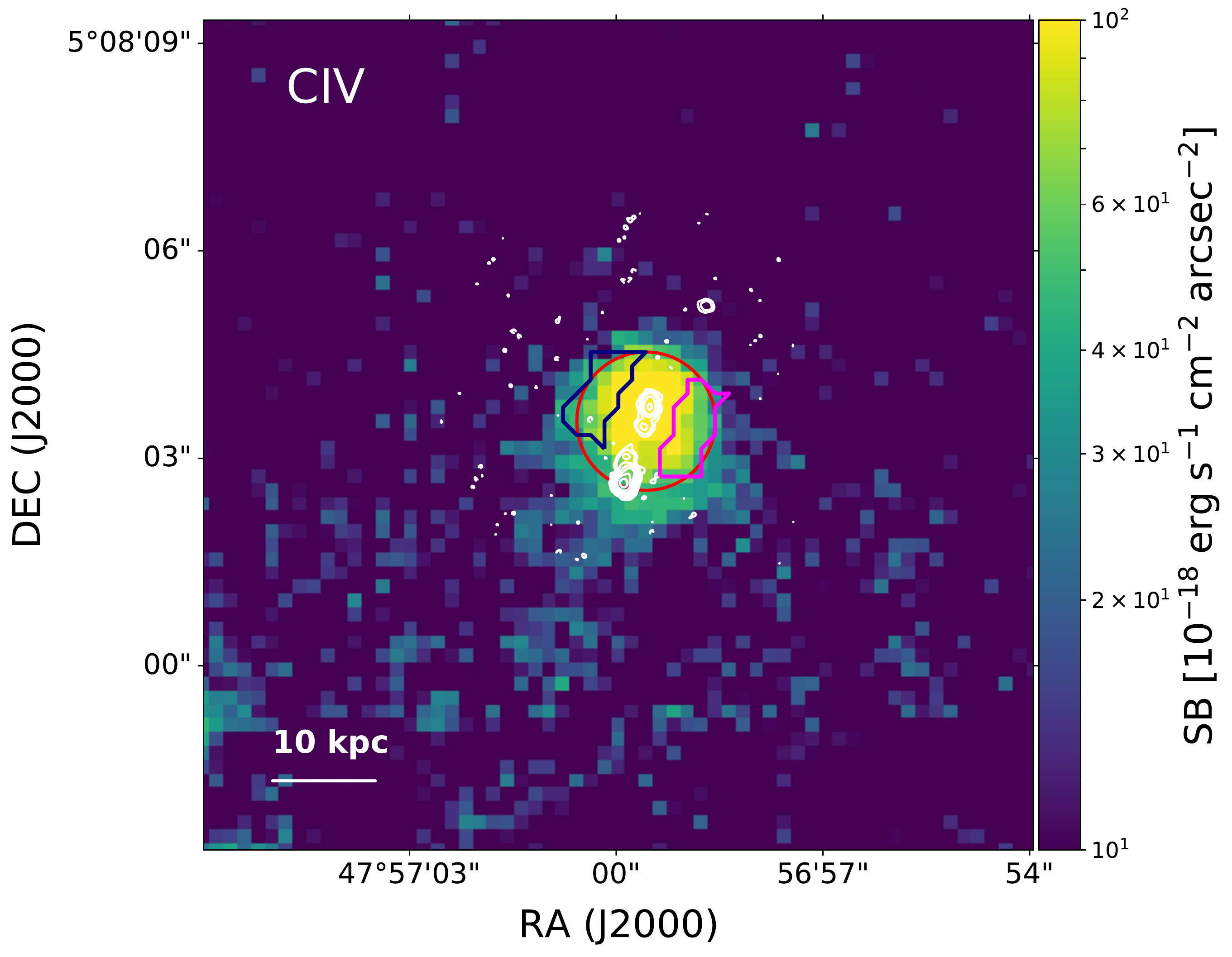}
      \caption{\ion{C}{iv} broadband image clasped between $8400-8600\,\text{\AA}$ in the observed frame. The white contour traces the radio jet, while the red circle marks the position where the spectrum analyzed in Sect. \ref{sec:fitresult} is extracted (see the caption of Fig. \ref{fig:moment0}). The dark blue and magenta regions show the apertures from which the spectra used to studied the spatial features of \ion{C}{iv} are extracted. The spectrum from the dark blue (magenta) region is marked as NE (SW). }\label{fig:civmoment0}
   \end{figure}


   \begin{figure*}
   \centering
   \includegraphics[width=16.4cm,clip]{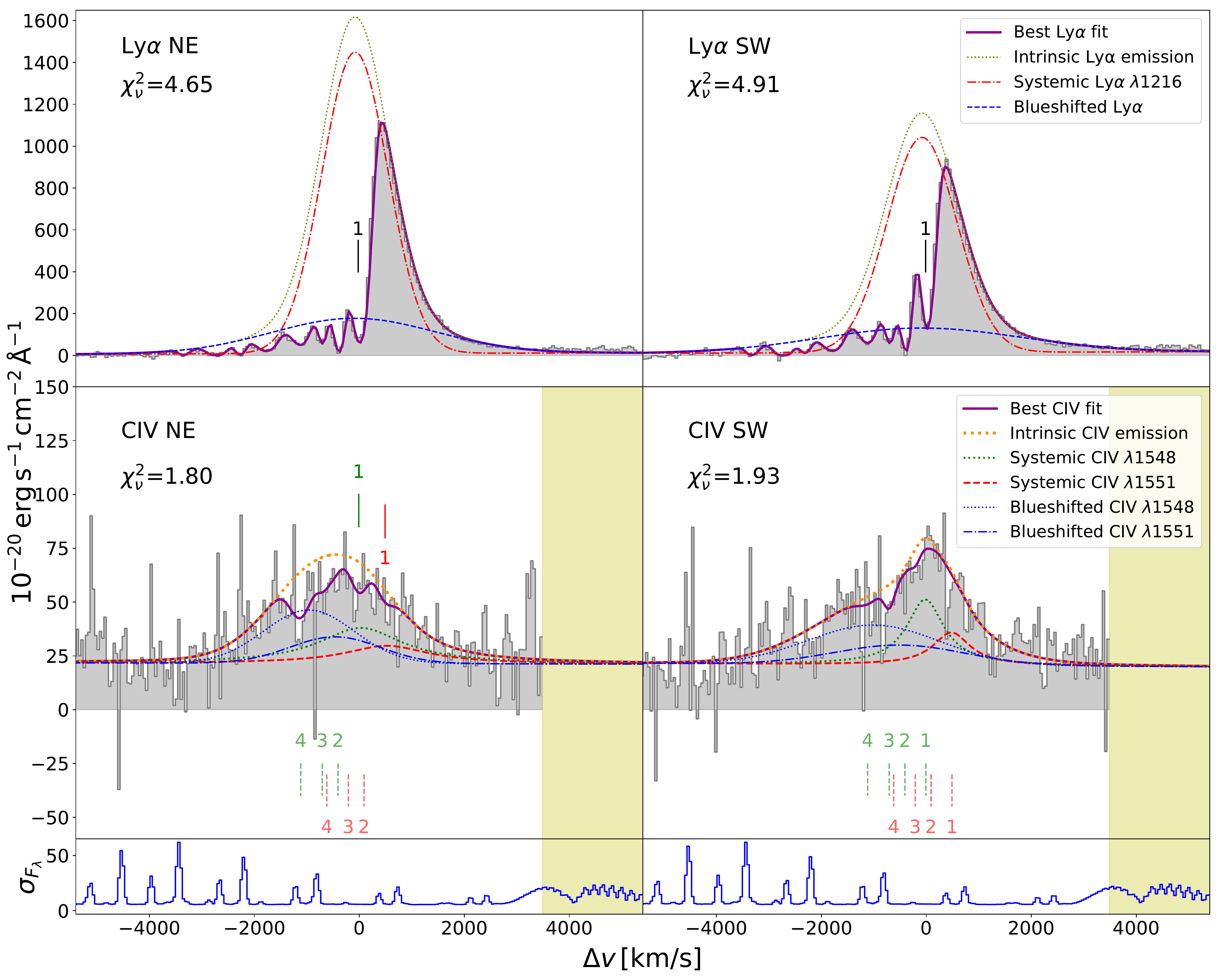}
      \caption{Spectra and fit of Ly$\alpha$ (top row) and \ion{C}{iv} (middle row) and noise spectra of \ion{C}{iv} (lower row) extracted from the dark blue (NE) and magenta (SW) regions shown in Fig. \ref{fig:civmoment0}. The positions of \ion{H}{i} absorber \#1 are marked with black bars in the top row. The velocity shift is relative to the systemic redshift, $z=4.5077$, fitted from \ion{He}{ii} (Sect. \ref{sec:civheii}). The line styles used to show the fitting results are the same ones as that of the master Ly$\alpha$ (Fig. \ref{fig:lyafit}) and \ion{C}{iv} (Fig. \ref{fig:civheiifit}). We note that except for \ion{C}{iv} absorber \#1 detected in the NE region, the positions for the other \ion{C}{iv} absorbers are marked in dashed bars, with lighter colors indicating that they are only marginally constrained (see text), which is consistent with the "master \ion{C}{iv}" presentation (Fig. \ref{fig:civheiifit}). The intrinsic \ion{C}{iv} emission is also shown in orange dotted lines for the two spectra in the middle row. The panels in the bottom row show the standard deviations (noise) of the \ion{C}{iv} spectra derived from the variance extension of the data cube, which are used as fitting weights. They are shown in the same units as the data spectra and can be used to show the quality of the spectra and trace the positions of skylines. }\label{fig:civlya2spatialNS}
   \end{figure*}


   \begin{figure}
   \centering
   \includegraphics[width=\hsize]{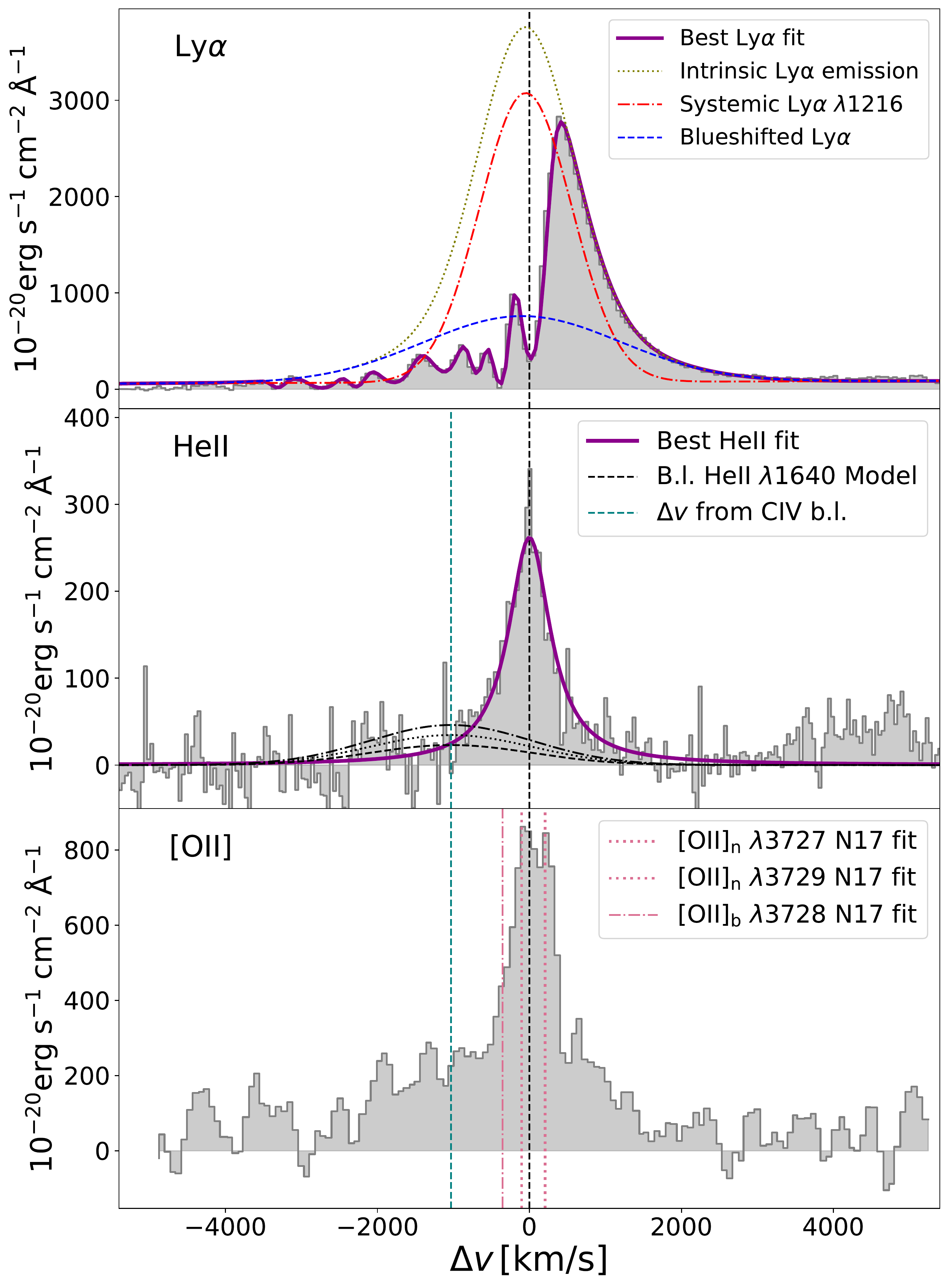}
      \caption{Comparison between the Ly$\alpha$, \ion{He}{ii,} and [\ion{O}{ii}] rest-frame spectra. The Ly$\alpha$ and \ion{He}{ii} presented here are the same ones analyzed in Sects. \ref{sec:lya} and \ref{sec:civheii} in velocity scale. We subtract the continuum from the \ion{He}{ii} here to better present the low flux region of the emission. The black lines are the same ones as shown in Fig. \ref{fig:civheiifit} for the blueshifted \ion{He}{ii} models, with the dashed, dotted, and dash-dotted lines indicating the line flux of 0.2, 0.3, and 0.4 of the total fitted flux of the blueshifted \ion{C}{iv}.  The [\ion{O}{ii}] is taken from SINFONI \citep[][]{2017A&A...599A.123N}. In each panel, the black dashed lines indicate the zero velocity. In the last two panels, the dashed vertical teal blue lines show the velocity shift of the \ion{C}{iv} blueshifted component. For Ly$\alpha$ and \ion{He}{ii}, the best fit models from this work are shown. We also mark the fitted line centers of the [\ion{O}{ii}] narrow doublet from \citet{2017A&A...599A.123N} in pink dotted lines and the broad component line center in a pink dash-dot line.}\label{fig:emissioncomparison}
   \end{figure}


\subsection{\texorpdfstring{\ion{C}{iv}}{Civ} and \texorpdfstring{\ion{He}{ii}}{HeII}}\label{sec:civheii}

The first line we focus on is \ion{He}{ii,} which is the brightest nonresonant line often used for determination of the systemic redshift of HzRGs observed in optical band \citep[e.g.,][]{2015MNRAS.449.1298S,2019A&A...625A.102K}. The nonresonant photons are produced through the cascade recombination of $\rm He^{+}$; they are not energetic enough to induce other transitions and suffer less from scattering than resonant lines (e.g., Ly$\alpha$). Previous work \citep[][]{2006AstL...32..433K} determined the redshift of \object{4C04.11} from the resonant Ly$\alpha$ line, which also heavily suffers from absorption (see Sect. \ref{sec:lya}). Hence, our fitting of the \ion{He}{ii} will provide a better estimate of the systemic redshift. 

\ion{C}{iv} and \ion{He}{ii} are located in the wavelength range that is affected by many strong skylines \citep[Fig. \ref{fig:civheiifit}, $8200-9300\,\text{\AA}$, see][for skylines observed at Paranal]{2003A&A...407.1157H}. Additionally, this wavelength range is near the edge of spectral coverage of MUSE. To obtain better results from these two low S/N lines, we have to reduce the number of free parameters used during the fitting. For this purpose, we (i) fit \ion{C}{iv} together with \ion{He}{ii} and constrain the line center of the systemic \ion{C}{iv} component with the redshift determined from \ion{He}{ii}; (ii) fix the continuum to a first-order polynomial during the emission and absorption fitting; (iii) use a Lorentzian profile for the systemic \ion{He}{ii} and \ion{C}{iv} to avoid an additional Gaussian component; (iv) include only 4 \ion{C}{iv} absorbers and fix the Doppler parameters and redshifts of absorber \#2 and \#3 (further descriptions are presented in Appendix \ref{apd:fittingnotescivheii}).  

The best-fit model of \ion{He}{ii} and \ion{C}{iv} are presented in Fig. \ref{fig:civheiifit} while fitted parameters are shown in Table \ref{tab:lineemissionfit} and \ref{fig:civheiifit} for emission and absorption (only \ion{C}{iv}), respectively. The systemic redshift calculated from the intrinsic \ion{He}{ii} emission is 4.5077 $\pm$ 0.0001, which is a significant improvement compared to \citet{2006AstL...32..433K} ($\sim10\,\text{\AA}$, in observed frame or $-1888\,\rm km\,s^{-1}$ difference of the \ion{He}{ii} center wavelength). 
We detect and report a blueshifted \ion{C}{iv} emission component (blue dot-dash and dotted line in Fig. \ref{fig:civheiifit}) with a relatively high velocity shift of $\Delta v = -1026 \pm 112 \rm \, km\, s^{-1}$. The high blueshifted velocity component is also detected in \ion{N}{v} (Sect. \ref{sec:nv}, further discussion in Sect. \ref{sec:emissionstructure}). The intrinsic \ion{C}{iv} (and \ion{He}{ii}) emission is shown in thick yellow dotted line from which it is clear that the absorption is needed to describe the line profile, especially the trough around 8500 \AA. The standard deviation derived from the data reduction is shown in the lower panel of Fig. \ref{fig:civheiifit}, which is used as weight in the fitting as well as tracer of the skylines. We excluded several regions (shaded yellow) that are affected heavily by skyline residuals during the fit. We note that there are two regions of skylines (overlap with \ion{C}{iv} absorbers \#1 and \#4) that are already given a low weight during the fit. Hence, we do not mask them in order to avoid complicating the absorption fit.

In Appendix \ref{apd:civheiimaster}, we present the corner plot (Fig. \ref{fig:civheiicorner}) and acceptance fraction plot (Fig. \ref{fig:civheiiaccept}) which traces the correlations between each pair of fitted parameters and quality of the MCMC run, respectively. From the corner plot, we notice that \ion{C}{iv} absorbers \#2 and \#3 are loosely constrained. Therefore, we only consider the column densities of absorbers \#2 and \#3 as upper limits (see Appendix \ref{apd:fittingnotescivheii} and \ref{apd:civheiimaster} for a further discussion). To visually distinguish them from the better constrained absorbers \#1 and \#4, we use the dashed bars and lighter colors for absorbers \#2 and \#3 in Fig. \ref{fig:civheiifit}. 

\citet{2017A&A...599A.123N} analyzed \object{4C04.11} with SINFONI observation \citep[the Spectrograph for INtegral Field Observations in the Near Infrared;][]{2003SPIE.4841.1548E,2004Msngr.117...17B} which reported the detection of the [\ion{O}{ii}]$\rm \lambda \lambda 3726, 3729$ ([\ion{O}{ii}]), a nonresonant line, with good S/N (Fig. \ref{fig:emissioncomparison}). The redshifts reported by  \citet{2017A&A...599A.123N} based on [\ion{O}{ii}] fitting are 4.5100 $\pm$ 0.0001 and 4.5040 $\pm$ 0.0002 for the two narrow Gaussian components used, respectively. Our fitted systemic redshift is in between these two values, which we consider to be reasonable and consistent with the near infrared observation (see Fig. \ref{fig:emissioncomparison}). In addition, the authors detect and include a broad blueshifted component ($\Delta v \simeq -240\,\rm km\,s^{-1}$, $FWHM \simeq 1400\,\rm km\,s^{-1}$). We use the Lorentzian profile for the systemic \ion{He}{ii} because the S/N in this wavelength range of our data is not enough to constrain the fitting with two Gaussian (Appendix \ref{apd:fittingnotescivheii}). We further discuss this in Sect. \ref{sec:emissionstructure}.

The blue wing of \ion{He}{ii} is too noisy to constrain whether there is a blueshifted component. We present three emission models of the blueshifted \ion{He}{ii} in Fig. \ref{fig:civheiifit} (black dashed, dotted and dash-dotted lines) with velocity shift and \textit{FWHM} fixed to the blueshifted component of \ion{C}{iv}. The line flux of this components are set to be 0.2 (dashed), 0.3 (dotted) and 0.4 (dash-dotted) of the total fitted flux of the blueshifted \ion{C}{iv}. From this we can estimate a lower limit of $F_{\rm \ion{C}{iv},b.l.}/F_{\rm \ion{He}{ii},b.l.} \gtrsim 3.3$. We discuss this result further in Sect. \ref{sec:emissionstructure}.

\subsection{\texorpdfstring{\ion{N}{v}}{Nv}} \label{sec:nv}
To fit the low S/N \ion{N}{v} on top of the broad wing of Ly$\alpha$ and relatively high continuum, we fix the Ly$\alpha$ to the one derived in Sect. \ref{sec:lya} and use a constant continuum during the \ion{N}{v} fitting. The fitting procedure is then carried out following Sect. \ref{sec:fitprocedure} (see Appendix \ref{apd:fittingnotesnv} for details of the \ion{N}{v} fitting ).

We show the best-fit \ion{N}{v} model in Fig. \ref{fig:nvfit} and the fitted parameters in Table \ref{tab:lineemissionfit} (emissions) and \ref{tab:absorberfit} (absorption). The blueshifted emission component is at $\sim -1587 \,\rm km\,s^{-1}$ which is consistent with the \ion{C}{iv} blueshifted component \footnote{Though this reported velocity shift is bluer than the one of \ion{C}{iv} blueshifted component taking uncertainty into account, the value of $\sim -1500\,\rm km\,s^{-1}$ will also give \ion{C}{iv} a good fit. See Appendix \ref{apd:fittingnotesnv} for more details.}. The large value of Doppler parameter, $b \simeq \rm 387\, km\,s^{-1}$, could be due to unresolved redshifted \ion{H}{i} absorber(s) and/or it is influenced by the skyline subtraction. However, we cannot constrain more without deeper and higher-resolution data. We remind the readers that this fit is limited by the low S/N of the data and depends strongly on the Ly$\alpha$ broad wing and it should be treated with caution. In Fig. \ref{fig:nvfit}, the positions of the marginally constrained absorber are shown. Given the degeneracy between $b$ and $N$ \citep[e.g.,][]{2018MNRAS.474.3649S}, the $N_{\ion{N}{v},1}$ should be treated as lower limit. The black dashed line in Fig. \ref{fig:nvfit} shows the combined emission structures from all sources (lines and continuum) without absorption. It is clear that at least \ion{N}{v} absorber \#1 is necessary to fit the data. We note the presence of skylines overlapping with \ion{N}{v} (lower panel in Fig. \ref{fig:nvfit}), which are already given a low weight in the fitting. Hence, we do not mask them in order to avoid complicating the absorption fitting. We further discuss the interpretation of the emission and absorption results in Sects. \ref{sec:emissionstructure}  and \ref{sec:metalabsorbers}, respectively.

In Appendix \ref{apd:nvmaster} we present the corner plot (Fig. \ref{fig:nvcorner}) and acceptance fraction plot (Fig. \ref{fig:nvaccept}), which traces the correlations between each pair of fitted parameters and quality of the MCMC run, respectively.

\subsection{\texorpdfstring{\ion{O}{iii}}{OIII}]} \label{sec:oiii}
For 4C04.11, the \ion{O}{iii}] doublet is detected. Although the \ion{O}{iii}] is near the \ion{He}{ii}, we fit them separately in order to avoid introducing more free parameters into the \ion{C}{iv} + \ion{He}{ii} fit, which is one of the major focuses of this work. The fit is preformed following Sect. \ref{sec:fitprocedure} with the line centers and underlying continuum fixed to the systemic redshift implied from \ion{He}{ii} (Sect. \ref{sec:civheii}) and to the model derived in Sect. \ref{sec:civheii}, respectively. We present the result of the \ion{O}{iii}] fitting in Fig. \ref{fig:oiiifit} and Table \ref{tab:lineemissionfit}. The corner plot and the acceptance rate are shown in Appendix \ref{apd:o3master}.

\section{Spatial mapping}\label{sec:spatialmapresults}

In this section we present the spatial mapping results for Ly$\alpha$ and \ion{C}{iv}. The Ly$\alpha$ emission is analyzed by following the method described in Sect. \ref{sec:spatialmethod} and can be studied in detail in both emission and absorption. As mentioned in Sect. \ref{sec:spatialmethod}, \ion{C}{iv} is detected at low S/N and its quality suffers from skyline contamination. For \ion{C}{iv}, we therefore only focus on the results from two larger spatial regions in Sect. \ref{sec:civspatial}. It is impossible to fit the \ion{N}{v} spatially due to its extremely low S/N even in the master spectrum.

\subsection{Spatially resolved Ly\texorpdfstring{$\alpha$}{a} signatures}\label{sec:lyaspatial}
We first present the morphological and kinematic features of Ly$\alpha$ emission derived from the spatially resolved fitting analysis (Fig. \ref{fig:lyaspatial1}). In each panel, we show the measured parameters in 64 spatial bins identified through our binning method in Sect. \ref{sec:spatialmethod}. In Fig. \ref{fig:lyaspatial1}a, we show the intrinsic Ly$\alpha$ surface brightness (SB) map. It is important to note that this shows the integrated Ly$\alpha$ flux derived from the Gaussian emission model (summation of the two components; see Sect.Sect. \ref{sec:fitprocedure}), after correction for the \ion{H}{i} absorption. The extended emission to the north, encompassing the northern jet hotspot, is due to the large size of the bin (see Fig. \ref{fig:binum_map}, bin 59). The position of the SB peak coincides with the radio core (central green contours). In all panels of Fig. \ref{fig:lyaspatial1} and \ref{fig:lyaspatial2}, we overplot this intrinsic Ly$\alpha$ SB as black contours. In Fig. \ref{fig:lyaspatial1}b, we present the $W_{80}$ map generated from the unabsorbed Ly$\alpha$ emission. $W_{80}$ is a nonparametric measurement of the velocity width of emission lines \citep[e.g.,][]{2013MNRAS.436.2576L}. We notice the $W_{80}$ peaks close to the southern jet hotspot, which is likely the approaching jet because of its clumpier morphology, which in turn could be caused by Doppler beaming \citep[][]{2014MNRAS.439.2314P}. While we consider that result as tentative, it may be a signature of jet-gas interactions \citep[e.g.,][]{2006MNRAS.369.1103H,2017A&A...599A.123N}. We present the $v_{50}$ map in Fig. \ref{fig:lyaspatial1}c, which is a nonparametric measurement of the velocity shift of the emission profile \citep[e.g.,][]{2013MNRAS.436.2576L} independent of interpreting the individual Gaussian components added to the fit. The result suggests the existence of blueshifted Ly$\alpha$ emission. However, since the map is based on fitting and estimating the intrinsic, unabsorbed Ly$\alpha$ emission (i.e., it is not fully nonparametric), we do not interpret it further.

Figure \ref{fig:lyaspatial2} shows the column density and velocity shift maps for absorbers \#1 and \#2, which are the two prominent absorbers detected in every spatial bin suggesting high areal fractions. We note that we show here the velocity shift with respect to the mean velocity of the respective absorber $\Delta v$ as derived from the from the master spectrum (see Table \ref{tab:absorberfit}).
In Fig. \ref{fig:lyaspatial2}a, we identify a column density gradient in absorber \#1 from southwest (SW) to northeast (NE), which is roughly in the perpendicular direction to the radio jet axis with an increasing of 1 dex in 24 kpc. We consider this as a robust detection after checking the associated uncertainties in each bin. In Fig. \ref{fig:lyaspatial2}b, we identify a small velocity gradient for absorber \#1 along the direction of the jet increasing from $\sim-50\,\rm km\,s^{-1}$ in the southeast (SE) to $\sim35\,\rm km\,s^{-1}$ in the northwest (NW) in 20 kpc. In Sect. \ref{sec:hi1spatial}, we discuss possible explanations for these observations. 

We do not observe such gradients or spatial variations in the column density and velocity shift maps of absorber \#2 in Fig. \ref{fig:lyaspatial2}c and \ref{fig:lyaspatial2}d. We note that the fitting uncertainties for absorber \#2 are larger compared to those for absorber \#1, such that any small variations would not show up in our analysis. 
This also demonstrates that our observations only provide us with enough sensitivity to study the spatial properties of \ion{H}{i} absorber \#1. Hence, we do not show the maps for absorbers \#3$-$8. During the analysis of the spatial properties of the \ion{C}{iv} absorbers, we also extract and fit the Ly$\alpha$ spectra from the two spatial apertures (see Sect. \ref{sec:civspatial} for details) that partly constrain the high-velocity shift of the \ion{H}{i} absorbers at different positions. The results of the fitted parameters are presented in Table \ref{tab:NESWabsorption}. Though the S/N is low and some absorbers are only partially constrained, we do not observe any significant changes in column density for absorber \#3-8 in the two regions. We notice that we do not observe any strong velocity gradients for any of the absorbers. We therefore exclude the possibility of absorber spatial blending, that is, absorbers identified at the same wavelength position could not be different absorbers in different spatial bins.

In Appendix \ref{apd:lyaspatialspectra}, Figs. \ref{fig:lyaindividualspatialplots1} $-$ \ref{fig:lyaindividualspatialplots4}, we present the fitting results of 64 individual Ly$\alpha$ spectra from the 64 spatial bins.

\subsection{Spatially resolved \texorpdfstring{\ion{C}{iv}}{Civ} signatures}\label{sec:civspatial}
We present the spatial analysis of \ion{C}{iv} in this section. As mentioned in Sect. \ref{sec:spatialmethod}, the S/N of \ion{C}{iv} makes it difficult to study its spatial variations in as much detail as Ly$\alpha$. However, since we observe a column density gradient of \ion{H}{i} absorber \#1 (Sect. \ref{sec:lyaspatial}), it is worthwhile to investigate whether the \ion{C}{iv} shows similar features. We manually set two regions from which we extract spectra (Fig. \ref{fig:civmoment0}): NE where the column density of \ion{H}{i} absorber \#1 is higher and SW where the column density of \ion{H}{i} absorber \#1 is lower. When selecting the apertures, we keep the same number of spaxels (30 spaxels or 1.2 arcsec$^{2}$) in these two regions and avoid the impact of the jet. Most of the spaxels in these two regions are covered in the master aperture (red circle in Fig. \ref{fig:civmoment0}) in order to be consistent with the 1D spectrum analysis.

For the spectral fitting, we follow the similar strategy described in Sects. \ref{sec:fitprocedure} and \ref{sec:civheii}. The fitting results from these two regions are presented in Table \ref{tab:NESWemission} for emissions and Table \ref{tab:NESWabsorption} for absorption. In Fig. \ref{fig:civlya2spatialNS}, we show the best-fit models of the two \ion{C}{iv} lines. The intrinsic \ion{C}{iv} emission shown in the figure indicates that the absorption is indeed needed to better describe the line profile, especially in the NE. The quality of the \ion{C}{iv} fits is affected by their low S/N partly due to smaller aperture from which the spectra are extracted and the influence of skylines. 
To avoid over-fitting, we fix the Doppler parameters, $b$, and redshifts, $z$, of all absorbers in the two regions and refer to the column density results as upper limits. The exception is the column density of \ion{C}{iv} absorber \#1 in the NE region, which has a well-defined probability distribution and is considered a detection (See Appendix \ref{apd:civspatialresults} and Fig. \ref{fig:CIVNESWdis} for more details). In Fig. \ref{fig:civlya2spatialNS}, we mark the positions for the un-constrained absorbers in dashed bars with lighter colors to visually distinguish them from absorber \#1 in the NE region.

In addition, we extract the Ly$\alpha$ spectra from these two regions and perform the fitting analysis (Fig. \ref{fig:civlya2spatialNS}) with the goal to compare the column density ratio of the \ion{C}{iv} and Ly$\alpha$ absorber \#1 (results shown in Table \ref{tab:NESWemission} and Table \ref{tab:NESWabsorption}).  We measure $N_{\ion{C}{iv},\,{\rm NE}}/N_{\ion{H}{i},\,{\rm NE}} = 0.11 \pm 0.04$, and $N_{\ion{C}{iv},\,{\rm SW}}/N_{\ion{H}{i},\,{\rm SW}} <0.04$ and further interpret this result in Sect. \ref{sec:discussioncivspatial}.

We also perform the similar analysis for another two regions along the radio jet for completeness (not shown in Fig. \ref{fig:civmoment0}). The two regions are chosen to be the similar as the previous dark blue and magenta ones but rotated 90$^{\circ}$ clockwise with respect to their geometric center. The column density variation in  \ion{C}{iv} absorber \#1 is also tentatively identified along this direction as well as the \ion{H}{i}-\ion{C}{iv} ratio (SE-NW) with NW region having a higher value. Specifically, the \ion{C}{iv} absorber \#1 is only marginally fitted in the SE region with its result can only be used as upper limit. We also check the velocity shift of the \ion{H}{i} absorber \#1 in the two regions (SE-NW) along the radio axis and confirm the gradient observed in Fig. \ref{fig:lyaspatial2}. 

 \section{Discussion}\label{sec:discussion}

\subsection{Emission line properties}\label{sec:emissionstructure}

Emission line fluxes, flux ratios and spatial locations of individual kinematic components provide powerful diagnostics of gas properties and ionization source. In this work, we detect five UV lines, namely Ly$\alpha$, \ion{N}{v}, \ion{C}{iv}, \ion{He}{ii} and \ion{O}{iii}]. Ly$\alpha$ is a resonant line that suffers heavily from scattering, making it difficult for us to trace its intrinsic velocity structures \citep[e.g.,][]{2014PASA...31...40D}. Although we detect a blueshifted broad component, we refrain from assigning it a physical meaning and do not to compare it with the blueshifted components seen in \ion{N}{v} and \ion{C}{iv}. 

\subsubsection{Emission line characteristics}

We first compare the emission line properties for Ly$\alpha$ and \ion{He}{ii} detected with MUSE in this work and the [\ion{O}{ii}] from SINFONI \citep[][]{2017A&A...599A.123N}. In Fig. \ref{fig:emissioncomparison}, all lines are shown within the velocity range where the zero point is set by the systemic redshift derived from the \ion{He}{ii} fit. For Ly$\alpha$ and \ion{He}{ii}, their best fits from this work are shown. In addition, we include the fitted line centers of the narrow component of the [\ion{O}{ii}] doublet (pink dotted lines) and the broad component (pink dot-dash line) \citep[][]{2017A&A...599A.123N}, respectively. We note that the wavelength calibration for SINFONI is done using the vacuum wavelength while MUSE uses air wavelengths. To eliminate this discrepancy, we apply the equation from \citet{2000ApJS..130..403M} to convert all wavelengths into air wavelengths \footnote{$\lambda_{\rm air} = \lambda_{\rm vac}/n$, where $n=1+8.34254  \times 10^{-5} + 2.406147\times 10^{-2}/(130 - s^{2})+1.5998\times10^{-4}/(38.9-s^{2})$, $s=10^{4}/\lambda_{{\rm vac}}$ and $\lambda_{{\rm vac}}$ in the unit of \AA}. We did not correct the difference in wavelength due to the heliocentric frame used in SINFONI observation, which is $\sim$  $30\,\rm km \, s^{-1}$. 

In the \ion{He}{ii} panel, we show again the three blueshifted models with the line center fixed to the value obtained from \ion{C}{iv} fit as in Fig. \ref{fig:civheiifit}. The velocity of the \ion{C}{iv} blueshifted component is indicated by the vertical dashed blue line in \ion{He}{ii} and [\ion{O}{ii}] panels. From this figure, we conclude that the broad blueshifted component observed in [\ion{O}{ii}] ($\Delta v \simeq -240\,\rm km\,s^{-1}$, $FWHM \simeq 1400\,\rm km\,s^{-1}$) is not seen in \ion{He}{ii}. Though affected by resonant scattering, the blueshifted Ly$\alpha$ component is consistent with the blueshifted component seen in [\ion{O}{ii}] and they both may trace emission from the same potential outflow. The high-velocity blueshifted component (as seen in \ion{C}{iv}), however, is possibly also present in [\ion{O}{ii}]. We discuss this in Sect. \ref{sec:lineratio} together with \ion{N}{v} and \ion{C}{iv}.  
The marginally detected continuum in \citet{2017A&A...599A.123N} is consistent with our MUSE observation.  


\subsubsection{Emission line ratios and sources of ionization}\label{sec:lineratio}

We next investigate the emission line flux ratios for the individual kinematic components that we observe for \ion{N}{v}, \ion{C}{iv,} and \ion{He}{ii} in order to determine the ionization mechanism. In Sect. \ref{sec:fitresult}, we report the fitted intrinsic emission line fluxes of these three lines. The derived flux ratios are presented in Table \ref{tab:uvlinediag}. For \ion{N}{v} and \ion{C}{iv}, the flux ratio between their systemic emission line components is $F_{\rm \ion{N}{v},sys}/F_{\rm \ion{C}{iv},sys} = 0.32 \pm 0.09$ \citep[which is comparable to other HzRGs; e.g.,][]{2000A&A...362..519D}. The ratio between the systemic \ion{C}{iv} and \ion{He}{ii} components is $F_{\rm \ion{C}{iv},sys}/F_{\rm \ion{He}{ii}} = 0.55 \pm 0.14$. 

For the blueshifted components, the velocity shifts (with respect to the zero point set by the systemic \ion{He}{ii}) of the \ion{C}{iv} ($-1026 \pm 112\, \rm km\,s^{-1}$) and \ion{N}{v} ($\sim -1587\,\rm km\,s^{-1}$) are roughly consistent and we therefore assume that they are tracing the same kinematic component of the gas (more detail on Appendix \ref{apd:fittingnotesnv}). The flux ratio between the blueshifted components has a value of  $F_{\rm \ion{N}{v},b.l.}/F_{\rm \ion{C}{iv},b.l.} \simeq 0.7$. We do not clearly observe a blueshifted component in \ion{He}{ii}. This is a somewhat different situation compared to observations in other HzRGs. For example, in \object{MRC 0943-242} \citep[a HzRG in our MUSE+ALMA sample,][]{2019A&A...625A.102K} a blueshifted component is observed in \ion{C}{iv} ($E_{\rm\ion{C}{iv}}=64.5\,\rm eV$) and \ion{He}{ii} but not \ion{N}{v}. Nevertheless, in order to constrain its flux, we plot three models of the blueshifted \ion{He}{ii} with velocity shift and \textit{FWHM} fixed to the values of blueshifted \ion{C}{iv} and having flux 0.2, 0.3 and 0.4 of $F_{\rm \ion{C}{iv},b.l.}$ (Sect. \ref{sec:civheii}, Fig. \ref{fig:civheiifit}). From this, we can set a lower limit of $F_{\rm \ion{C}{iv},b.l.}/F_{\rm \ion{He}{ii},b.l.}\gtrsim 3.3$.

\begin{table}
\caption{Line flux ratios and equivalent width.}\label{tab:uvlinediag}
\centering
\begin{tabular}{lcccc}
\hline\hline 
&$F_{\rm \ion{C}{iv}}/F_{\rm \ion{He}{ii}}$&$F_{\rm \ion{N}{v}}/F_{\rm \ion{C}{iv}}$&$F_{\rm \ion{N}{v}}/F_{\rm \ion{He}{ii}}$ &EW [\AA]\\
\hline
sys.& 0.55 $\pm$ 0.14 & 0.32 $\pm$ 0.09 & 0.17$\pm 0.02$& $\sim$ 12 \\
b.l.& $\gtrsim 3.3$ & 0.7 $\pm$ 0.1 &$\gtrsim 2$&$\sim$ 26 \\
\hline
\end{tabular}
\tablefoot{ The line flux of \ion{C}{iv} and \ion{N}{v} are summations of the doublet. The sys. and b.l. indicate that the derived values are for the systemic or blueshifted, respectively. We note that the EW derived should be treated as lower limit as the continuum level may be overestimated.}
\end{table}

\citet{Feltre_2016} presents emission-line diagnostics at ultraviolet wavelengths of photoionization models of active and inactive galaxies with the aim is to identify new line-ratio diagnostics to discriminate between gas photoionization by AGN and star formation. According to their models \citep[Fig. 5 and 7 in ][]{Feltre_2016} the ionization source for the systemic kinematic component that we observe in \ion{N}{v}, \ion{C}{iv} and \ion{He}{ii} (Table \ref{tab:uvlinediag}) is consistent with photoionization from an AGN, though the \ion{C}{iii}] data are unavailable. This is also consistent with the diagnostic from \citet{2018A&A...612A..94N}, which involves the equivalent width of \ion{C}{iv} (EW(\ion{C}{iv}$_{\rm sys})\simeq\,$12\,\AA) and $F_{\rm \ion{C}{iv}}/F_{\rm \ion{He}{ii}}$.  As for the ionization source of the blueshifted component, the diagnostic from \citet{Feltre_2016} indicate it to be due to star formation only with our derived upper limit of $F_{\rm \ion{C}{iv},b.l.}/F_{\rm \ion{He}{ii},b.l.}$. Using EW$(\rm \ion{C}{iv}_{b.l.}) \simeq\,$26\,\AA\, and $F_{\rm \ion{C}{iv},b.l.}/F_{\rm \ion{He}{ii},b.l.}\gtrsim3.3$ and comparing to Fig. 11 in \citet{2018A&A...612A..94N}, the diagnostics are consistent with the region where ionization from both AGN and star formation are possible. This is surprising given even extreme star formation processes are unlikely to drive such a high-velocity outflow \citep[see][]{Heckman_2015}. 

High-velocity shocks (due to the radio jets) may be another possible solution to explain the blueshifted emission line component. \citet{1995ApJ...455..468D,1996ApJS..102..161D} modeled the shock ionization process and provided spectral line diagnostics that can be applied to narrow line regions (NLRs) of AGN. \citet{2000A&A...362..519D,2008MNRAS.383...11H} used these models to analyze samples of HzRGs and suggested some limitations of these models. \citet{2008ApJS..178...20A} extended the \citet{1995ApJ...455..468D,1996ApJS..102..161D} models to embrace larger parameter ranges. Due to the limited number of available spectral lines for 4C04.11 (e.g., lacking useful diagnostic lines [\ion{O}{iii}]$\lambda\lambda$4959, 5007, \ion{C}{iii}]$\lambda\lambda$1906, 1908 and \ion{C}{ii}]$\lambda$2326), we cannot draw strong conclusions on shock ionization scenarios. Nevertheless, with the inferred high $F_{\rm \ion{C}{iv},b.l.}/F_{\rm \ion{He}{ii},b.l.}$ and $F_{\rm \ion{N}{v},b.l.}/F_{\rm \ion{He}{ii},b.l.}$ (Table \ref{tab:uvlinediag}), the blueshifted emission is not inconsistent with being due to shocks. This is also consistent with [\ion{O}{ii}] if the flux excess seen at $\sim -1000\,\rm km\,s^{-1}$ (Fig. \ref{fig:emissioncomparison}) comes from the same gaseous component with \ion{C}{iv} and \ion{N}{v}.

We remind the reader that the uncertainties associated with our flux and flux ratio measurements are non-negligible and deeper data are needed to investigate the true nature of the individual gaseous components. Our observations nevertheless indicate that the blueshifted kinematic component observed in \ion{N}{v} and \ion{C}{iv} traces a metal-enriched (see Sect. \ref{sec:absorptionfrommetallines}) gaseous outflow within the ISM of \object{4C04.11} that is distinct in both kinematics and ionization mechanism from the systemic component.



We further investigate the differences between the blueshifted- and systemic components by assessing their respective spatial locations. Usually, the broad component will have compact (often un-resolved) spatial distribution if it is AGN-driven. 
We compare the spatial locations of these two components from pseudo-narrowband images of \ion{C}{iv} focused on its blue wing (8400$-$8500\AA, $-4464<\Delta v< -948\,\rm km\,s^{-1}$) and on its red wing (8500$-$8600\AA, $-948<\Delta v< 2567\,\rm km\,s^{-1}$), respectively. The S/N of the \ion{N}{v} is too low to preform this check. We do not observe a significant spatial difference as the two components are located around the center of the Ly$\alpha$ SB peak with a extension of $\sim3$ arcsec ($\sim$2 for the blue wing), which is larger than the seeing element. 
The large detected line widths of the blueshifted components (Table \ref{tab:lineemissionfit}) could also represent a set of individual clouds that are not spatially nor spectrally resolved in our data leaving the possibility open for the AGN being the primary ionization source.  



\subsection{\texorpdfstring{\ion{H}{i}}{HI} absorbers}\label{sec:otherHIabsorbers}

When fitting the absorption features in Sect. \ref{sec:fitprocedure}, we work with the assumption where several extended screens of gas are responsible for the absorption troughs. This assumption is justified as we coherently observe the signatures of \ion{H}{i} absorbers \# 1, 2, 3, and 4 across large spatial scales, which indicates large areal fractions. The spatial extent for absorbers \#1 and 2 is $\sim30\times30\,\rm kpc^{2}$ (Fig. \ref{fig:lyaspatial2}) and $\sim16\times16\,\rm kpc^{2}$ for \#3 and 4 whose maps are not shown in this paper due to their low S/N. For clarification, the presence of absorber \#3 and 4 (and further) cannot be obviously identified in the tessellation bins 50$-$64 (see Figs. \ref{fig:binum_map}$-$\ref{fig:lyaindividualspatialplots4}) based on which their spatial extent is determined. The screens may be part of a shell similar to the shell models proposed by many theoretical works, for example \citet{2006A&A...460..397V} and \citet{2015ApJ...812..123G}. 
For absorber \#1, with the highest S/N in our data, we furthermore observe a significant column density and velocity gradient (Sect. \ref{sec:lyaspatial}), which we discuss separately in Sect. \ref{sec:hi1spatial}. However, our observations are not sensitive enough to probe the spatial (morphological and kinematic) details of absorber \#2 (Sect. \ref{sec:hi1spatial}) or any of the higher-velocity absorbers.  

As for absorbers \#5$-$8 (which have velocity shifts with respect to the systemic redshift of $-1791\pm9$, $-2306\pm 10$, $-2748\pm 13$ and $-3348 \pm 15\,\rm km\,s^{-1}$, respectively), their spatial distributions are difficult to identify since they are located in the blue, low S/N wing of Ly$\alpha$ and are therefore only observed in the high surface brightness regions of Ly$\alpha$ close to the center of the host galaxy. They may have a larger spatial extent but this cannot be constrained without deeper observations. Additionally, we note that there is a large velocity shift difference between \ion{H}{i} absorber \#4 and \#5, $\sim 600\rm \,km\,s^{-1}$. Absorbers \#5$-$8 are therefore likely intervening absorbers between the radio galaxy and the observer beyond the galaxy potential well. The reason that many of these intervening absorbers have large $b$ values when compared to related works about Ly$\alpha$ forest absorption \citep[e.g.,][]{1998ARA&A..36..267R,2000MNRAS.318..817S,2007A&A...461..847F} is probably due to the spectral resolution of MUSE not resolving individual components of connect narrower absorbers \citep[e.g.,][]{1997A&A...317..358V,2003MNRAS.338..263J}.

If the velocity shift for absorbers \#2$-$4 corresponded to a cosmological redshift difference as is probably the case for absorbers \#5$-$8, we can calculate the physical separation between central radio galaxy and the absorber. For absorber \#2, which has the smallest shift, the luminosity distance difference between it and the systemic redshift is 84 Mpc, much larger than the virial radius of the host galaxy, $R_{{\rm vir}} \simeq 175\,$kpc (Sect. \ref{sec:stellarmass}). Hence, if the physical distance was the reason for the velocity shifts, all absorbers would be gravitationally unbound to the host galaxy. In contrast, if the velocity shift was caused by the kinematics of an outflowing shell, the absorption troughs can and should be observed on large spatial scales, which they are. Given the velocity offset of absorbers \#2, 3, and 4 derived from the master spectrum and their spatial extent (i.e., large areal fraction), we therefore conclude that they are likely outflowing gas shells potentially driven by the AGN. 

While the other absorbers (\#5-8) are very likely intervening absorbers, we cannot fully exclude from our data that they may represent fast-outflows. For example, \citet{2018ApJ...859...94K} investigates ultrafast X-ray outflows (UFOs) seen in AGN in absorption and their relation with the associated \ion{H}{i} and other lower-ionization ions, such as \ion{C}{iv}. These UFOs with $v_{{\rm out}} \gtrsim 0.1c$, where $c$ is the speed of light, are much more extreme cases of outflows compared to our observations. Absorber \#8, if it was an outflow, would have a velocity of $0.01c$. In addition, the \ion{H}{i} absorption widths predicted by the UFOs are much wider than our observations ($FWHM \sim 1000\,\rm km\,s^{-1}$), their column densities are lower ($< 10^{14}\,\rm cm^{-2}$) and the required ionization parameter is higher \citep[e.g., compared to absorption studies in other HzRGs;][]{2019A&A...625A.102K}. All this indicates that the UFOs seen in X-ray and Ly$\alpha$ absorption studied by \citet{2018ApJ...859...94K} trace a different scenario than the absorbers in \object{4C04.11}. Even if there was UFO-associated \ion{H}{i} absorption for \object{4C04.11}, it would be located at much shorter wavelengths where the continuum level is too faint to allow them to be detected.


\subsection{Metal absorption}\label{sec:absorptionfrommetallines}

\subsubsection{Metal absorbers in the master spectrum}\label{sec:metalabsorbers}

\begin{figure*}
   \centering
   \includegraphics[width=16.4cm]{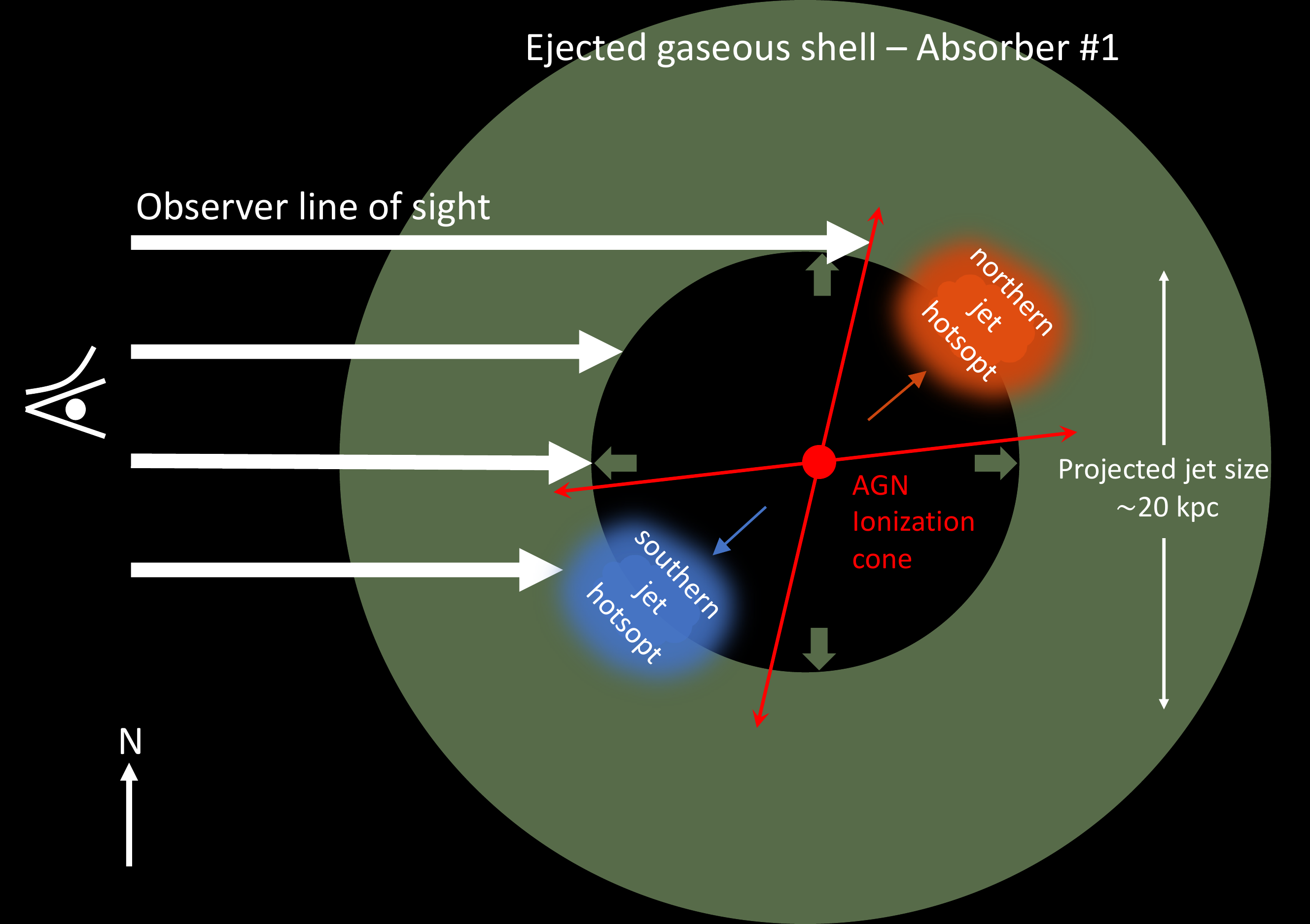}
      \caption{Schematic view of the proposed outflowing shell model in Sect. \ref{sec:outflowshell}. The large dark green annulus represents the outflowing gaseous shell that could be the absorbing cloud of absorber \#1. The blue and orange regions mark the southern (approaching) and northern (receding) jet hotspot interacting with the previous ejected shell, respectively. We note that the morphology of the gaseous shell is not necessarily in a circular shell as shown here, and we do not have information for the shell at the backside of the AGN. The red lines in the annulus center indicate the region of the AGN ionization cone that could have a wider opening angle than the jet beam (see text). The column density gradient we observed in Sect. \ref{sec:lyaspatial} in the S-N (SW-NE, due to the orientation on the sky plane, which is not shown here) direction could simply be explained by the different lengths of the observer line of sight intersecting with the gaseous shell at different spatial locations (see text). This process is shown with the length of white arrows intersecting with the dark green annulus in the figure. For the column density decreasing after passing the midplane, which cannot be explained by the geometry setting, the southern jet (blue region) interaction with the ejected gaseous shell could cause the decreasing of column density through instabilities and/or partially ionizing the gas. Though the rough projection size of the jet is shown, we note that other parts of this sketch are not to scale.}\label{fig:ejectedshellmodel}
   \end{figure*}

Relative column density ratios between different elements can provide information on the enrichment of the gas assuming an ionization parameter. The underlying assumption for the metal absorbers analyzed here is that they are ionized by the central AGN \citep[e.g.,][]{2019A&A...625A.102K}. Constraining whether this AGN photoionization is geometrically possible is beyond the scope of this work given the resolution of the data and the limited knowledge of the evolution state and ionization episode of the radio galaxy. Hence, we do not discuss more about the source(s) of ionization for the metal absorbers and proceed with the discussion of the following implication with the assumption of central AGN ionization. One hint on the AGN ionization could be due to the wide ionization cone that covers some fraction of the absorbing gas seen (e.g., Fig. \ref{fig:ejectedshellmodel}). Nevertheless, we remind the reader that the shocks or a hard source of ionization (for example, AGN of meta-galactic background) inferred by the presence of a high column density \ion{N}{v} absorber (see blow) could also be possible.

In this work, we identified the absorbers around the systemic redshift of \ion{H}{i}, \ion{C}{iv}, and \ion{N}{v} in the master spectrum, which we assume belong to the same cloud. The corresponding ratios are $N_{\ion{C}{iv},1}/N_{\ion{H}{i},1} = 0.12 \pm 0.05$ and $N_{\ion{N}{v},1}/N_{\ion{H}{i},1} = 1.4 \pm 0.2$. We remind the readers that the $N_{\ion{N}{v},1}$ should be treated as a lower limit given that the Doppler parameter associated with it hits the upper boundary (Appendix \ref{apd:fittingnotesnv}). 
Comparing this with \textsc{cloudy} \citep[spectral synthesis code,][]{2017RMxAA..53..385F} models \citep[the same models as Fig. 17 in][]{2019A&A...625A.102K}, we can roughly estimate that absorber \#1 has (super) solar metallicity ($Z \gtrsim 1\,Z_{\odot}$) independent of a specific assumption for the ionization parameter. The derived $N_{\ion{N}{v},1}$ value is consistent with the conclusion. This suggests strongly that absorber \#1 has an origin inside the ISM of the radio galaxy. Given the age of the Universe at $z=4.5077$, it is unlikely that absorber \#1 is the infalling material that has been enriched by a previous outflow and is now recycled through a galactic fountain mechanism. The column density, $N_{\ion{N}{v}} \sim 14.99\,\rm cm^{-2}$, is relatively high. As discussed in \citet{2019A&A...625A.102K}, the secondary nitrogen production is responsible for the nitrogen column density enhancement of the absorber if the gas has (super) solar metallicity \citep[also][]{1993ApJ...418...11H}. Hence, though the CNO cycle for the secondary carbon and nitrogen can produce solar N/C ratio over a large range in metallicities \citep[e.g.,][]{2017MNRAS.466.4403N}, $Z \gtrsim 1\,Z_{\odot}$ is needed, which is consistent with our conclusion here.  We further discuss the nature of absorber \#1 in Sect. \ref{sec:hi1spatial} combining the metallicity and spatial features observed in \ion{H}{i}.

For absorber \#2 and \#3, their column density results can only be treated as upper limits. The corresponding ratios of \ion{H}{i} absorbers are $N_{\ion{C}{iv},2}/N_{\ion{H}{i},2} \sim 3\times10^{-4}$ and $N_{\ion{C}{iv},3}/N_{\ion{H}{i},3} \sim 0.02$. In addition, absorber \#4 has the ratio of $N_{\ion{C}{iv},4}/N_{\ion{H}{i},4} = 0.25 \pm 0.05$. It is difficult to confine the metallicities of these three absorbers without the data from \ion{N}{v}. Nevertheless, the ratios are indicative of the absorbers having sub-solar metallicity according to the \textsc{cloudy} models ($Z \lesssim 0.5\,Z_{\odot}$), except for absorber \#4, which could have solar metallicity. From the discussion in Sect. \ref{sec:otherHIabsorbers}, we consider these three absorbers as outflowing gaseous clouds with the one having the highest velocity shift (absorber \#4) being the most recently ejected. Combined with the proposed scenario of absorber \#1 in Sect. \ref{sec:outflowshell}, the low metallicity of absorbers \#2$-$\#4 can be explained: Absorber \#1 was ejected first and carried most of the metal elements produced in the early star formation activities. The timescale between the ejection of absorber \#1 and the later ejected clouds (absorbers \#2$-$\#4) was then not large enough to enrich the gas again to solar like metallicity. If we take it further that this may explain the latest absorber \#4 has higher metallicity than absorber \#2 and \#3 because it has the longest time to be enriched, though this is impossible to be proved and has large uncertainties.

The reason we do not observe \ion{C}{iv} absorber\#5$-$\#8 could be that they are located in the low S/N emission wing. Considering the discussion in Sect. \ref{sec:otherHIabsorbers}, however, this could be alternatively explained as they are intervening absorbers outside the potential well of the central radio galaxy (IGM) and they consist of cold, un-enriched pristine gas. 

\subsubsection{Spatial distribution}\label{sec:discussioncivspatial}

As derived in Sect. \ref{sec:civspatial}, the column density ratios between the \ion{H}{i} and \ion{C}{iv} absorber \#1 are $0.11\pm0.04$ and $<0.04$ for the $N_{\ion{C}{iv},\,{\rm NE}}/N_{\ion{H}{i},\,{\rm NE}}$ in the NE and $N_{\ion{C}{iv},\,{\rm SW}}/N_{\ion{H}{i},\,{\rm SW}}$ in the SW, respectively. This is considered to be consistent with the observed \ion{H}{i} column density in absorber \#1.  We compare these two ratios with the \textsc{cloudy} models and find that, despite the lack of spatial information of \ion{N}{v}, the NE cloud may have solar metallicity ($Z_{\rm NE} \gtrsim 0.5\,Z_{\odot}$), while the metallicity of the SW cloud may be sub-solar ($Z_{\rm SW} \lesssim 1.0\,Z_{\odot}$). Readers should bear in mind that the inferred metallicity of absorber \#1 in these two regions are derived from low S/N spectra and should be treated with caution. 

If we assume these two regions have similar metallicity, this is consistent with the analysis in Sect. \ref{sec:metalabsorbers} and Sect. \ref{sec:outflowshell} that absorber \#1 is an outflowing shell that is enriched homogeneously by star formation activities in the ISM prior to the launch of the outflow.

On the other hand, if we assume the NE region has higher metallicity than the SW region, it suggests a spatially inhomogeneous enrichment. This could indicate merger activities or unevenly dilution due to cosmic accretion of metal-poor gas. Given that the S/N of the two \ion{C}{iv} spectra (especially the SW one) , we do not further interpret the result.


\section{Large-scale characteristics of \texorpdfstring{\ion{H}{i}}{HI} absorber \#1}\label{sec:hi1spatial}

In this section we discuss several possible explanations for the observed properties (areal fraction, kinematics, column density gradient) of the \ion{H}{i} absorber \#1 (Sect. \ref{sec:lyaspatial}). The \ion{H}{i} absorbers \#1 and 2 are the only two absorbers in our analysis with a high enough S/N to perform the spatial analysis. We identify a significant column density gradient of absorber \#1, but not for absorber \#2 (Fig. \ref{fig:lyaspatial2}) suggesting a relatively uniform distribution within the uncertainties of our fitting analysis. The data quality does not allow us to spatially map any of the higher-velocity absorbers and we briefly discuss their potential nature in Sect. \ref{sec:otherHIabsorbers}.
We argue that the absorbing gas responsible for absorber \#1 is spatially detached from the Ly$\alpha$ emitting gas since we do not observe any similarity between the kinematics of the absorbing and emitting components (Fig. \ref{fig:lyaspatial1}b and c, Fig. \ref{fig:lyaspatial2}b). The opposite situation may be seen in the blue absorber in \object{TN J1338-1942} (a HzRG in our ALMA-MUSE sample) where the velocity shift maps of the emission and absorption show resemblance \citep[Fig. 4 in ][]{2015MNRAS.449.1298S} and indicate that emitting and absorbing gas are mixed.

Recently, \citet{2016ApJ...833L..26G,2017A&A...607A..71G} showed simulations of Ly$\alpha$ radiative transfer through a medium where detailed subparsec structures are considered, which is a more realistic model than a continuous shell model. It should be noticed that our spatial mapping of the column density of \ion{H}{i} absorber \#1 is far from sensitive enough to probe the details in the absorption medium. In addition, \citet{2017A&A...607A..71G} concluded that the resultant spectral features of their clumpy model are similar to a shell model, that is, the observational features are not affected much by the absorbing medium being either clumpy or a continuous shell. Hence, we only present simple explanations to the reported column density and velocity gradient (Fig. \ref{fig:lyaspatial2}) and use the simulation works to assess them to first-order, leaving sophisticated modeling for future works. For example, \citet{2019ApJ...873..129P} simulated the CGM on a sub-kiloparsec scale and find that the absorbing gas responsible for observed feature around several hundreds of $\rm km\,s^{-1}$ in velocity shift range may span hundreds of kiloparsecs in space. It is impossible even for the state-of-the-art instrument like MUSE to probe the real morphology of the CGM gas given that we can only observe features in projection. 

\subsection{Outflowing shell}\label{sec:outflowshell}
We first consider the model of an outflowing gas shell to explain the column density and velocity shift gradient we identify for \ion{H}{i} absorber \#1. We remind the reader that absorber \#1 is at ${\sim\,0\,\rm km\,s^{-1}}$ and covers a large area of the sky (Sects. \ref{sec:lya} and \ref{sec:lyaspatial}) approximately on scales of $30\times30\,\rm kpc^{2}$. We identify a column density gradient along the SW-NE direction increasing over 1 dex in 24 kpc. We also check Ly$\alpha$ spectra extracted from the two spatial regions along the radio jet in the SE-NW direction (the two regions for completeness study of \ion{C}{iv} spatial mapping in Sect. \ref{sec:civspatial}) where we also marginally identify the column density gradient. 

Following \citet{2000A&A...356...23B,2006A&A...459...31B,2002A&A...386L...1K,2005A&A...436..845K}, we propose that the absorbing material of absorber \#1 is a wind-driven gaseous shell. The wind may have been powered by stellar feedback and/or AGN activity several tens of megayears before the radio jet was launched. The wind traveled isotropically with its speed decreasing from thousands of $\rm km\,s^{-1}$ within 1 kpc to a few $\rm km\,s^{-1}$ at 10s of kpc \citep[may even halt or fall back;][]{2005A&A...436..845K,2013ApJ...763L..18W,2018MNRAS.478.3100R}. Tens of megayears after the beginning of the shell expansion, the relativistic jet is launched. The jet that travels at a few tenths of the speed of light catches up with the previously ejected shell in a few megayears \citep[][]{2014MNRAS.439.2314P}, accelerates and disturbs it. The age of the jet could be even smaller ($\sim$1 Myr) in the scenario described here if an earlier galactic wind has cleared the surrounding leading to fewer interactions of the jet traveling inside the wind-driven shell. The velocity gradient along the radio axis we see in Fig. \ref{fig:lyaspatial2}b could be a hint for this jet gas interaction. Using polarization measurements, \citet{2014MNRAS.439.2314P} reported that the southern (northern) jet is very likely the approaching (receding) one, which agrees with absorber \#1 velocity shift map (the southern part of absorber \#1 is the blue-shifted part). In Fig. \ref{fig:ejectedshellmodel}, we show a schematic presentation of the proposed scenario where the SW-NE column density gradient may be explained by the spatially different intersected length between the observational line of sight (thick white arrows on the left of the figure) and the gaseous shell (absorber \#1, dark green annulus). 
This is a similar situation to the sunlight traveling a longer path through the Earth's atmosphere when the altitude of the Sun is low and causing more scattering. For the column density decreasing in the northern half of the shell, the gradient may be explained by this geometry. The southern jet, which we believe is the approaching one, may be responsible for the observed column density keeping decreasing in the southern half of the gaseous shell. The approaching and receding jet-gas interaction hotspots are marked in Fig. \ref{fig:ejectedshellmodel} as blue and red regions, respectively. As the figure shows, the approaching jet catches and disturbs the gaseous shell probably through Kelvin–Helmholtz instability in its surrounding \citep[e.g.,][]{2020MNRAS.499..681M}. The interaction will cause a decrease in the particle number density \citep[][]{2020MNRAS.499..681M} in the immediate vicinity of the jet hotspot (or the jet may even ionize a part of the cooled gas). Hence, the combination of the two aforementioned effects will result in the observed column density and velocity gradient. Though we mark the approximate projected size of the observed jet, readers should bare in mind that the Fig. \ref{fig:ejectedshellmodel} is not to scale. We note that the red lines mark the regions of the AGN ionization cones as the radio jets are narrow collimated streams \citep[i.e., opening angles; e.g.,][]{2012A&A...548A..45D,2016MNRAS.456.2861O} of the ionization cones are suggested to be wider than the jet beams. Besides, the jet-gas interaction hotspots are smaller regions compared to the gaseous shell; nevertheless, we emphasize them in Fig. \ref{fig:ejectedshellmodel} with larger symbols. 

Several simulation works have shown the possibility of AGN wind expelling medium to kiloparsec scales \citep[e.g.,][for an AGN driven wind accelerating the surrounding medium to $\sim1000\, \rm km\,s^{-1}$ within 1 kpc]{2013ApJ...763L..18W}. \citet{2020MNRAS.491.2939O} specifically studied the impact of AGN feedback on the CGM. This research shows that feedback can drive out the metal elements, which could explain the metal-enrichment of absorber \#1 in our observations. The authors reported that the expulsion of metal elements beyond the virial radius, for example like \ion{C}{iv} in our case, takes longer with a timescale of 0.5$-$2.5 Gyr. Hence, on smaller timescales, such as a few hundred megayears, the CGM can be enriched with the gaseous metal-enriched cloud still within the galactic potential well like the case of 4C04.11.  Furthermore, \citet{2018MNRAS.478.3100R} showed that swept up gas can efficiently cool within an outflow, which could be one possible origin for the neutral gas that absorber \#1 consists of. 

In Sect. \ref{sec:absorptionfrommetallines} we discussed the enrichment of absorber \#1, which is also observed in \ion{N}{v} and \ion{C}{iv,} suggesting that the absorbing cloud has super solar metallicity. This supports the proposed scenario in which the outflowing shell is launched from within the galaxy where the gas has been enriched through star formation activities. It furthermore suggests that we are observing the redistribution of metals through feedback processes and the enrichment of both the ISM and the CGM. Our previous analysis in Sect. \ref{sec:absorptionfrommetallines} is based on the assumption that the metal absorbers (\ion{N}{v} and \ion{C}{iv}) are ionized by the AGN. This could be possible if the opening angle of the ionization cone is wide enough to cover some fraction of the absorber \#1 gaseous shell (like the scenario shown in Fig. \ref{fig:ejectedshellmodel}).

It may be too coincident for absorber \#1 to be at ${\Delta v \sim 0\,}$ km s$^{-1}$. Even considering the reported 1$\sigma$ fitting uncertainty, the range of the velocity shift is still close to 0. We notice that the spectral resolution of MUSE, $\sim100$ km s$^{-1}$, is much larger than the MCMC reported probability distribution range. Hence, the absorption could have some intrinsic velocity with respect to the systemic redshift of the radio galaxy, which would need to be verified with high-resolution spectroscopy.  

\subsection{Absorption by large-scale CGM gas}\label{largerscaleabsorption}

\citet{2019ApJ...873..129P} showed the simulation of the FOGGIE project (Figuring Out Gas \& Galaxies in Enzo), which focuses on the CGM. In their work, the authors presented the absorption characteristics of the CGM gas on scales of hundreds of kiloparsecs with considerable column density for both \ion{H}{i} and metals (for example, \ion{C}{iv}). An alternative scenario for the characteristics of our absorber \#1 is therefore that it consists of gas in the CGM, which extends tens to hundreds of kiloparsecs and has a complex sub-structure beyond the detectability of MUSE. This gaseous cloud is the surrounding medium unrelated to the ejected material by the central radio galaxy. In this scenario, the column density gradient could be due the uneven concentration nature of the CGM gas as shown in \citet{2019ApJ...873..129P}. The tentative velocity gradient may be invoked through a rotation of the large-scale medium. For a system of virial mass on the order of $10^{13}\,\rm M_{\odot}$, the virial velocity is around 500 km s$^{-1}$ (Sect. \ref{sec:stellarmass}), which is larger than the value we observe. While unlikely, this inconsistency between the observed velocity gradient ($-50\,\rm km\,s^{-1}$ to $35\rm \,km \,s^{-1}$, Fig. \ref{fig:lyaspatial2}b) of absorber \#1 and the virial velocity could be due to a combination of the MUSE spectral resolution and projection effects.

This gaseous halo could be on its way to being accreted onto the central galaxy to feed the SMBH and/or star formation activities. The observed radio jet with a projection length of $\sim20\,\rm kpc$ \citep[][]{2014MNRAS.439.2314P} could be well within the giant CGM gas halo and unrelated if the aforementioned scenario is the case. The high column density part in the north may be related be the inner part of inflow, which is denser according to the simulation presented by \citet{2020MNRAS.498.2415M}. The spatial extent of this inflow could be up to $\sim 60\,$kpc in our case, which covers the detected  absorber \#1 well \citep[$0.5R_{{\rm vir}}$; see Sect. \ref{sec:stellarmass};][]{2020MNRAS.498.2415M}. 

Although this scenario can explain the \ion{H}{i} column density and the surrounding medium can be enriched to a small extent, it is difficult to reconcile super solar metallicty with this scenario (Sect. \ref{sec:metalabsorbers}, and especially the high column density of \ion{N}{v}). Hence, we consider the model proposed in Sect. \ref{sec:outflowshell} as the more probable situation.

\subsection{Alternative models}\label{alternativemodels}
Alternatively, the double-peak structure of the Ly$\alpha$ spectrum (which we believe is due to the \ion{H}{i} absorber \#1) with the trough at $\sim0\,\rm km\,s^{-1}$ could be explained by other numerical models.

The absorbing gas could be entrained within the jet and detached from the jet path, which could explain the tentative velocity gradient. After the detachment, the gas will gradually slow down, which could explain its low velocity shift around 0 km s$^{-1}$.  This model, however, has problems to reproduce the observed column density gradient in the direction roughly perpendicular to the radio jet. 

The absorbing \ion{H}{i} gas we see may be the product of the "positive feedback" from the jet interaction with the ISM/CGM \citep[e.g.,][]{ 2006ApJ...647.1040C,2012MNRAS.425..438G,2017ApJ...850..171F}. In this situation, we could ignore the self-gravity of the gas due to the dark matter dominated potential \citep[][]{2017ApJ...850..171F}, which could explain the velocity shift of around 0 km$\,\rm s^{-1}$ of the \ion{H}{i} absorber. The jet compresses the CGM gas on its path. The higher \ion{H}{i} absorber \#1 column density in the NE could be that the line of sight passing longer length in the northern part given the jet orientation (similar to the geometric effect proposed in the outflowing model, Sect. \ref{sec:outflowshell}). This could also in principle explain the observed tentative velocity shift gradient. This explanation again, however, has the shortcoming to reproduce the metal enrichment of absorber \#1.

\section{Conclusions}\label{sec:conclusions}
In this paper we present MUSE observations of the CGM of a radio galaxy, \object{4C04.11}, at $z=4.5077$. Particularly, we focus on the absorption in the halo and its spatial properties. The main conclusions of this work are summarized as follows:  
   \begin{enumerate}

      \item The Ly$\alpha$ emission halo is detected on scales of $70\times30$ kpc$^2$ (more extended low surface brightness regions are not shown in the presented narrowband image in Fig. \ref{fig:moment0}). We model the Ly$\alpha$ profile using a double Gaussian and report on a blueshifted component at $\sim \rm -102 \,km\,s^{-1}$ whose nature is still debatable. The map of the Ly$\alpha$ velocity width (Fig. \ref{fig:lyaspatial1}) may indicate signatures of jet-gas interactions.
      
      \item The systemic redshift of \object{4C04.11} is derived from the brightest nonresonant line, \ion{He}{ii}, of $4.5077\pm0.0001$. This is consistent with the near-infrared observation of [\ion{O}{II}] \citep[][]{2017A&A...599A.123N} and a large improvement compared to previous work using Ly$\alpha$ \citep[][]{2006AstL...32..433K}.
      
      \item Metal emission lines, \ion{C}{iv}, \ion{N}{v,} and \ion{O}{iii}] are also detected;  \ion{C}{iv} in particular can be spatially mapped (Fig. \ref{fig:civmoment0}). This suggests that the CGM is largely metal enriched. Both the \ion{C}{iv} and \ion{N}{v} lines show blueshifted emission components with consistent velocity shifts ($\Delta v_{\rm \ion{C}{iv}, b.l.} = -1026 \pm 112\, \rm km\,s^{-1}$ and$\Delta v_{\rm \ion{N}{v}, b.l.} \sim -1587\, \rm km\,s^{-1}$). This component may have a different ionization mechanism than the systemic emission and could provide evidence for a star formation and/or AGN-driven outflow (Sect. \ref{sec:emissionstructure}). 
      
      \item  We identify at least eight \ion{H}{i} absorbers with a velocity shift range of $-3345\sim0\,\rm km\,s^{-1}$. The column density of these eight \ion{H}{i} absorbers are around $10^{14.8}\,\rm cm^{-2}$, and their Doppler parameters, $b$, have a range of $40-271\,\rm km\,s^{-1}$ (Table \ref{tab:absorberfit}). We infer the presence of two \ion{C}{iv} absorbers, which are believed to be associated with \ion{H}{i} absorbers \#1 and \#4 and have a column density of $\sim 10^{14} \, \rm cm^{-2}$. The column densities of \ion{C}{iv} absorbers \#2 and \#3 are only constrained to upper limits (Table \ref{tab:absorberfit}). The presence of absorber \#1 is also inferred in \ion{N}{v} with a relatively high column density, $\sim 10^{14.99}\,\rm cm^{-1}$. This suggests that the first four absorbers are within the potential well of the host galaxy, while absorbers \#5$-$8 are likely intervening absorbers (Sects. \ref{sec:otherHIabsorbers} and \ref{sec:absorptionfrommetallines}).
      
      \item We spatially map the \ion{H}{i} absorbers and identify a column density gradient of absorber \#1  in the SW-NE direction (increases 1 dex in 24 kpc; Fig. \ref{fig:lyaspatial2}). The velocity map of \ion{H}{i} absorber \#1 shows a tentative gradient along the radio jet axis, with the blueshifted part in the south. This is spatially coincident with the approaching radio jet \citep[][]{2014MNRAS.439.2314P}. Absorber \#1 is also detected in \ion{C}{iv;}  we can measure its column density in two distinct regions, and we identify a column density gradient similar to that of \ion{H}{i} \#1, albeit with large uncertainties (Sect. \ref{sec:civspatial}). We propose and discuss several possible models to explain the observed features. We conclude that absorber \#1 likely represents a metal-enriched expelled gaseous shell that is disturbed by the jet that was launched later (Sect. \ref{sec:outflowshell}). 
      
      \item Our observations suggest that we are observing the redistribution of metals through feedback processes and the enrichment of both the ISM and the CGM.

   \end{enumerate}

This work represents a pilot study and showcases the power of IFS instruments like MUSE for studying the absorbing "invisible" CGM gas and its enrichment and interplay with AGN and star formation activity in and around massive active galaxies in the early Universe. We will perform a similar analysis to our full sample of eight HzRGs with redshift 2.9$\sim$4.5, whose SFRs span a wide range \citep[$84-626\,\rm M_{\odot}\, yr^{-1}$;][]{2019A&A...621A..27F}. Although our targets are rare in terms of number density predicted from the galaxy mass function, they are unique representatives for studying the early stellar mass assembly, the feedback process, and the baryon cycle.

\begin{acknowledgements}
We thank the anonymous referee for the valuable comments, which improved this manuscript. All of the MUSE data used in this paper are available through the ESO science archive. This work was based on a Master Thesis supervised by Benjamin P. Moster. We thank his support for the thesis work. We thank Dr. Thorsten Tepper Garc\'ia for pointing us to their erratum and sharing the original numerical code of Voigt Hjerting function. DW is supported by through the Emmy Noether Programme of the German Research Foundation. AH was supported by Fund\c{c}\~ao para a Ci\^encia e a Tecnologia (FCT) through the research grants UIDB/04434/2020 and UIDP/04434/2020, and an FCT-CAPES Transnational Cooperation Project "Parceria Estrat\'egica em Astrof\'isica Portugal-Brasil". MVM acknowledges support from grant PGC2018-094671-BI00 (MCIU/AEI/FEDER,UE). Her work was done under project No. MDM-2017-0737 Unidad de Excelencia "Mar\'ia de Maeztu" Centro de Astrobiolog\'ia (CSIC-INTA).\\

This work uses the NASA's Astrophysics Data System and a number of open source software other than the aforementioned ones such as Jupyter notebook \citep[][]{Kluyver2016jupyter}; \texttt{matplotlib} \citep[][]{Hunter:2007}; \texttt{SciPy} \citep[][]{2020NatMe..17..261V}; \texttt{NumPy} \citep[][]{harris2020array}; \texttt{Astropy} \citep[][]{2018AJ....156..123A}; and \texttt{LMFIT} \citep[][]{2016ascl.soft06014N}.
\end{acknowledgements}

%
%
\bibliographystyle{aa}
\bibliography{references}

\input{appendix.tex}

\end{document}

%% file: appendix.tex
\begin{appendix} 




\section{Line fitting procedure}\label{apd:fitequation}
In this work we use both Gaussian (Ly$\alpha$, \ion{C}{iv} and \ion{C}{iv}) and Lorentzian (\ion{He}{ii} and \ion{C}{iv}) functions to fit the emission of the spectrum lines. The Gaussian emission model is expressed as 
\begin{equation}\label{eq:gaussian}
F_{\rm \lambda,G} = \frac{F}{\sigma_{\lambda}\sqrt{2\pi}}\exp\left[-\frac{1}{2}\left(\frac{ \lambda - \lambda_{0}}{\sigma_{\lambda}}\right)^{2}\right],
\end{equation}
and the Lorentzian emission profile is defined as 
\begin{equation}\label{eq:lorentzian}
F_{\rm \lambda,L} =  \frac{F}{\pi} \frac{\frac{1}{2}\Gamma_{\lambda}}{(\lambda - \lambda_{0})^{2}+\left(\frac{1}{2}\Gamma_{\lambda}\right)^{2}},
\end{equation}
where the $F$ is the integrated emission flux of the line, $\lambda_{0}$ is the line center and $\lambda$ is the wavelength at which the flux density, $F_{\rm \lambda,G}$ or $F_{\rm \lambda,L}$, is calculated. The $\sigma_{\lambda}$ in Eq. \ref{eq:gaussian} is the line width while the $\Gamma_{\lambda}$ in Eq. \ref{eq:lorentzian} is the Full Width at Half Maximum (FWHM) of the line. 

The absorption can be described as $\exp\left(-\tau_{\lambda}\right)$ by the radiation transfer theory. The parameter, optical depth $\tau_{\lambda}$, is approximated by the Voigt-Hjerting (Voigt for short) function, 
\begin{equation}\label{eq:tau}
\tau_{\lambda} = \frac{\sqrt{\pi} e^{2} f_{i} \lambda^{2}_{0}}{\Delta \lambda_{\rm D} m_{\rm e} c^{2}} \times N \times H(a,x),
\end{equation}
where $N$ is the column density, $e$ is the electron charge, $m_{\rm e}$ is the electron mass, $c$ is the speed of light and $f_i$ is the oscillator strength. In this work, we adopt the atomic data from \citet{2017ApJS..230....8C} and \cite{NIST_ASD}. The $\Delta \lambda_{\rm D}$ is defined as $\Delta \lambda_{\rm D} = \frac{b}{c}\lambda_{0}$, where $b$ is the Doppler parameter. $H(a,x)$ is the Hjerting function in the following definition: 
\begin{equation}\label{eq:H}
H(a,x) \equiv \frac{a}{\pi} \int \limits_{-\infty}^{+\infty} \frac{\exp\left(-y^2\right)}{\left(x-y\right)^2+a^2} dy.
\end{equation}
In this approximation, $x \equiv \frac{(\lambda - \lambda_{0})}{\Delta \lambda_{\rm D}}$ and the constant $a$ is defined as 
\begin{equation}\label{eq:a}
a \equiv \frac{\lambda_{0}^{2}\Gamma_{i}}{4\pi c \Delta \lambda_{\rm D}} ,
\end{equation}
where $\Gamma_{i}$ is the Lorentzian width. We use the approximation of $H(a,x)$ in this work adopted from \citet{2006MNRAS.369.2025T,2007MNRAS.382.1375T} for system whose column density $\rm < 10^{22}\,\rm cm^{-2}$, which has the form
\begin{equation}\label{eq:HTG}
H(a,x) = H_{0} - \frac{a}{\sqrt{\pi}x^2} \times \left(H_0 \times H_0 \times \left(4x^4+7x^2+4+Q\right)-Q-1\right),
\end{equation}
where $H_0 = \exp\left(-x^2\right)$ and $Q = 1.5x^{-2}$. The calculated Voigt profile by the aforementioned equations, $\tau_{\lambda}$, is then combined with other profiles, which represent different absorbers seen in one emission line using the radiation transfer equation. Then this convolves with the line spread function (LSF) of MUSE to match the observed resolution,
\begin{equation}\label{eq:cv}
CV =  e^{-\left(\sum \limits_{i=1}^{n}\tau_{\tau_{\lambda,n}}\right)} \circledast LSF\left(\lambda\right),
\end{equation}
where the $CV$ is the acronym of convolved Voigt and n is the number of absorbers.
The LSF is described by a Gaussian model with a full width of half maximum ($FWHM$) of 2.65 \AA. We note that the LSF varies with observed wavelength, the location on the charge-coupled device and many other factors \citep[][]{2020A&A...641A..28W}. We determine the mean $FWHM$ using the intermediate production in the data reduction process that contains the LSF profile. This is consistent with the MUSE LSF approximated by polynomial in other work \citep[e.g.,][]{2018A&A...611A..95W}. This is also the LSF value used in \citet{2019A&A...625A.102K}, who study the MUSE observations of another HzRG in our sample. The fitting procedure is implemented in Python by the package \texttt{LMFIT}. The convolution is realized through the fast-Fourier transform method in the package \texttt{SciPy} \citep[][]{2020NatMe..17..261V} following \citet{2018ascl.soft11016K}. The final fitted function is 
\begin{equation}\label{eq:fitfunction}
F_{\lambda} = \left(\sum \limits_{j=1}^{m} F_{\lambda,\rm G\,or\,L}\right) \times CV,
\end{equation} 
where $m$ is the number of emission components (Gaussian or Lorentzian). Both the fits of \ion{C}{iv} and Ly$\alpha$ need to include an additional blueshifted emission component (see Sect. \ref{sec:fitresult}).

\section{SED fitting}\label{apd:sed}
\begin{table}
 \caption{Photometric results used for the SED fitting.}\label{tab:sedphotometry}
 \centering
\begin{tabular}{c c c }
\hline
\hline
Band & $S_{\nu}$ [mJy] & Ref. \\
\hline
0.5$-$7 keV & $4.2\pm0.8\times10^{-7}$ & S20\\
 $B$    & $<6.4\times10^{-4}$          & P14\\
 $V$    & $<7.6\times10^{-4}$          & P14\\
 $R$    & $(5\pm1) \times10^{-3}$      & P14\\
 $I$    & $(3.6\pm0.4)\times10^{-3}$   & P14\\
 $K$    & $(2.1\pm0.2)\times10^{-2}$   & P14\\
 IRAC 1 & $<8.67\times10^{-2}$         & [1]\\
 IRAC 2 & $(7.71\pm0.05)\times10^{-2}$ & [1]\\
 WISE 3 & $<0.56$                      & [2]\\
 WISE 4 & $3.04\pm 1.16$               & [2]\\
\hline   

 \hline

\end{tabular}
\tablefoot{The X-ray $0.5-7.0$ keV photometry is reported in \citet[][S20]{2020ApJ...899..127S}. The $BVRIK$ and $K$ band photometric results are taken from \citet[][P14]{2014MNRAS.439.2314P}. [1] The \textit{Spitzer} IRAC 1 and 2 data are from Program ID 70135 \citep[PI: D.Stern, see][for data reduction and flux measurement]{2013ApJ...769...79W,2014ApJ...786...17W}. [2] The WISE 3 and 4 are archival ALLWISE data \citep[][]{2010AJ....140.1868W,2011ApJ...731...53M}\footnote{\url{https://irsa.ipac.caltech.edu/Missions/wise.html}}.}
\end{table}


  \begin{figure*}
  \centering
        \includegraphics[width=16.4cm,clip]{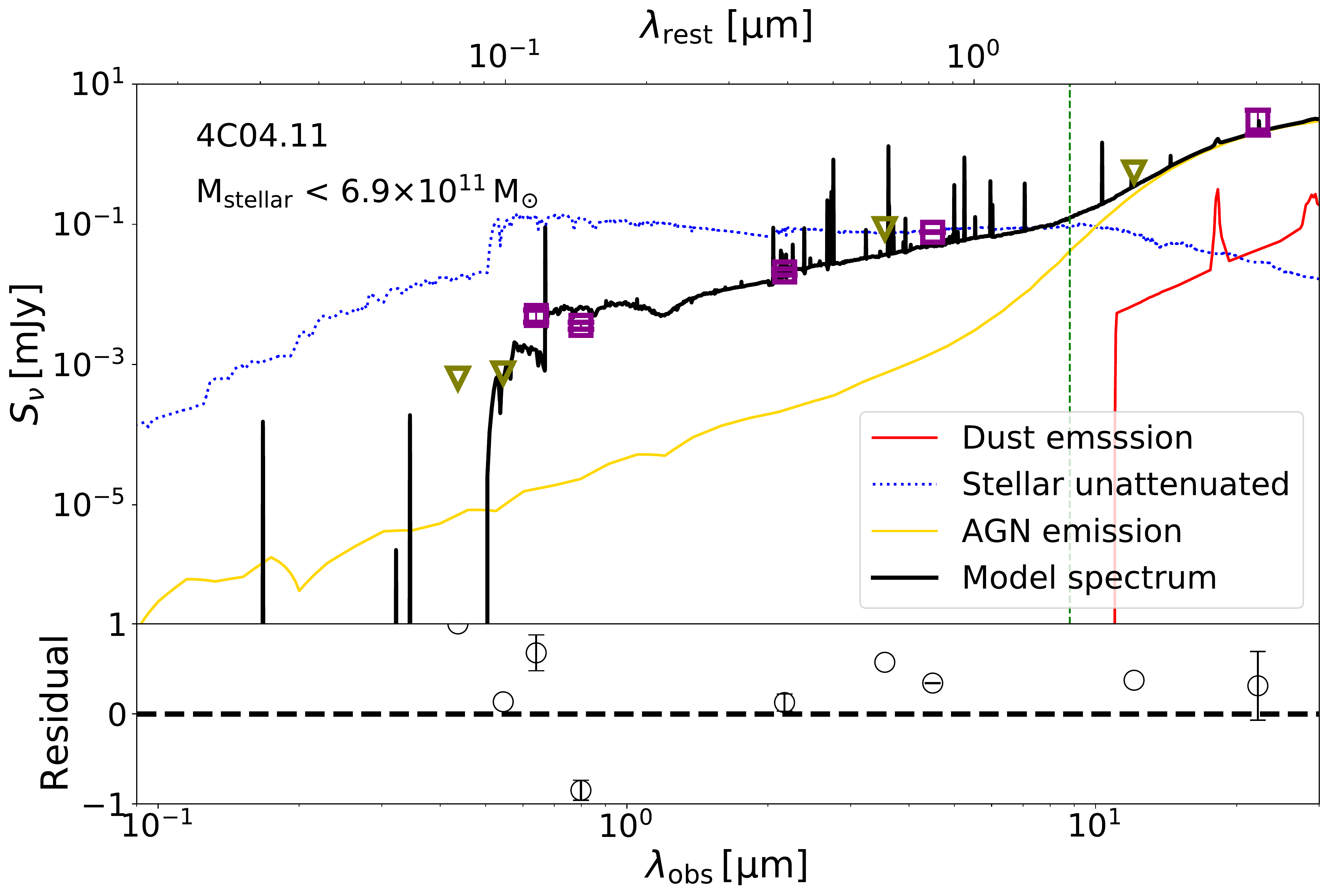}
      \caption{SED fitting model and photometric data. In the \textit{upper panel}, we show the fitted SED model spectrum from X-CIGALE with a black curve. In addition, dust and unattenuated stellar and AGN emissions are shown in red, blue dotted, and yellow curves, respectively. The input observed photometry flux densities are marked in dark magenta boxes and olive triangles (upper limits). The X-ray data are not shown as they do not constrain the stellar component. The green vertical dashed line is the position of rest frame 1.6 $\rm \mu m$ from which the unattenuated stellar flux is adopted for $\rm M_{stellar}$ estimation. In the \textit{lower panel}, we present the relative residuals, $\frac{S_{\nu,\rm obs}-S_{\nu,\rm mod}}{S_{\nu,\rm obs}}$, where $S_{\nu,\rm obs}$ and $S_{\nu,\rm mod}$ are the observed and model flux densities, respectively.}
         \label{fig:sed}
  \end{figure*}


In this appendix, we present the photometric data used for the SED fitting (Table \ref{tab:sedphotometry}) and the fitting result (Fig. \ref{fig:sed}) of 4C04.11. The listed photometric data are given in flux densities in this paper.  As stated in \citet{2014MNRAS.439.2314P}, the $BVRI$ bands used are closer to the Johnson-Kron-Cousins systems \citep{1990PASP..102.1181B} with which we convert the magnitudes to flux densities. The $K$ band magnitude is calibrated using 2MASS \citep[Two-micron All-Sky Survey;][]{2006AJ....131.1163S} sources \citep{2014MNRAS.439.2314P} using which we convert it to flux density. The IRAC 1 is treated as upper limit in this SED fitting as it is contaminated by the H$\alpha$ line. Since the WISE bands 1 and 2 are closer to IRAC 1 and 2, we only use the high S/N IRAC data.  X-ray photometry from \citet[][]{2020ApJ...899..127S} is converted using the function provided with X-CIGALE \citep{2020MNRAS.491..740Y}. We also include the detected systemic emission fluxes of Ly$\alpha$, \ion{C}{iv} and \ion{He}{ii} in this work into the X-CIGALE to better constrain the nebular emission component.

The fitted SED model and dust, AGN and unattenuated stellar emission components are shown in Fig. \ref{fig:sed}. From this fit, we extract the rest frame 1.6 $\rm \mu m$ flux for stellar mass estimation (Sect. \ref{sec:stellarmass}). As shown in the figure, the stellar flux is the dominating emission component at this wavelength, that is to say, the flux at this sweet spot will offer relatively accurate stellar mass estimation \citep[][]{2007ApJS..171..353S}. We should, however, bare in mind that this should be treated as upper limit as (i) the AGN may contribute more flux and (ii) the photometry data point, WISE 3, which constrains the flux at this wavelength more, is an upper limit.

\section{Notes on the master spectra fitting}\label{apd:masterfitnotes}

\begin{table}
 \caption{Boundary constrains for the parameters of Ly$\alpha$ fitting using MCMC method.}\label{tab:lyafitcon}
 \centering
\begin{tabular}{ l l }
\hline
\hline
Fit parameters & Constraints \\
\hline
 {\it Gaussian emission:} &    \\

Line center, $\lambda_{0}\,\left[\text{\AA}\right]$ & $\Delta \lambda = 4 $ \\

Line center (b.l.), $\lambda_{0,\, \rm blue}\,\left[\text{\AA}\right]$\tablefootmark{a} &  $\lambda_{\rm in} - 8.93 \sim \lambda_{\rm in} - 2.23$ \\

Line flux, $F\,\rm \left[erg \, s^{-1} \, cm^{-2}\right]$ & $(90\% - 120\%) F_{\rm in, \, m}$ \\

Line width, $\sigma\,\left [\text{\AA}\right]$\tablefootmark{b} & $(80\% - 120\% \text{/}  150\%) \sigma_{\rm in, \, m}$ \\

\hline
{\it Voigt absorption:} &  \\
Column density $\log(N_{\rm H}\, \rm /cm^{-2})$ & $13-20$ \\ 

Doppler parameter $b\,\left[\rm km\,s^{-1}\right]$  & $40-400$\\

Absorber redshift $z$\tablefootmark{c}  & $\Delta z = 0.004$  \\
Absorber redshift $z$\tablefootmark{d}  & $\Delta z = 0.005$ \\

 \hline
\end{tabular}

\tablefoot{The lower index "in" stands for initial. $\lambda_{\rm in}$ is the observed Ly$\alpha$ wavelength calculated using systemic redshift derived from \ion{He}{ii} fitting (see Sect. \ref{sec:civheii}). The $F_{\rm in, \, m}$ and $\sigma_{\rm in, \, m}$ are line flux and width derived from primary fitting of the emission using only the red wing as input. The lower index "m" indicates both the systemic and blueshifted components have the similar boundary setup. The boundaries of the redshift, $z$, of different absorbers are customized according to their sensitivities to the parameters tested in running the fitting.\\
\tablefoottext{a}{The wavelength range corresponds to $-400 \sim -100\, \rm km \, s^{-1}$.}
\tablefoottext{b}{120\% for the narrow and 150\% for the broad component.}
\tablefoottext{c}{This set of constraints is for absorbers \#1-3, 5 and 8.}
\tablefoottext{d}{This set of constraints is for absorbers \#4, 6 and 7.}
}
\end{table}
\begin{table}
\caption{MCMC fitting constrains of \ion{C}{iv} + \ion{He}{ii} and \ion{N}{v}.}\label{tab:civheiinvconstrain}
\centering
\begin{tabular}{lc}
\hline\hline
Fit parameters&Constrains\\
\hline
{\it Emission:} & \\

\ion{C}{iv} systemic line center & $\frac{\lambda_{\rm 0, \ion{C}{iv}}}{\lambda_{\rm 0,\ion{He}{ii}}}=\frac{\lambda_{\rm \ion{C}{iv}}}{\lambda_{\ion{He}{ii}}}$ \\

blueshifted doublet line center\tablefootmark{a} & $\frac{\lambda_{0,1}}{\lambda_{0,2}}=\frac{\lambda_{1}}{\lambda_{2}}$\\
doublet line width\tablefootmark{a,b} & $\frac{\sigma_{1}}{\sigma_{2}} = \frac{\lambda_{1}}{\lambda_{2}}$\\
doublet line flux\tablefootmark{a} & $\frac{F_{1}}{F_{2}} = \frac{f_{1}}{f_{2}}$\\
\hline
{\it Voigt absorption:} & \\
Redshift & $z_{1}=z_{2}$\\
Doppler parameter & $b_{1}=b_{2}$\\
Column density & $N_{1} = N_{2}$\\

\hline
\end{tabular}
\tablefoot{The systemic \ion{C}{iv} emission line centers (both of the doublet lines), $\lambda_{\rm 0, \ion{C}{iv}}$,  are constrained to the line center of \ion{He}{ii,} which is the systemic redshift, i.e., the fitted line center ratio, $\frac{\lambda_{\rm 0, \ion{C}{iv}}}{\lambda_{\rm 0,\ion{He}{ii}}}$ is set equal to the rest frame line ratio $\frac{\lambda_{\rm \ion{C}{iv}}}{\lambda_{\ion{He}{ii}}}$. The lower index "1" and "2" used in this table indicate the blue and red component of the doublet line, respectively. The blueshifted doublet line center ratio, $\frac{\lambda_{0,1}}{\lambda_{0,2}}$, is only constrained to the ratio in rest frame, $\frac{\lambda_{1}}{\lambda_{2}}$, i.e., the velocity shift of blueshifted doublet is leaving free. The line widths of the doublet are set to be equal to each other in velocity space. Hence, the ratio of the line width in wavelength space, as the direct fitting parameter in this work, is proportional to the ratio of doublet line center in rest frame, $\frac{\lambda_{1}}{\lambda_{2}}$. The line flux ratio of the doublet, $\frac{F_{1}}{F_{2}}$, is set to be equal to the ratio of its oscillator strength, $\frac{f_{1}}{f_{2}}$. Using data from \citet{2017ApJS..230....8C}, we fix the $\frac{f_{1}}{f_{2}}$ to be approximately 2 for both the \ion{C}{iv} and \ion{N}{v} doublets. The absorption fit parameters are set to be the same for the doublet. \\
\tablefoottext{a}{The constrains here apply to both \ion{N}{v} and \ion{C}{iv}.}\tablefoottext{b}{The fitted line width of systemic \ion{C}{iv} component is $\Gamma$ (FWHM) in the Lorentzian model. For others, they are the $\sigma$ for Gaussian model.}}
\end{table}

\subsection{Fitting procedure implementation}\label{apd:fitingprodecurenotes}

During the process of implementing the MCMC fitting, we notice that the numerical approximation of the Voigt profile by \citet{2006MNRAS.369.2025T,2007MNRAS.382.1375T} may not behave well at the center, that is, the Voigt function (Eq.\ref{eq:HTG}) will return a double-peak feature when $x \xrightarrow[]{} 0$. Hence, we manually set \[\lim_{x \to 0} H(a,x) = 1-2a/\sqrt{\pi}\] (T. Tepper-Garc\'ia, priv. comm.). We also test the possibility using a more sophisticated function, the Faddeeva function, to approximate the Voigt function following \citet{2019A&A...623A..43B}. It, however, does not perform well to produce the expected result, probably because the resolution of MUSE does not allow such a delicate function to work. Hence, we keep the \citet{2006MNRAS.369.2025T,2007MNRAS.382.1375T} approximation (Eq. \ref{eq:HTG}), which has proven to be successful on MUSE data \citep[][]{2019A&A...625A.102K}.
\subsection{\texorpdfstring{Ly$\alpha$}{a} fitting notes}\label{apd:lyamasterfitnotes}

In this appendix, we discuss the details and uncertainties run into during the Ly$\alpha$ fitting. For the continuum with relatively low flux underneath the Ly$\alpha$, we decide to fit it using a first-order polynomial function and let the slope and intercept as free parameters during the further Gaussian+Voigt fitting. We describe how sensitive the \ion{H}{i} absorption fitting results are with respect to different continuum fitting strategies in Appendix \ref{apd:lyacontinuum}. 

As shown in Fig. \ref{fig:lyafit}, the Ly$\alpha$ line is asymmetric and highly absorbed. It can be fitted with the systemic redshift determined from \ion{He}{ii} (Sect. \ref{sec:civheii}). To fit this complicated line with Gaussian+Voigt profile with many free parameters, it is challenging without any prior-knowledge \citep[unlike][which has a previous high spectral resolution UV spectrum analysis as guidance]{2019A&A...625A.102K}. Since all absorbers identified here are located at the blue wing (one near the $v = 0\,\rm km\,s^{-1}$) of the Ly$\alpha$ line, we fisrt use only the red wing of Ly$\alpha$ to constrain the unabsorbed, intrinsic emission. Then, we add Voigt profiles to model the \ion{H}{i} absorbers following the procedure described in \ref{sec:fitprocedure}. However, it is still impossible to fit all eight absorbers simultaneously without the initial values of $N_{\rm H}$ and $b$ confined to an accurate range. Hence, we first start with fitting the first four absorbers with the data input just covering their spectral range and add more absorbers when satisfied with the previous step. There are at least eight \ion{H}{i} absorbers. We decide not to include further ones due to the low S/N and their large velocity shifts indicating them being outside the galactic potential well. As discussed in Sect. \ref{sec:fitprocedure}, the primary fit is performed using the least-squares method and is then changed to MCMC later using the results from least-squares as initial inputs for accurate results and uncertainties. The boundary conditions applied to the Ly$\alpha$ fitting are shown in Table \ref{tab:lyafitcon}. We follow \citet{2019A&A...625A.102K} to constrain the \ion{H}{i} column density to be within $\rm 10^{13}-10^{20} \, cm^{-2}$. 

The Ly$\alpha$ blueshifted Gaussian component is at $\sim -102\, \rm km \, s^{-1}$, which is the boundary manually set. If no constraints are applied in the final fit, both Gaussian emission components would be at $\sim0\,\rm km\,s^{-1}$, which would lead to an underestimation of the flux on the blue wing. This is probably due to the red wing being un-absorbed to which the algorithm gives high weight. We limit the line center ($<-100\,\rm km\,s^{-1}$) of the broad component to be blueshifted to account for the flux excess between absorbers \#7 and \#8 (Fig. \ref{fig:lyafit}). 

We note that the Doppler parameters of absorbers \#4 and \#5 have large values exceeding 200 $\, \rm km \, s^{-1}$, which may indicate that we are observing two or more spectrally unresolved absorbers with similar velocity shifts. This may be a the similar situation as observed for \object{MRC 0200+015} using low- and high-resolution spectrographs \citep[][]{1997A&A...317..358V,2003MNRAS.338..263J}. We test this by including two secondary absorbers (\#4a and \#5a) close to absorbers \#4 and \#5 when performing the fitting. There is no significant improvement and values of the fitted parameters of absorbers \#4, \#4a, \#5, and \#5a are not well constrained, and we therefore do not further regard this option. This issue may be revisited in the future using higher spectral resolution data.

\subsection{\texorpdfstring{\ion{C}{iv}}{CIV} and \texorpdfstring{\ion{He}{ii}}{HeII} fitting notes}\label{apd:fittingnotescivheii}
We describe several strategies used in Sect. \ref{sec:civheii} to fit the \ion{C}{iv} and \ion{He}{ii} lines, which have low S/N. In this appendix, we present details and reasons for the adjustment and discuss some uncertainties faced in running the fitting. In particular, we adjust the fitting procedure described in Sect. \ref{sec:fitprocedure} in the following eight aspects.

First, we fit \ion{C}{iv} and \ion{He}{ii} simultaneously. Because \ion{C}{iv}, which is also a doublet, suffers from absorption and may contain several kinematic emission components, it is better to fix the line center of its intrinsic emission with the redshift determined by the \ion{He}{ii}.
    
    Second, we fixed the continuum level underneath these two lines, which is determined beforehand with emissions lines masked. Third, we excluded the wavelength ranges from the fitting where the contribution of skylines is significant (marked as yellow shaded regions in Fig. \ref{fig:civheiifit}).

Fourth, we removed the potentially blueshifted broad component of the \ion{C}{iv} intrinsic emission, which is probably heavily absorbed. This could be the same emission component we included in the Ly$\alpha$ fitting (Sect. \ref{sec:lya}). 

Fifth, to alleviate the removal of this blueshifted component, we adopted a Lorentzian profile instead of a single Gaussian to account for the broad wings of both \ion{C}{iv} and \ion{He}{ii}. The Lorentzian may not be the best physical description of the underlying emission profile, but it is the best solution given the limited S/N to allow for absorption fits in the \ion{C}{iv} blue side.
Sixth, since the \ion{C}{iv} is very broad (Fig. \ref{fig:civheiifit}), we included an additional set of Gaussian to account for the extreme ($> 1000\, \rm km\, s^{-1}$) blueshifted component. 

Seventh, to account for the absorption, we included four \ion{C}{iv} absorbers that we assumed to be the same ones causing the Ly$\alpha$ absorption (Sect. \ref{sec:lya}). The reason we only included four absorbers instead of all eight is 
    that the positions of absorber \#5 and beyond are in the low flux and S/N part of the \ion{C}{iv} lines, which also suffers from skyline contamination and cannot be robustly fitted. We notice that we allow the redshift of the absorbers to be free within a limited range ($\Delta z = 0.006$ for absorber \#1, $\Delta z = 0.004$ for absorber \#4) following \citet{2019A&A...625A.102K} who argues that fixing the redshift of absorption caused by different species is un-physical given their different ionization energies.

Finally, we fixed the Doppler parameters and redshifts of absorbers \#2 and \#3 to the values derived from \ion{H}{i} absorbers \#2 and \#3. We also set the ranges of the column densities to $10^{11.5}-10^{12}\,\rm cm^{-1}$ and $10^{12}-10^{14}\,\rm cm^{-1}$ for absorbers \#2 and \#3, respectively. These are implemented due to the large overlapping of these two absorbers (with others), which leads to a failure of fitting without further constrains.

The initial guess for the redshift (\ion{He}{ii} line center), 4.514, is adopted from \citet{2006AstL...32..433K} and is used only in the least-squares fitting. After the redshift constrained to a relatively satisfied range, we allow it to vary within $\pm 2$ \AA  during the MCMC fitting. The velocity shift of the blueshifted \ion{C}{iv} component is left relatively free with a broad range, $-3000\sim-500\,\rm km\,s^{-1}$. The results from $\sim -1500\,\rm km\,s^{-1}$ to $\sim -900\,\rm km\,s^{-1}$ will all give us satisfied overall fit. This is also indicated by the distribution of $L_{\ion{C}{iv},2}$, line center of the blueshifted component, seen in Fig. \ref{fig:civheiicorner}, which has a long tail toward the lower values. Based on this, we argue in Sect. \ref{sec:nv} that the blueshifted component of \ion{C}{iv} and \ion{N}{v} likely result from the same gaseous cloud. The boundaries of the Doppler parameters, $b$, and column density, $N_{\ion{C}{iv}}$, are set to $40-400\,\rm km\,s^{-1}$ and $10^{13}-10^{16}\,\rm cm^{-2}$, respectively (stricter settings for absorber \#2 and \#3 as mentioned above) following \citet{2019A&A...625A.102K}. Table \ref{tab:civheiinvconstrain} summarizes the constraints used in the \ion{C}{iv} and \ion{He}{ii} fitting. The rules for line center, line width and line flux of the doublet are presented as well as the constraints of the absorption parameters of the doublet. 

The major result we interested in from the \ion{He}{ii} fitting is the systemic redshift. We test the possibility of using a double Gaussian to fit the \ion{He}{ii}. The result is similar ($4.5079 \pm 0.0001$ and $4.5077\pm 0.0001$) to the Lorentzian fit and shows that the algorithm will fit both of the line centered at $\Delta v \sim 0 \rm \, km\,s^{-1}$. This further indicates that the S/N of the line is not enough to fit the two Gaussian. 

We test the possibilities of using (i) only absorbers \#1 and \#4 or (ii) absorbers \#1, \#3, and \#4 to fit the \ion{C}{iv,} which is under the assumption that absorbers \#2 and/or \#3 have low metallicity. Both settings (i) and (ii) have similar results for absorbers \#1 and \#4. Hence, we include all four absorbers and avoid the strong assumption.

\subsection{\texorpdfstring{\ion{N}{v}}{NV} fitting notes}\label{apd:fittingnotesnv}

The position of the \ion{N}{v} is at the broad wing of Ly$\alpha$.
The flux excess we see on Fig. \ref{fig:lyafit} above the fitted Ly$\alpha$ model at around 5000 $\rm km\,s^{-1}$ is due to the contribution of the blueshifted \ion{N}{v} component. In addition, the low flux and highly absorbed systemic emission make \ion{N}{v} a non-obvious detection. 

As described in Sect. \ref{sec:nv}, we fix the best-fit Ly$\alpha$ model from Sect. \ref{sec:lya} and use a constant continuum when fitting the \ion{N}{v}. The reasons we do not apply the same continuum level for Ly$\alpha$ and \ion{N}{v} lines are: (i) we test in Appendix \ref{apd:lyacontinuum} that using a continuum determined from the red wing and free first-order polynomial have similar results for the \ion{H}{i} fitting (Appendix \ref{apd:lyacontinuum}); (ii) the first-order polynomial will slightly overestimate the continuum flux of Ly$\alpha$ red wing whose influence is probably negligible for Ly$\alpha,$ which is a broad line with high flux. For \ion{N}{v}, however, which is at longer wavelength than Ly$\alpha$ and has extremely low S/N, the continuum overestimation will affect the fit more. This is indeed the case when the first-order polynomial continuum is applied during the \ion{N}{v} fitting, which results in poor constraints.

As we see in Fig. \ref{fig:nvcorner}, the line center of the blueshifted component, $L_{\ion{N}{v},b}$, and Doppler parameter, $b_{1}$, are not fully constrained. For $L_{\ion{N}{v},2}$, we set the range $-1600\sim-500\,\rm km\,s^{-1}$ according to \ion{C}{iv} blueshifted component. The reason it hits the lower boundary is due to the influence of Ly$\alpha$ red wing. As mentioned in Appendix \ref{apd:civheiimaster}, the velocity shift down to $-1500\,\rm km\,s^{-1}$ will still give satisfying results for the \ion{C}{iv} fit. The main parameter in this fit, $N_{\ion{N}{v}}$, does not change much if we vary the given boundary of the line center of the blueshifted component. Besides, we test to leave it relatively free and end up with poor fit quality (unphysical result). Therefore, we manually set this lower boundary, which results in a satisfying fit given the low S/N. More importantly, this velocity shift shows the consistence between \ion{N}{v} and \ion{C}{iv,} which agrees with the hypothesis. For $b_{1}$, which hits the upper boundary, we tested fixing it to the value obtained from \ion{C}{iv} absorber \#1, which results in a poor fit quality as well. This could probably be due to (i) the low S/N of the \ion{N}{v}; (ii) the influence of the unresolved \ion{H}{i} redshifted absorber(s); (iii) the imperfect sky line subtraction. Hence, we present this fit and remind the readers to be cautious about the $N_{\ion{N}{v}}$, which should be considered as a lower limit given the well-known $b-N$ degeneracy \citep{2018MNRAS.474.3649S}.

\section{Ly\texorpdfstring{$\alpha$}{a} continuum sensitivity test}\label{apd:lyacontinuum}
   \begin{figure}
  \centering
        \includegraphics[width=\hsize]{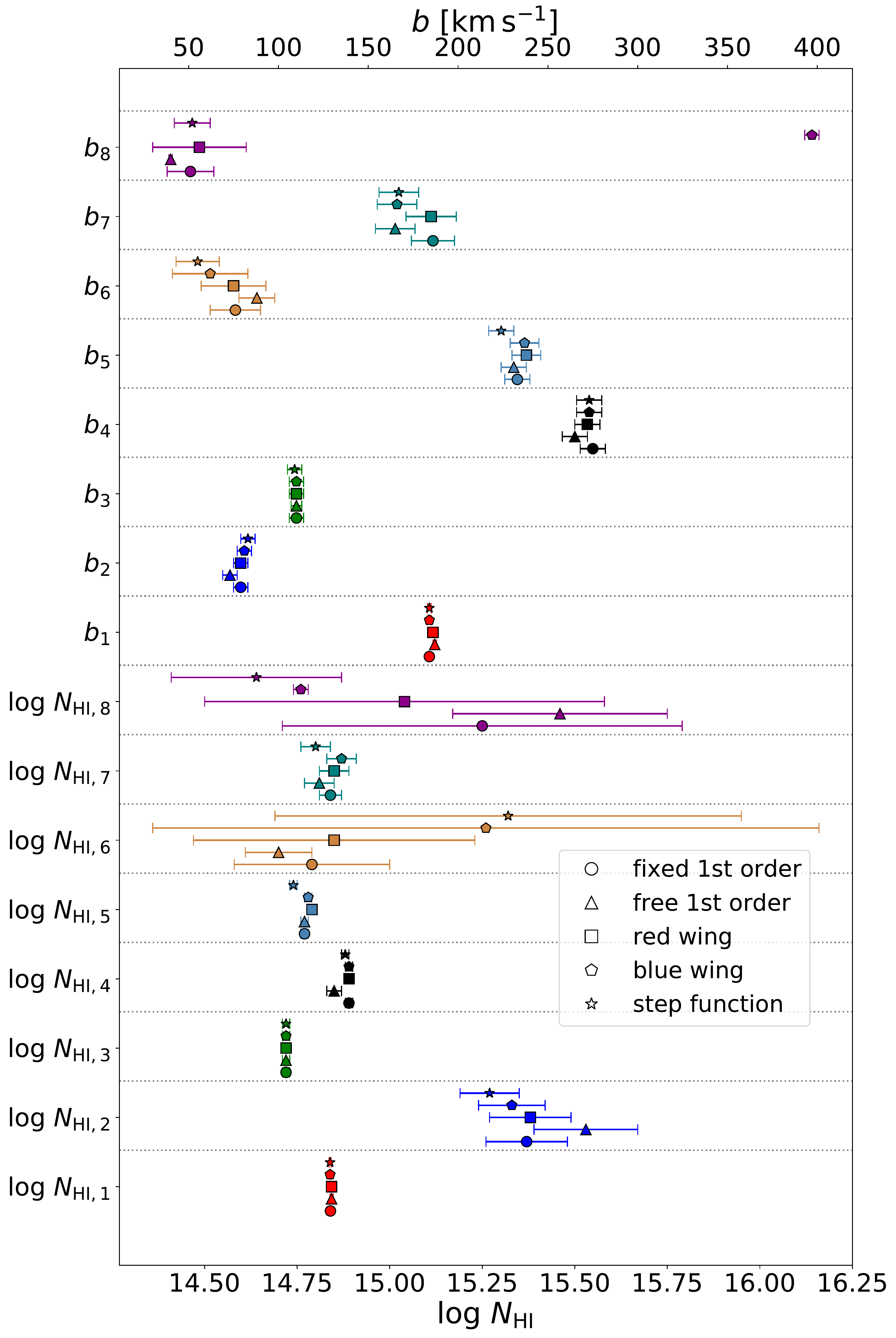}
      \caption{MCMC fitted Voigt parameters, $N_{\rm H}$ and $b$, of the eight \ion{H}{i} absorbers from five continuum fitting methods. The subscripts 1$-$8 represent the indices of the absorbers. The colors represent different parameters (the same color is used for the $N_{\rm H}$ and $b$ of the same absorber), while the mark styles indicate different continuum fitting methods. }
         \label{fig:continuumcheck}
   \end{figure}

The Ly$\alpha$ emission of quasars and galaxies at high redshift are highly absorbed by the intervening hydrogen clouds located between observer and source. This so-called Ly$\alpha$ forest \citep[e.g.,][]{ 2003ApJ...584...45A} heavily affects the blue wing of Ly$\alpha$ making it difficult for continuum fitting. For our MUSE observation, the spectral resolution is too low to resolve the narrow intervening absorbers, but we can clearly identify a change between the continuum of the blue and red side of the Ly$\alpha$, namely the continuum flux is lower in the blue than the red. Several potential methods can be used to fit this continuum. Hence, we run a test to see how significant the change that different continuum fitting strategies may cause the absorption fitting. We use five different methods, which are described below. 

      In the first method, the continuum is fitted using a first-order polynomial prior to the Gaussian+Voigt fitting and fixed during the following fitting. The wavelength range used in this method for the continuum is $6405-6986\,\text{\AA}$ ($-$13\,000 $\sim$ 13\,000 $\rm km\,s^{-1}$) with the emission region masked.
      
      For the second method, the continuum is fitted using a first-order polynomial together with the Gaussian+Voigt fitting. The values of slope and intercept from method 1 are used as initials.
      
      In the third method, the continuum is fitted using a zero order polynomial of the red wing and fixed during the Gaussian+Voigt fitting. The wavelength range used in this method for the continuum is $6880-6986\,\text{\AA}$ (8\,258 $\sim$ 13\,000 $\rm km\,s^{-1}$). We chose the wavelength range to be at extremely red wing of the Ly$\alpha$ to avoid the contamination of the \ion{N}{v} (Sect. \ref{sec:nv} and Appendix \ref{apd:fittingnotesnv}).
      
      For the fourth method, the continuum is fitted using a zero order polynomial of the blue wing and fixed during the Gaussian+Voigt fitting. The wavelength range used in this method for the continuum is $6405-6600\,\text{\AA}$ ($-13\,000$ $\sim$ $-4\,300$ $\rm km\,s^{-1}$).
      
      As for the fifth method, the continuum is fitted using a step function. The wavelength of the step is fitted together with the Gaussian+Voigt fitting. The left (right) value is set to the one from blue (red) wing zero order polynomial result and fixed during the fit.
      
 The result is shown in Fig. \ref{fig:continuumcheck} in which the fitted Voigt parameters using MCMC method, $N_{\rm \ion{H}{i}}$ and $b$, of the eight absorbers from different continuum fitting strategies are presented. It is intuitive to identify that most of the fitted values from different continuum methods are consistent, though some have relatively large scatters (e.g., absorbers \#6$-$8), which is understandable given that these absorbers are located at the low S/N tail. Therefore, we decided that the continuum fitting method has a minor effect on the absorption parameters, which is our primary focus in this work. By checking the corner plots produced from these five methods, we find that fitting the first-order polynomial continuum together with the Gaussian+Voigt model (method 2) constrains the probability distribution of the absorption parameters ($z$, $b,$ and $N_{\rm\ion{H}{i}}$) best. Hence, we use the method 2 when doing the Ly$\alpha$ fit. Using a first-order polynomial to fit the continuum is a commonly used method \citep[e.g.,][]{2019A&A...625A.102K}, and giving freedom to the slope and intercept will constrain better the line features.
 
 As for the spatially resolved Ly$\alpha$ analysis, we first fit the continuum level of the spectra in each spatial bin with emission part masked. We then keep the continuum fixed to this level when running the Gaussian+Voigt fitting. Because, for consistence, we use the same model for fitting each of the 64 spectrum (see Sect. \ref{sec:spatialmethod}) and some spectra have lower S/N ratios, which may affect the fitting if we add more free parameters (for the continuum).

\section{Auxiliary materials of MCMC fitting}\label{apd:auxmcmc}

In this appendix, we present some side products of the MCMC fitting, which can be used as auxiliary materials to better understand the fitting quality, namely the corner plot and acceptance fraction plot. The corner plot is a tracer for the probability distributions of the fitted parameters and correlation between each parameters; in other words, we can identify the degeneracy between fitted parameters from the corner plot. The acceptance fraction plot is used to check the acceptance fraction of each walker used in MCMC \citep[see][for details]{2013PASP..125..306F}. This can be used as an easy check of the performance of the MCMC run. Although some details are not clear, the acceptance fraction lies in 0.2$-$0.5 is thought to be an indication of a successful run \citep[][]{2013PASP..125..306F}.
   \begin{figure}
  \centering
        \includegraphics[width=\hsize]{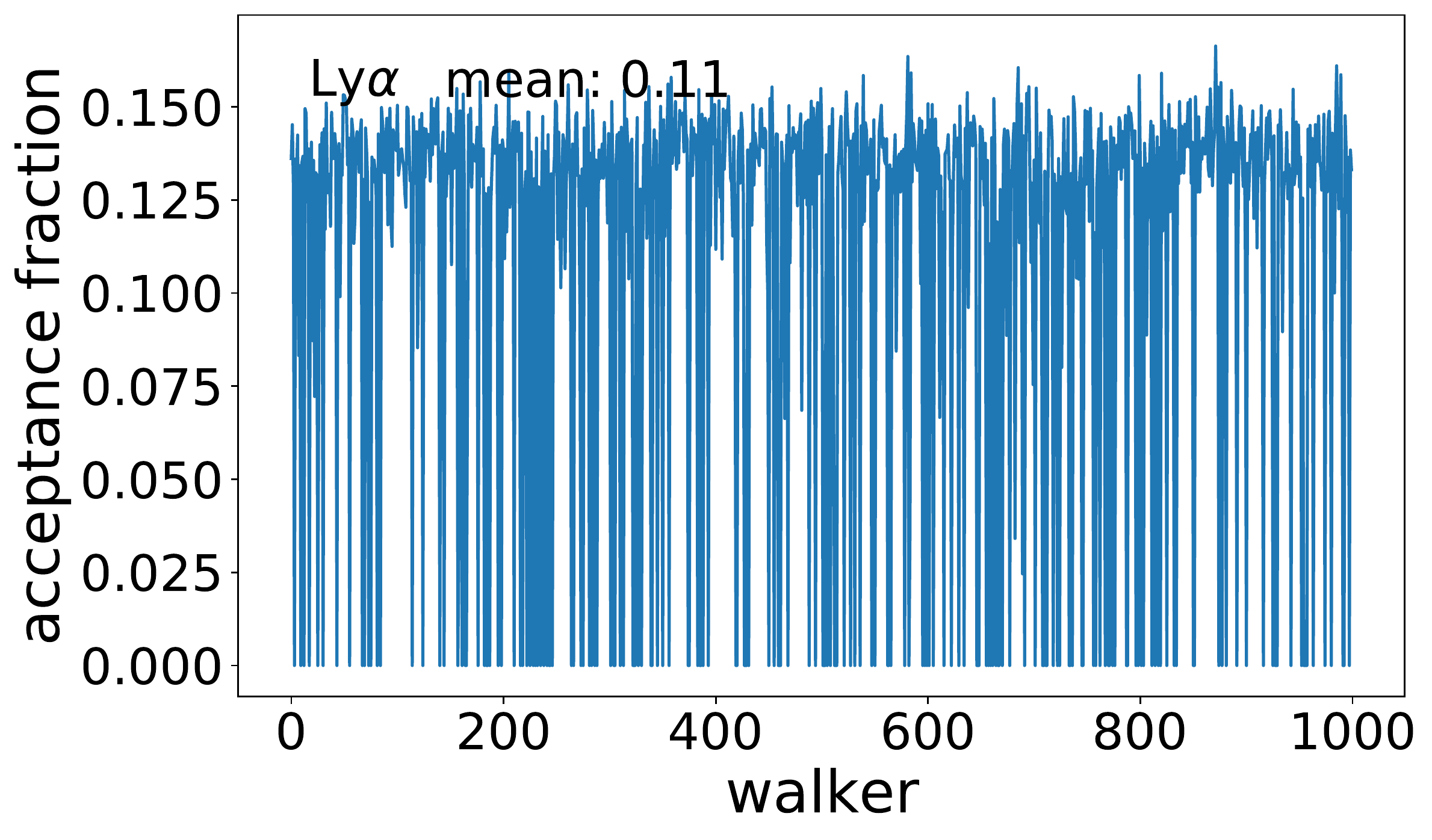}
      \caption{Acceptance fraction of each walker used in the Ly$\alpha$ MCMC fitting.}
         \label{fig:lyaaccept}
   \end{figure}

  \begin{figure*}
  \centering
        \includegraphics[width=16.4cm,clip]{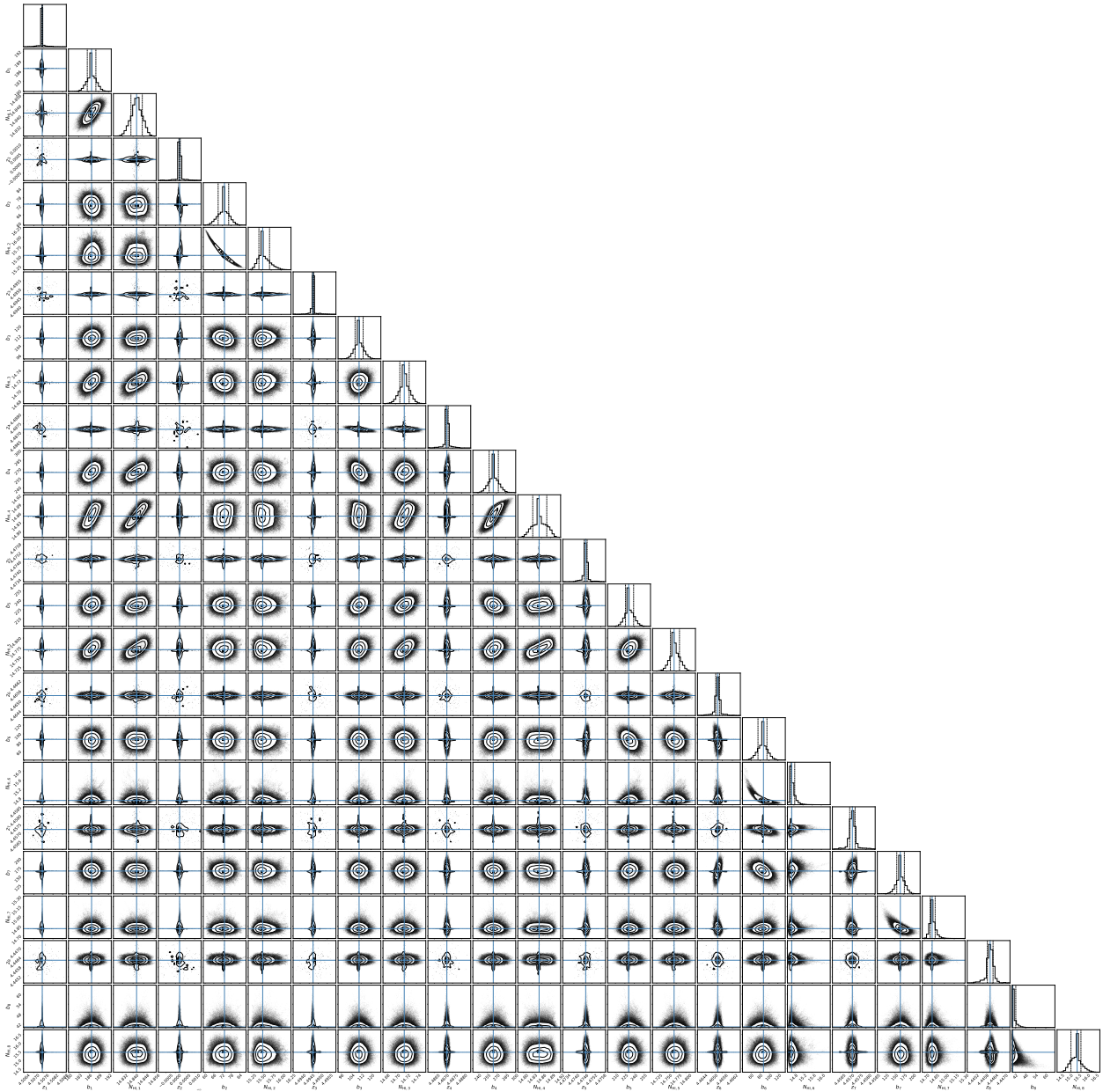}
      \caption{Corner plot derived from the MCMC fitting of the Ly$\alpha$ line (see Sect. \ref{sec:lya}). In this figure we only show the correlation between the absorption parameters, namely $z_{i}$, $b_{i}$, and $N_{\ion{C}{iv}i}$, which are the fitted redshifts, Doppler parameters (in units of $\rm km\,s^{-1}$), and column densities (logarithmic) of the eight \ion{H}{i} absorbers. The black dotted lines in each of the histograms represent 15.8 and 84.2 percentiles, which correspond to the reported uncertainty ranges. The blue solid lines mark the median and reported fit values, respectively.  }
         \label{fig:lyacorner}
  \end{figure*}

   \begin{figure}
  \centering
        \includegraphics[width=\hsize]{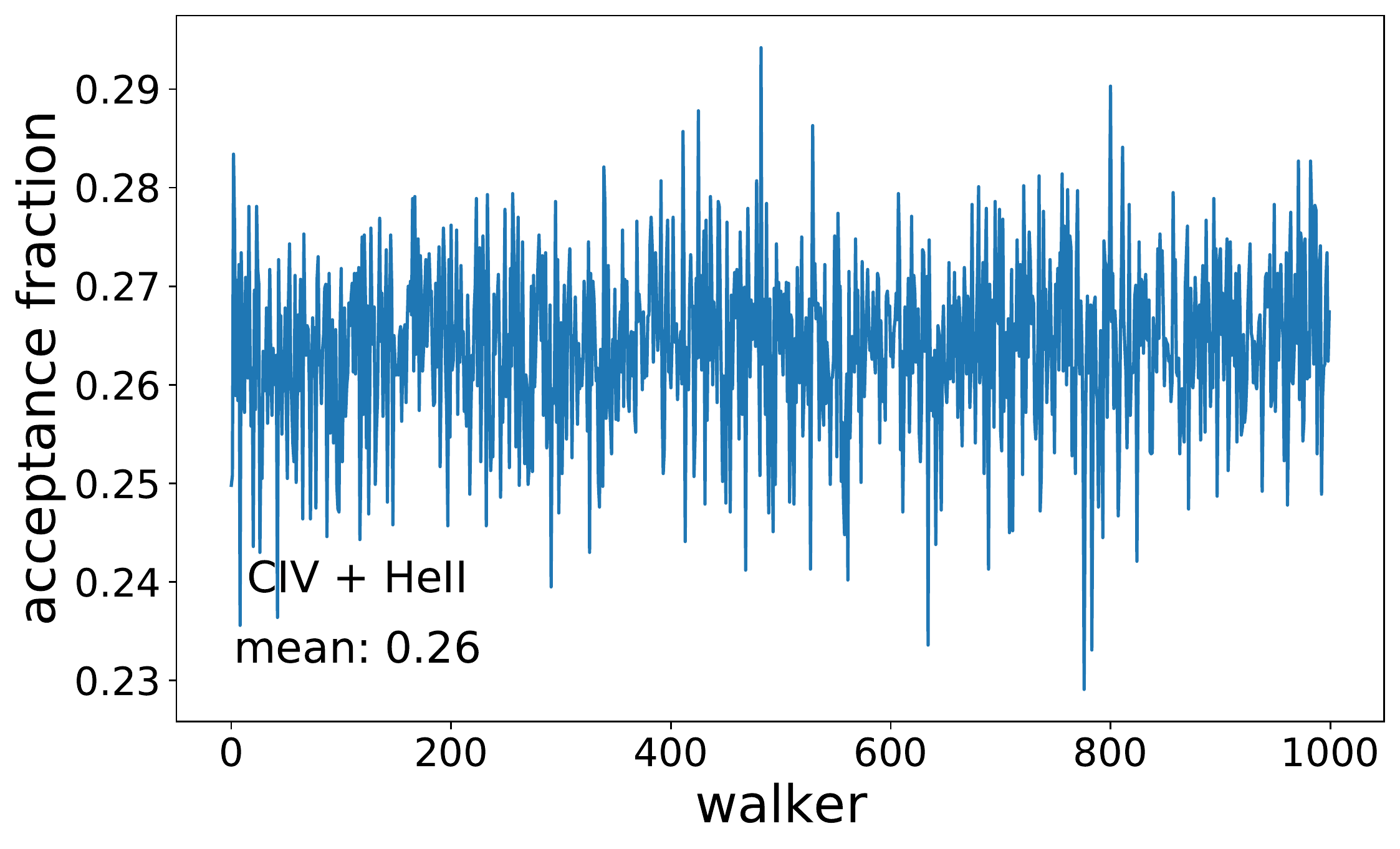}
      \caption{Acceptance fraction of each walker used in \ion{the C}{iv} and \ion{He}{ii} MCMC fitting.}
         \label{fig:civheiiaccept}
   \end{figure}

  \begin{figure*}
  \centering
        \includegraphics[width=16.4cm,clip]{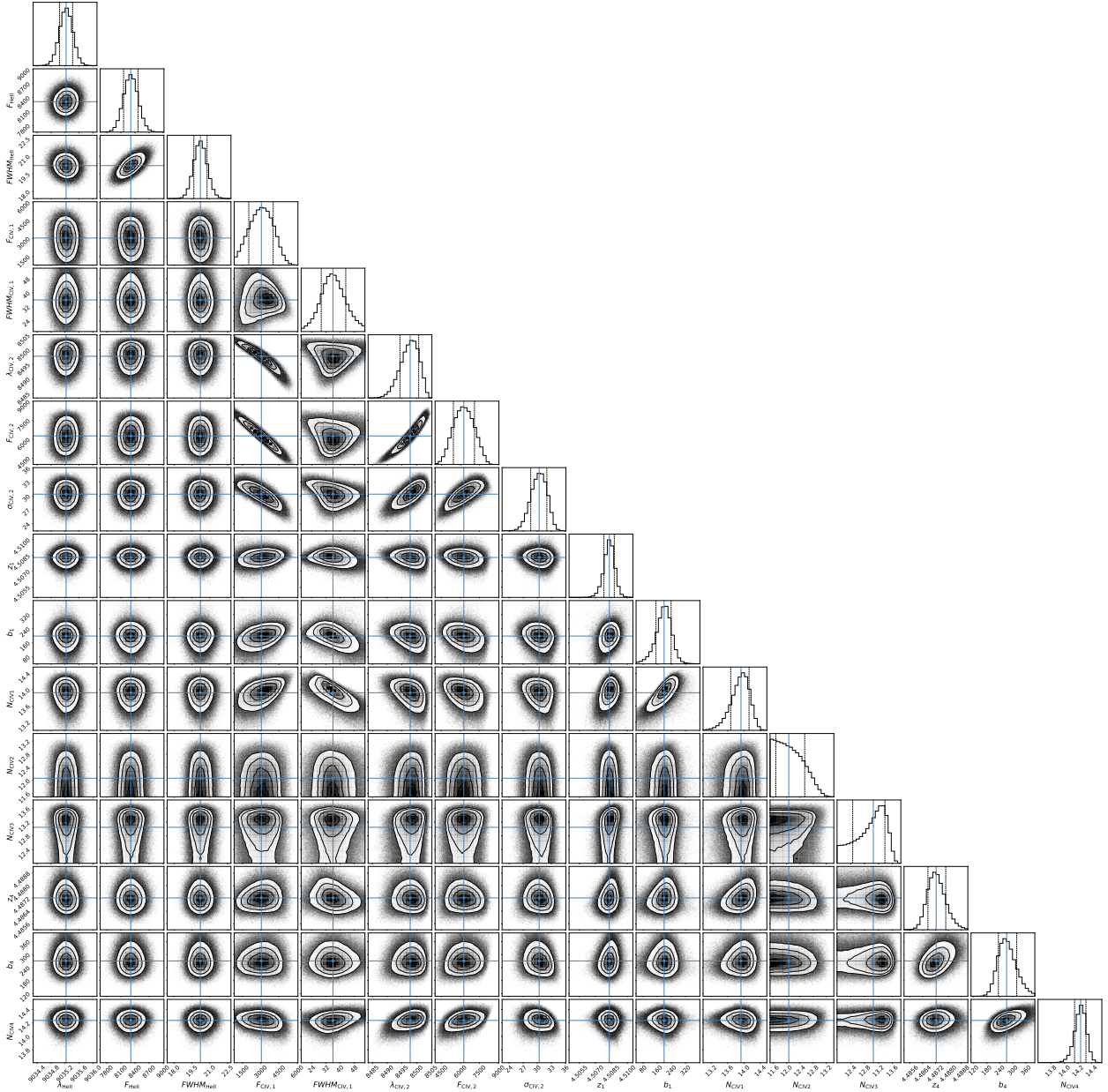}
      \caption{Corner plot derived from the MCMC fitting of the \ion{C}{iv} and \ion{He}{ii} (see Sect. \ref{sec:civheii}). In this figure we show the correlation between the free parameters in \ion{the C}{iv}$+$\ion{He}{ii} fit. The $\lambda_{\rm \ion{He}{ii}}$ and $\lambda_{\ion{C}{iv},2}$ are the line centers of the \ion{He}{ii} and blueshifted \ion{C}{iv} (for the 1548 \AA \ line) emissions in \AA, respectively. The $F_{\ion{He}{ii}}$, $F_{\ion{C}{iv},1}$, and $F_{\ion{C}{iv},2}$ are the integrated line fluxes of \ion{He}{ii}, systemic \ion{C}{iv,} and blueshifted \ion{C}{iv}  (only for the 1548 \AA\  line) emissions in $10^{-20}\,\rm erg\,s^{-1}\,cm^{-2}\,\AA^{-1}$, respectively. The $FWHM_{\ion{He}{ii}}$, $FWHM_{\ion{C}{iv}}$, and $\sigma_{\ion{C}{iv},2}$ are the line widths of \ion{He}{ii}, systemic \ion{C}{iv,} and blueshifted \ion{C}{iv}  (for the 1548 \AA \ line) emissions in \AA. $z_{i}$, $b_{i}$, and $N_{\ion{C}{iv}i}$, which are the fitted redshifts, Doppler parameters (in units of $\rm km\,s^{-1}$), and column densities (logarithmic) of the four absorbers. The black dotted lines in each of the histograms represent 15.8 and 84.2 percentiles, which correspond to the reported uncertainty ranges. The blue solid lines mark the median and reported fit values, respectively.}
         \label{fig:civheiicorner}
  \end{figure*}

   \begin{figure}
  \centering
        \includegraphics[width=\hsize]{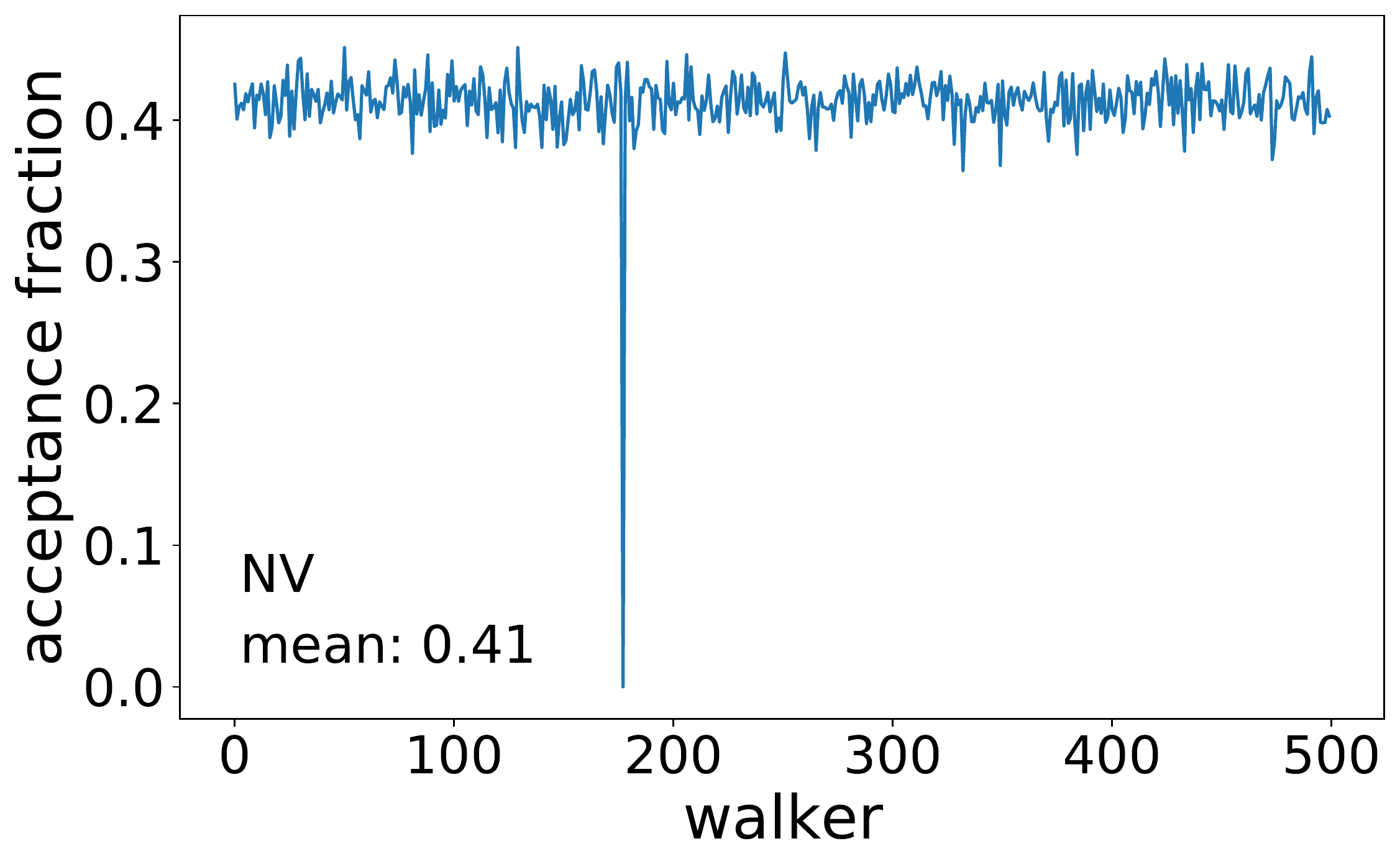}
      \caption{Acceptance fraction of each walker used in \ion{the N}{v} MCMC fitting.}
         \label{fig:nvaccept}
   \end{figure}

  \begin{figure*}
  \centering
        \includegraphics[width=16.4cm,clip]{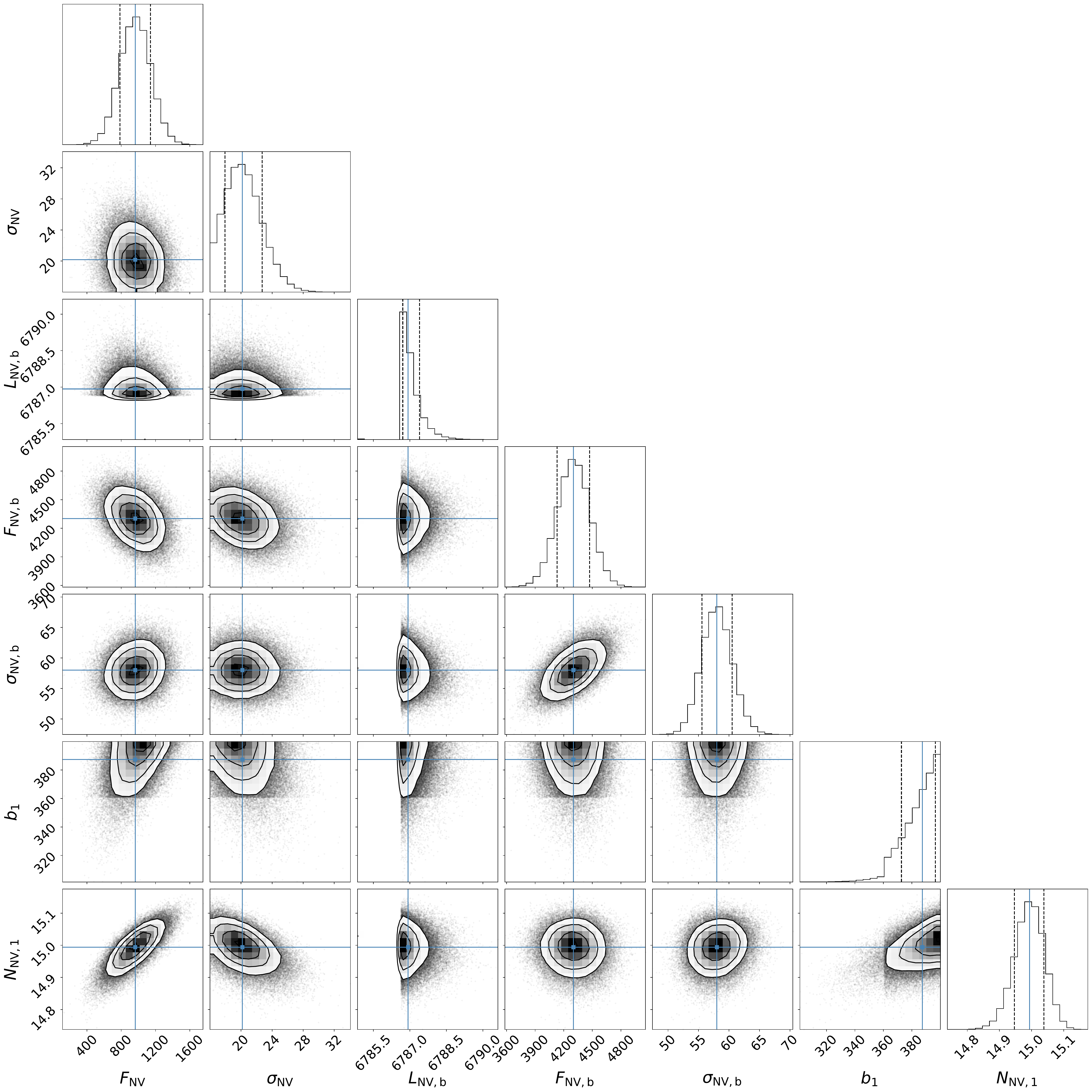}
      \caption{Corner plot derived from the MCMC fitting of the \ion{N}{V} (Sect. \ref{sec:civheii}). In this figure we show the integrated line flux ($F_{\rm \ion{N}{v}}$ in $10^{-20}\,\rm erg\,s^{-1}\,cm^{-2}\,\AA^{-1}$) and the Gaussian line width ($\sigma_{\rm \ion{N}{v}}$, in \AA) of the systemic \ion{N}{v} emission; the line center ($L_{\rm \ion{N}{v},b}$ in \AA), integrated line flux ($F_{\ion{N}{v},b}$), and Gaussian line width ($\sigma_{\ion{N}{v},b}$) of the blueshifted \ion{N}{v} component (for the 1238\AA line); and the Doppler parameter ($b_{1}$) and column density ($N_{\rm \ion{N}{V}}, 1$) of absorber \#1. The black dotted lines in each of the histograms represent 15.8 and 84.2 percentiles, which correspond to the reported uncertainty ranges. The blue solid lines mark the median and reported fit values, respectively. }
         \label{fig:nvcorner}
  \end{figure*}

   \begin{figure}
  \centering
        \includegraphics[width=\hsize]{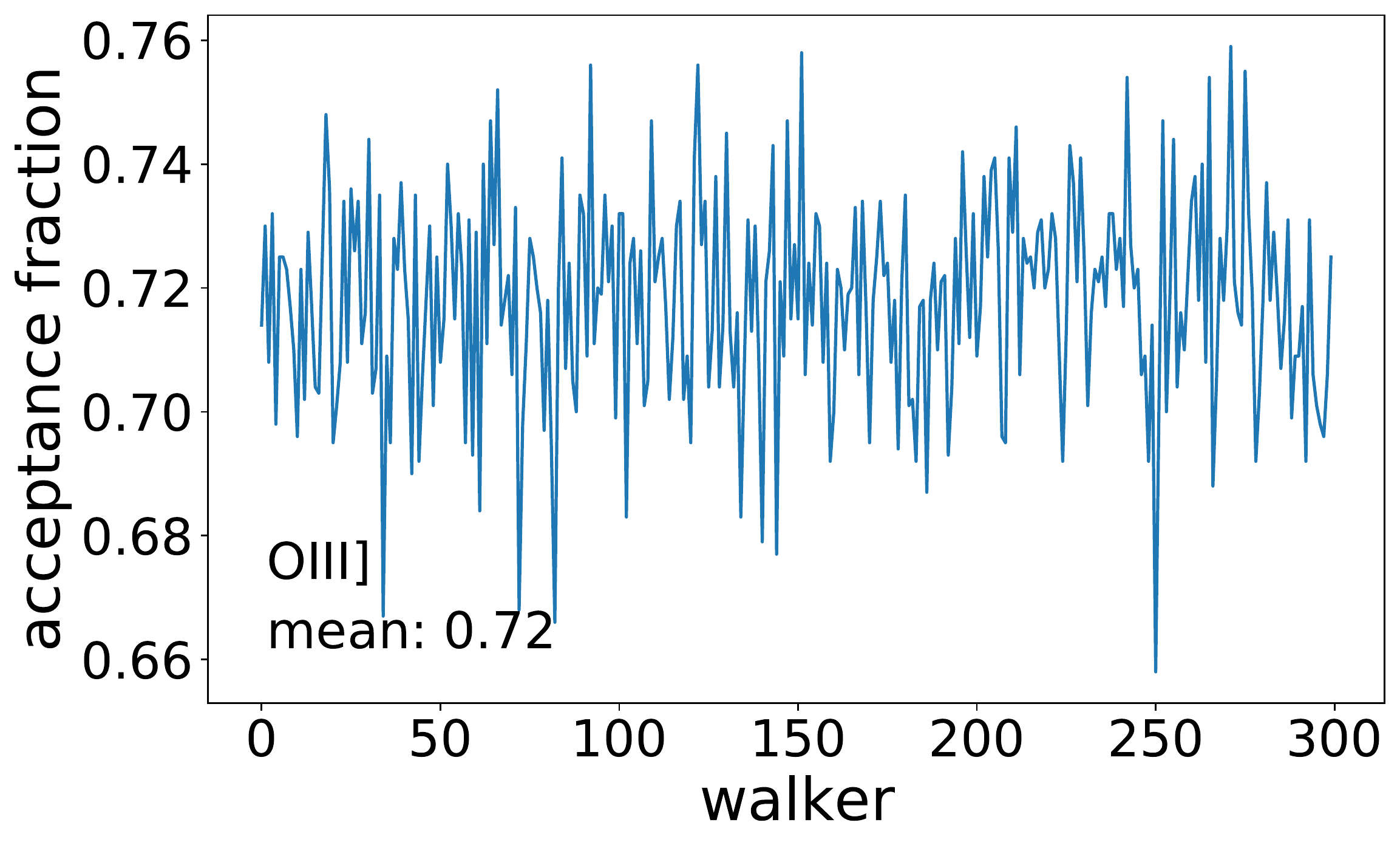}
      \caption{Acceptance fraction of each walker used in \ion{O}{iii}] MCMC fitting.}
         \label{fig:o3accept}
   \end{figure}

   \begin{figure}
  \centering
        \includegraphics[width=\hsize]{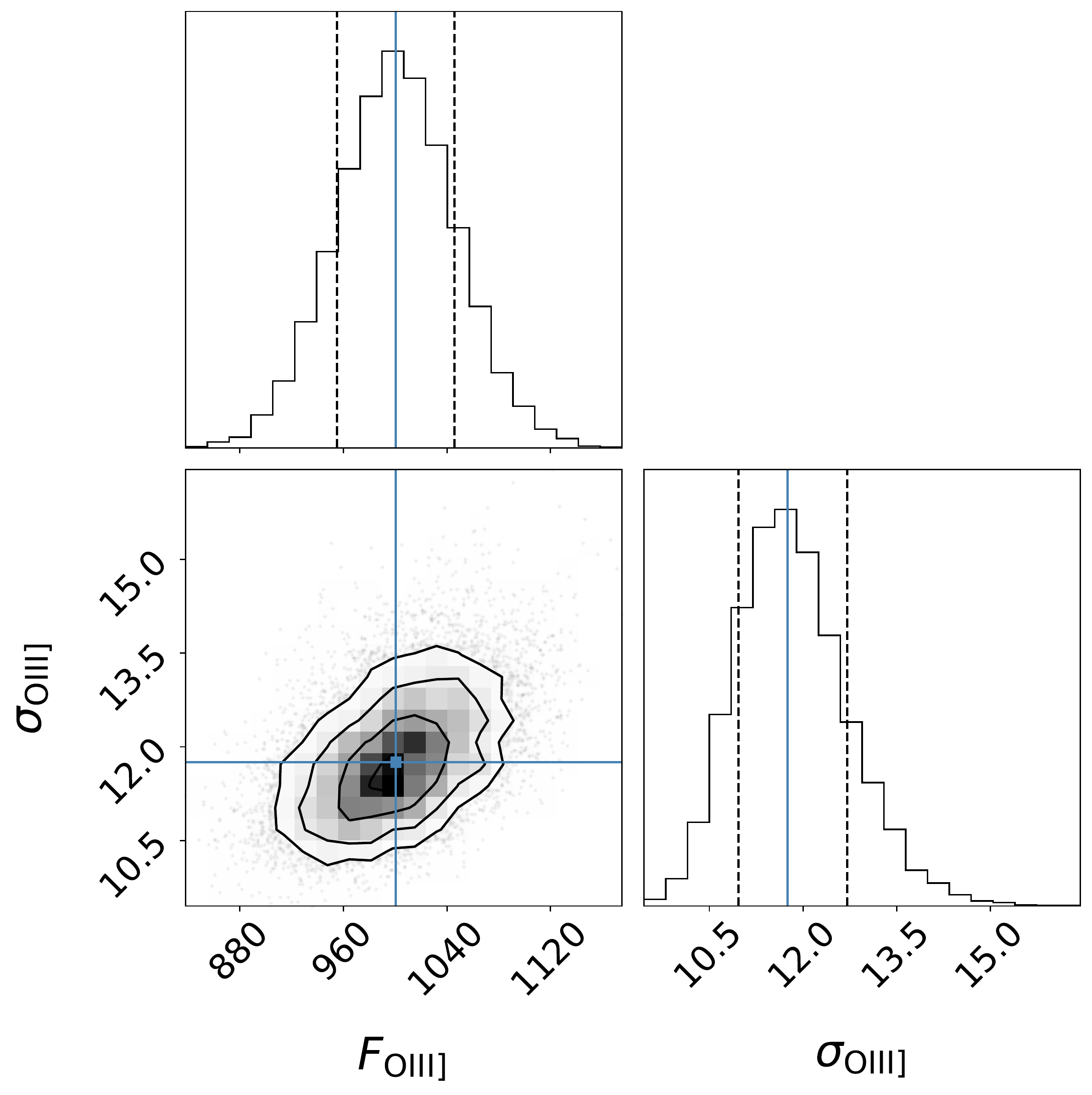}
      \caption{Corner plot derived from the MCMC fitting of the \ion{O}{iii}] (Sect. \ref{sec:oiii}). The $F_{\rm \ion{O}{iii}]}$ and $\sigma_{\ion{O}{iii}]}$ are the integrated line flux and line width of the 1660.81\AA\,doublet line, respectively. The black dotted lines in each of the histograms represent 15.8 and 84.2 percentiles, which correspond to the reported uncertainty range. The blue solid lines mark the median and reported fit values, respectively.}
         \label{fig:o3corner}
   \end{figure}


\subsection{Ly\texorpdfstring{$\alpha$}{a}}\label{apd:lyamaster}

The acceptance fraction and corner plot of MCMC fitting of master Ly$\alpha$ are shown in Fig. \ref{fig:lyaaccept} and \ref{fig:lyacorner}, respectively. In Fig. \ref{fig:lyaaccept}, we can see that a number of walkers are rejected with a mean acceptance fraction of 0.11. This is due to the larger number (32) of free parameters used in the Ly$\alpha$ fitting. In Fig. \ref{fig:lyacorner}, we identify several banana shapes of the correlation distribution (e.g., between $b_{\rm i}$ and $N_{\rm Hi}$). This is a well-known degeneracy in the Voigt fitting \citep[e.g.,][]{2018MNRAS.474.3649S} that a combination of larger $N$ value and smaller $b$ value can the overall similar result with a combination of smaller $b$ and larger $N$.

\subsection{\texorpdfstring{\ion{C}{iv}}{CIV} and \texorpdfstring{\ion{He}{ii}}{HeII}}\label{apd:civheiimaster}
The acceptance fraction and corner plot of MCMC fitting of master \ion{C}{iv} \ion{He}{ii} are shown in Fig. \ref{fig:civheiiaccept} and \ref{fig:civheiicorner}, respectively. In Fig. \ref{fig:civheiiaccept}, we can see that the acceptance fraction of all of the walkers are in the range of $0.2-0.5$ with a mean of 0.26. This is evidence indicating the success of the fitting.
In Fig. \ref{fig:civheiicorner}, we see that the probability distributions of $N_{\rm \ion{C}{iv}2}$ and $N_{\rm \ion{C}{iv}3}$ have tentative peak and un-defined tail to the lower value. This is more severe for absorber \#2 given that the lower boundary we apply ($10^{11.5}\,\rm cm^{-2}$) is extremely low. Based on the above, we decide to treat the column density of absorber \#2 and \#3 as upper limit.

\subsection{\texorpdfstring{\ion{N}{v}}{NV}}\label{apd:nvmaster}
The acceptance fraction and corner plot of the MCMC fitting of the master \ion{N}{v} are shown in Fig. \ref{fig:nvaccept} and \ref{fig:nvcorner}, respectively. In Fig. \ref{fig:nvaccept}, we can see that all walkers are in the range of 0.2$-$0.5 with a mean of 0.41, which indicates the MCMC fitting of \ion{N}{v} is successful. We discuss the uncompleted sampled distribution of $L_{\rm \ion{N}{v},b}$ and $b_{1}$ in Appendix \ref{apd:fittingnotesnv}.

\subsection{\texorpdfstring{\ion{O}{iii}}{OIII}]}\label{apd:o3master}
The acceptance fraction and corner plot of the MCMC fitting of master \ion{O}{iii}] are shown in Fig. \ref{fig:o3accept} and \ref{fig:o3corner}, respectively. We perform the MCMC fitting for this two-free-parameter model to be consistent with other fittings. In Fig. \ref{fig:o3accept}, we can see that all walkers are above 0.5 with a mean of 0.72, which indicates that the model is probably over-fitted.

\section{Individual Ly\texorpdfstring{$\alpha$}{a} spectra fitting}\label{apd:lyaspatialspectra}
   \begin{figure}
  \centering
        \includegraphics[width=\hsize]{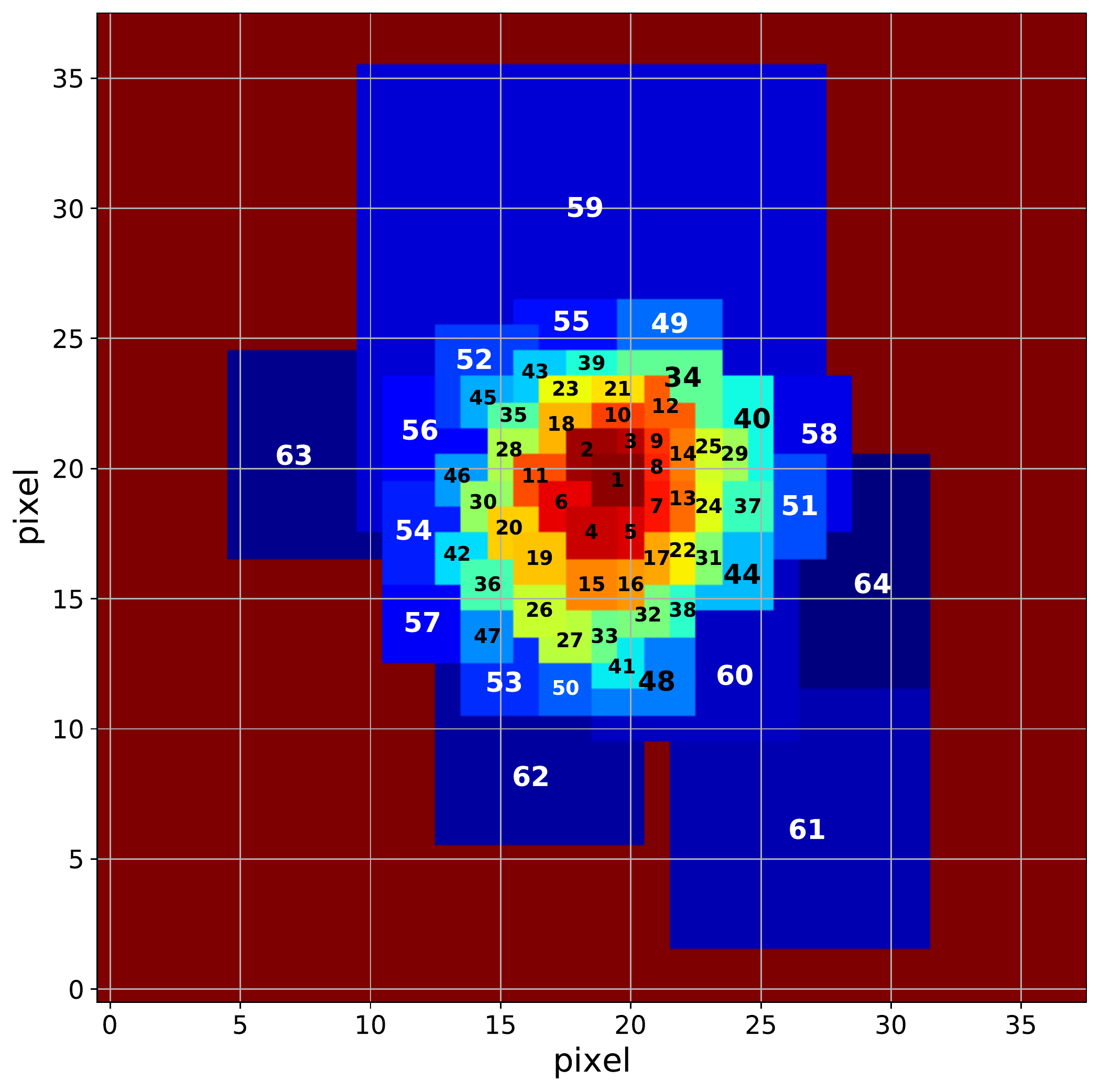}
      \caption{Spatial binning results from Sect. \ref{sec:spatialmethod}. The colors represents different bins. The bin numbers correspond to the numbers marked in Figs. \ref{fig:lyaindividualspatialplots1} - \ref{fig:lyaindividualspatialplots4}. The black or white colors of the bin number are given to better distinguish it from the color of the bin.   }
         \label{fig:binum_map}
   \end{figure}

\begin{figure*}
  \centering
        \includegraphics[width=16.4cm,clip]{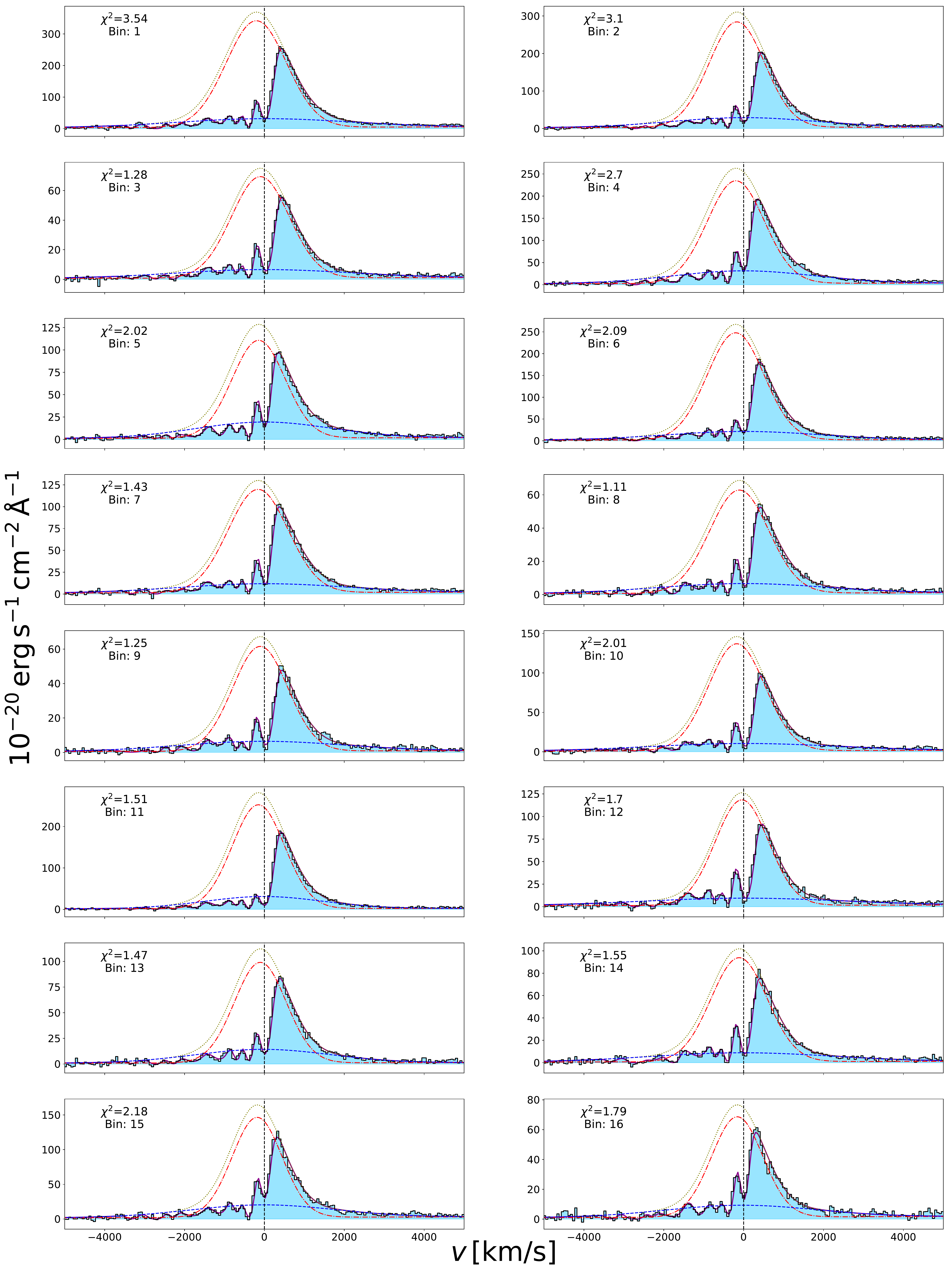}
      \caption{Spectra extracted from the 64 spatial bins (see Sect. \ref{sec:spatialmethod}) and Gaussian+Voigt fitting results. In each panel, the thick dark magenta line indicates the best fit model. The dot-dash red line and dashed blue line denote the narrow and broad emission components, respectively. The dotted olive lines are the summation of the two emissions. The $\chi^{2}_{\nu}$ calculated from each fit is shown as an indicator of the fit quality. We note that each spectrum is summed from the spatial bin, each of which contains a different number of spaxels. Hence, the fluxes of the spectra shown here are not to be compared directly.}
         \label{fig:lyaindividualspatialplots1}
  \end{figure*}
\begin{figure*}
  \centering
        \includegraphics[width=16.4cm,clip]{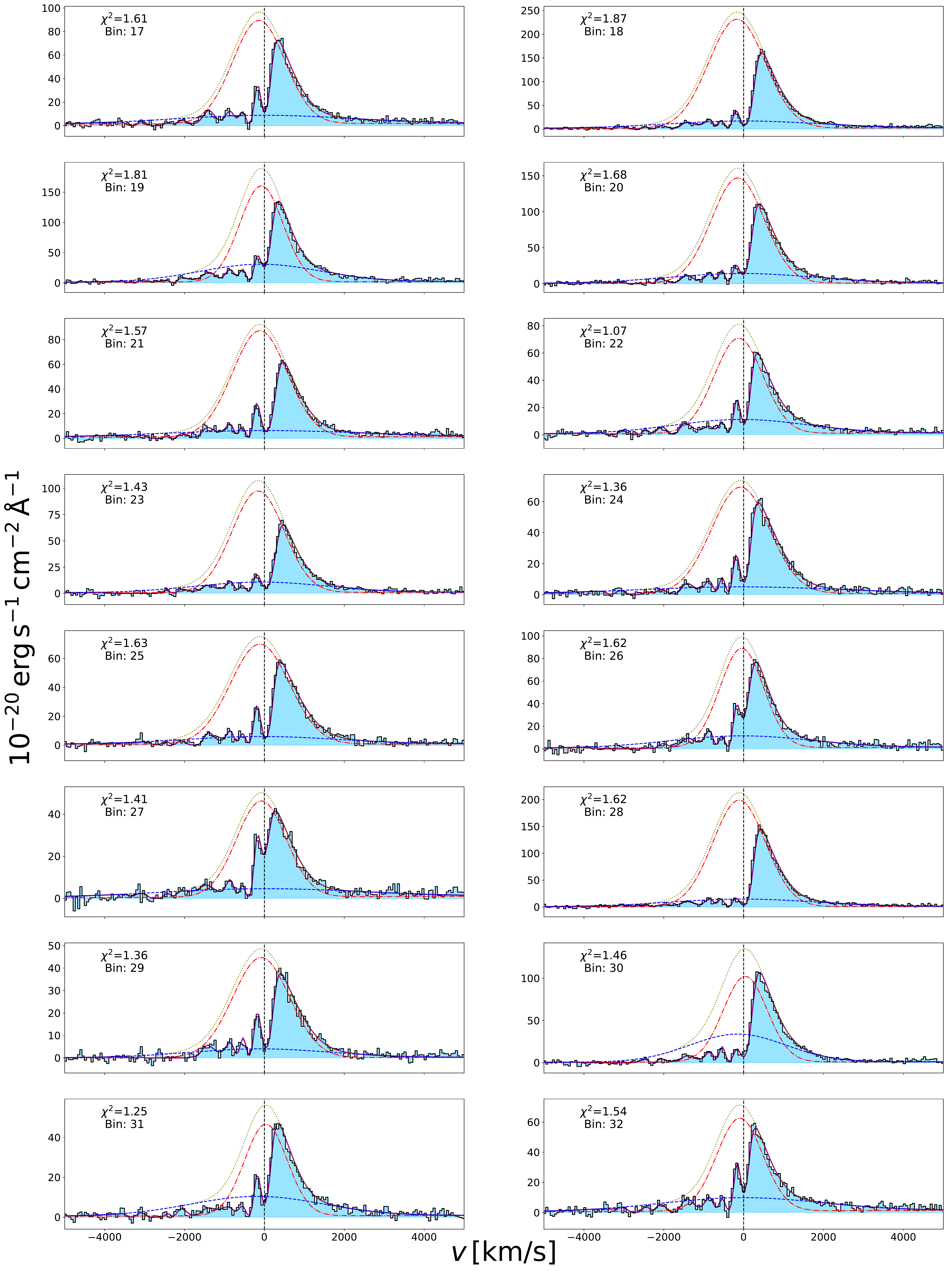}
      \caption{Figure \ref{fig:lyaindividualspatialplots1} continued.}
         \label{fig:lyaindividualspatialplots2}
  \end{figure*}
\begin{figure*}
  \centering
        \includegraphics[width=16.4cm,clip]{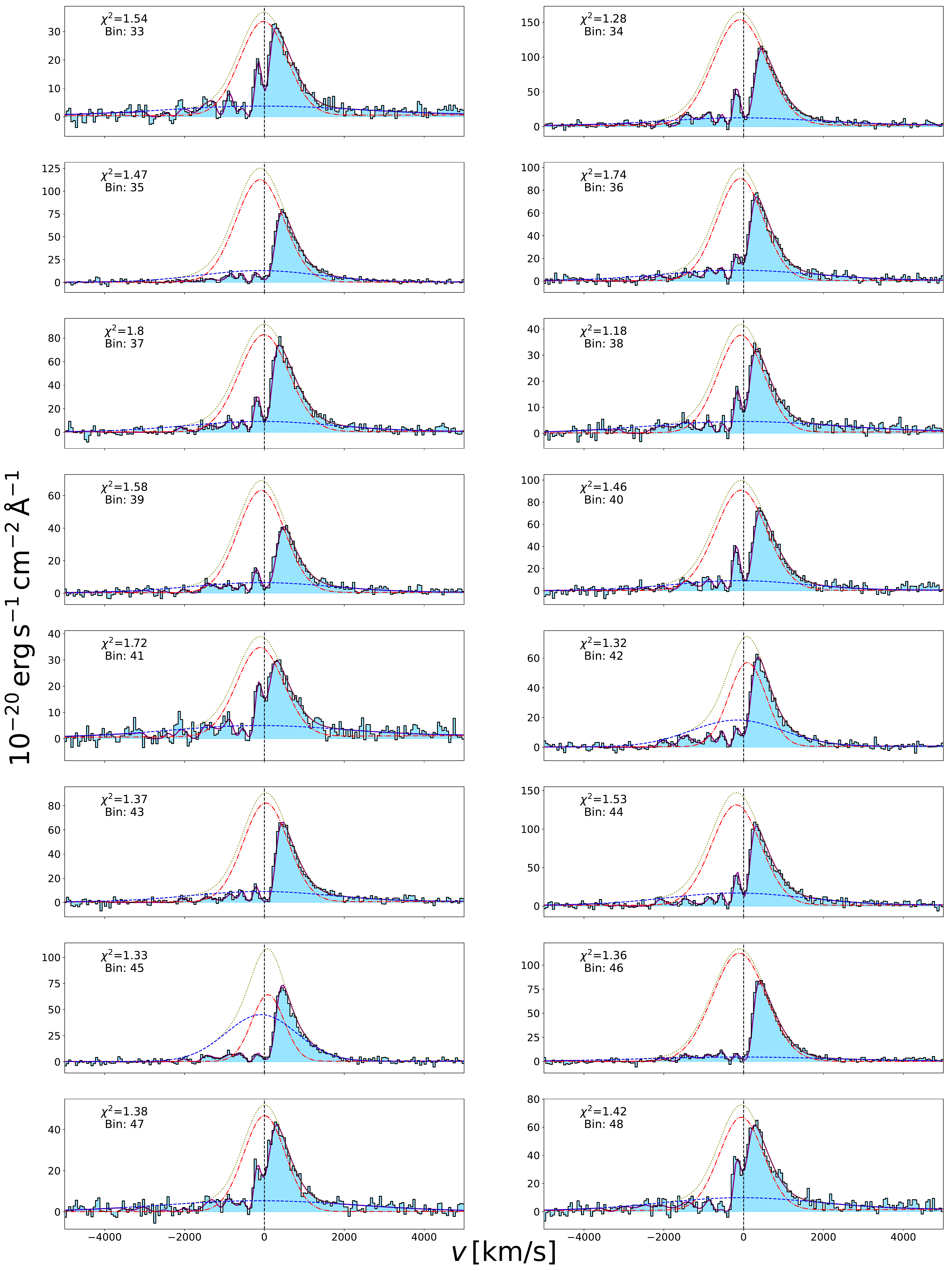}
      \caption{Figure \ref{fig:lyaindividualspatialplots2} continued.}
         \label{fig:lyaindividualspatialplots3}
  \end{figure*}
\begin{figure*}
  \centering
        \includegraphics[width=16.4cm,clip]{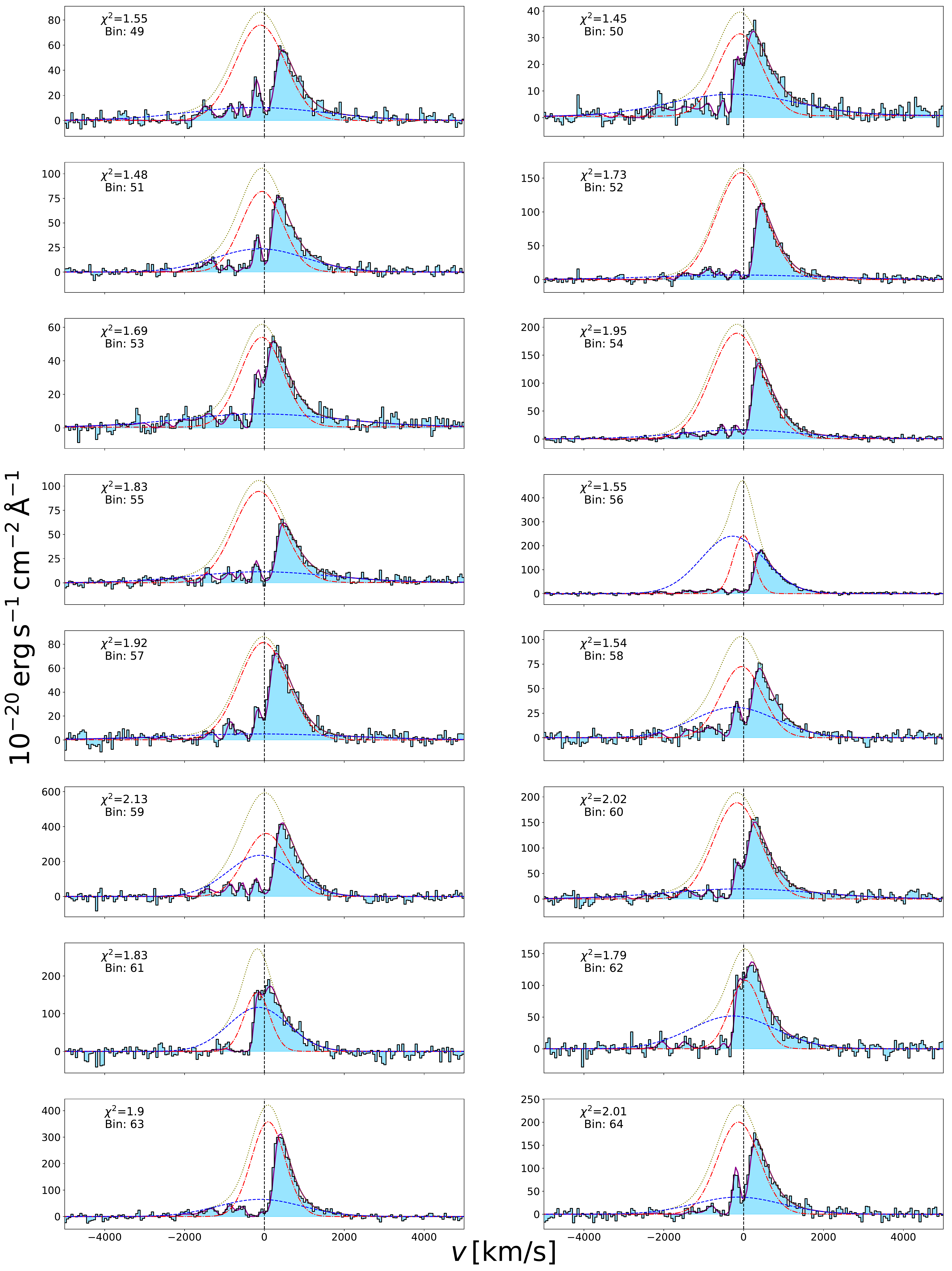}
      \caption{Figure \ref{fig:lyaindividualspatialplots3} continued.}
         \label{fig:lyaindividualspatialplots4}
  \end{figure*}

In this appendix, we present the Ly$\alpha$ fitting results from each individual bins described in Sects. \ref{sec:spatialmethod} and \ref{sec:lyaspatial} in Fig. \ref{fig:lyaindividualspatialplots1} to \ref{fig:lyaindividualspatialplots4}. The spatial region (bin) from which each of the presented spectrum is extracted is shown in Fig. \ref{fig:binum_map}. The spatial bins are numbered (and color-coded) and are the same as those marked in Figs. \ref{fig:lyaindividualspatialplots1} to \ref{fig:lyaindividualspatialplots4} in order to trace their spatial location. The details of the tessellation is described in Sect. \ref{sec:spatialmethod}.

\newpage

\newpage

\newpage

\newpage

\section{\texorpdfstring{\ion{C}{iv}}{CIV} spatial mapping}\label{apd:civspatialresults}

   \begin{figure}
  \centering
        \includegraphics[width=\hsize]{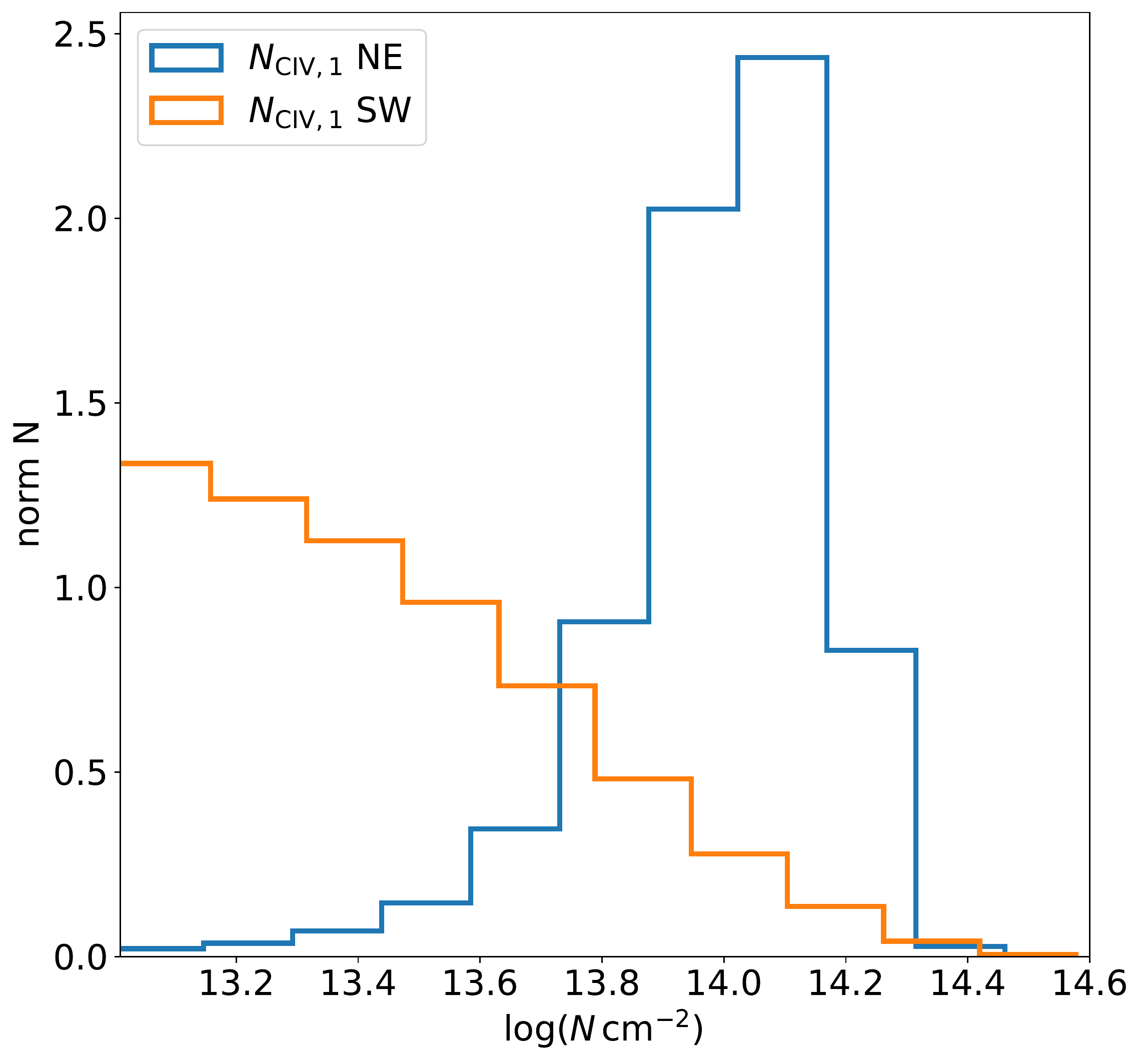}
      \caption{Probability distributions of the \ion{C}{iv} absorber \#1 column density  extracted from the corner plots. The blue and orange histograms represent the results of the NE and SW spectra, respectively. }
         \label{fig:CIVNESWdis}
   \end{figure}


In this appendix we present the parameters derived from the \ion{C}{iv} spatial fitting (see Sect. \ref{sec:civspatial}). The fittings are done for \ion{He}{ii} and \ion{C}{iv} simultaneously with the similar procedure described in Sects. \ref{sec:fitprocedure} and \ref{sec:civspatial}. The emission results for both \ion{He}{ii} and \ion{C}{iv} from NE and SW are presented in Table \ref{tab:NESWemission}. In Table \ref{tab:NESWabsorption}, we show the absorption fitting results of the 4 \ion{C}{iv} absorbers from these two regions. For comparison, we also show the emission and absorption fitting results of the Ly$\alpha$ lines from these two regions in the corresponding tables. Due to the low S/N of the two \ion{C}{iv} spectra and in order to avoid overfitting, we fix the $z$ and $b$ of the absorbers in the fit and use the derived column densities as upper limit. The exception is the column density of absorber \#1 of the NE spectrum, which has a well distributed probability extracted from the corner plot. The distribution is presented in Fig. \ref{fig:CIVNESWdis} in blue with a well defined peak. We also show the distribution of the same absorber from the SW spectrum for comparison in this figure in orange.  We note that the $z$ and $b$ for the absorbers in both of the NE and SW \ion{C}{iv} spectra are fixed to the corresponding values derived from master Ly$\alpha$ during the fitting. It will cause unsuccessful fitting if we use the fitted values from NE and SW Ly$\alpha$ to fix the corresponding \ion{C}{iv} fits. We argue that the adopted choice is reasonable because (i) large parts of the NE and SW regions are covered by the master aperture; (ii) the master Ly$\alpha$ has high S/N.

\begin{table*}
 \caption{Best fitted emission results of the 1D aperture-extracted spectrum from the NE and SW regions using the MCMC method.}\label{tab:NESWemission}
 \centering
\begin{tabular}{ l c c c c }
\hline
\hline

Ion & Line center (rest) & Line center (obs.) & Line flux & Line width \\
    & $ \lambda_{0}\,[\text{\AA}]$ & $\lambda\,[\text{\AA}]$ & $F$ [$\rm 10^{-17} \, erg \, s^{-1} \, cm^{-2}$] & $FWHM$ [$\rm km \, s^{-1}$] \\
    
\hline
\textit{NE} &&&&\\
\hline
Ly$\rm \alpha$        & 1215.67 & 6693.75 $\pm$ 0.24 & 50.51 $\pm$ 1.25 & 1474 $\pm$ 21\\
Ly$\rm \alpha$ (b.l.) & 1215.67 & $\sim$ 6693.20  & 14.68 $\pm$ 0.91 & 3669 $\pm$ 150\\

\ion{C}{iv}           & 1548.20 & 8527.65 $\pm$ 0.46 & 1.43 $\pm$ 0.35 & 1933 $\pm$ 342\\
\ion{C}{iv} (b.l.)    & 1548.20 & 8499.69 $\pm$ 4.29 & 1.52 $\pm$ 0.31 & 2027 $\pm$ 297\\

\ion{C}{iv}           & 1550.77 & 8541.86 $\pm$ 0.46 & 0.71 $\pm$ 0.17 & 1930 $\pm$ 342 \\
\ion{C}{iv} (b.l.)    & 1550.77 & 8513.85 $\pm$ 4.30 & 0.76 $\pm$ 0.15 & 2024 $\pm$ 296  \\

\ion{He}{ii}          & 1640.47 & 9035.94 $\pm$ 0.48 & 2.39 $\pm$ 0.11 & 651 $\pm$ 43 \\
\hline
\textit{SW} &&&&\\
\hline
Ly$\rm \alpha$        & 1215.67 & 6693.72 $\pm$ 0.20 & 38.34 $\pm$ 0.69 & 1569 $\pm$ 17 \\
Ly$\rm \alpha$ (b.l.) & 1215.67 & $\sim$ 6693.23  & 12.58 $\pm$ 0.46 & 4522 $\pm$ 122\\

\ion{C}{iv}           & 1548.20 & 8526.41 $\pm$ 0.46 & 1.17 $\pm$ 0.50 & 857 $\pm$ 254 \\
\ion{C}{iv} (b.l.)    & 1548.20 & 8498.50 $\pm$ 9.11 & 1.63 $\pm$ 0.46 & 2935 $\pm$ 484 \\

\ion{C}{iv}           & 1550.77 & 8540.62 $\pm$ 0.46 & 0.58 $\pm$ 0.25 & 856 $\pm$ 253 \\
\ion{C}{iv} (b.l.)    & 1550.77 & 8512.66 $\pm$ 9.13 & 0.82 $\pm$ 0.23 & 2930 $\pm$ 484 \\

\ion{He}{ii}          & 1640.47 & 9034.63 $\pm$ 0.49 & 2.24 $\pm$ 0.11 & 646  $\pm$ 50  \\

 \hline
\end{tabular}
\tablefoot{The notations used are the same as in Table \ref{tab:lineemissionfit} (see table notes there).}

\end{table*}

\begin{table*}
 \caption{Best absorption fitting results of the 1D aperture-extracted spectrum from the NE and SW regions using the MCMC method.}\label{tab:NESWabsorption}
 \centering
\begin{tabular}{c c c c c c c}
\hline
\hline
Abs. & Ion & Redshift & Absorber wav. & Velocity & Column density & Doppler \\
\#    & & $z$        & $\lambda\,[\text{\AA}]$ & $\Delta v\,[\rm km\,s^{-1}]$ & $\log(N/ \rm cm^{-2})$ & $b\,[\rm km\,s^{-1}]$ \\
\hline
\textit{NE} &&&&&\\
\hline   

 1 & Ly$\alpha$ & 4.5073 $\pm$ 0.0001& 6695.08 $\pm$ 0.06 & $-21$ $\pm$ 3 & 14.99 $\pm$ 0.01& 199 $\pm$ 3 \\ 
   & \ion{C}{iv}& $-$  & 8526.70  & $-$  & 14.02 $\pm$ 0.17  & $-$\\ 
  
 2& Ly$\alpha$  & 4.5001$\pm$ 0.0001 & 6686.31 $\pm$ 0.07 & $-414$ $\pm$ 3 & 15.1 $\pm$ 0.1 & 97 $\pm$ 10 \\  
   & \ion{C}{iv}& $-$  & 8515.37  & $-$  & $<$12.43  & $-$ \\
   
 3& Ly$\alpha$  & 4.4946 $\pm$ 0.0001 & 6679.64 $\pm$ 0.10 & $-712$ $\pm$ 4  & 14.72 $\pm$ 0.02 & 105 $\pm$ 10 \\ 
   & \ion{C}{iv}& $-$  & 8506.86  & $-$  & $<$13.04  & $-$ \\
   
 4& Ly$\alpha$  & 4.4872 $\pm$ 0.0001 & 6670.62 $\pm$ 0.15 & $-1116$ $\pm$ 7  & 14.97 $\pm$ 0.02 & 284 $\pm$ 18 \\
   & \ion{C}{iv}& $-$  & 8495.20  & $-$  & $<$14.4  & $-$\\
 5& Ly$\alpha$  & 4.4750 $\pm$ 0.0002 & 6655.77 $\pm$ 0.22 & $-1781$ $\pm$ 10  & 14.81 $\pm$ 0.02 & 241 $\pm$ 20 \\
 6& Ly$\alpha$  & 4.4657 $\pm$ 0.0002 & 6644.45 $\pm$ 0.21 & $-2288$ $\pm$ 10  & $<$15.81 & $\sim$54\\
 7& Ly$\alpha$  & $\sim$4.4572  & $\sim$6634.20 & $\sim -2747$  & $<$15.05 & $\sim$158 \\
 8& Ly$\alpha$  & $\sim$4.4457  & $\sim$6620.21 & $\sim -3373$  & $<$14.91 & $\sim$75 \\
 \hline

\textit{SW} &&&&&\\
\hline
 1 & Ly$\alpha$ & 4.5075 $\pm$ 0.0001& 6695.31 $\pm$ 0.03 & $-6$ $\pm$ 3 & 14.74 $\pm$ 0.01 & 156 $\pm$ 2 \\ 
   & \ion{C}{iv}& $-$  & 8526.70  & $-$ & $<$13.4  & $-$\\ 
  
 2& Ly$\alpha$  & 4.5002$\pm$ 0.0001 & 6686.47 $\pm$ 0.06 & $-427$ $\pm$ 10 & 15.16 $\pm$ 0.11 & 89 $\pm$ 7 \\  
   & \ion{C}{iv}& $-$  & 8515.37  & $-$ & $<$12.4  & $-$ \\
   
 3& Ly$\alpha$  & 4.4947 $\pm$ 0.0001 & 6679.74 $\pm$ 0.08 & $-750$ $\pm$ 26  & 14.70 $\pm$ 0.02 & 110 $\pm$ 9 \\ 
   & \ion{C}{iv}& $-$  & 8506.86  & $-$ & $<13.7$  & $-$ \\
   
 4& Ly$\alpha$  & 4.4871 $\pm$ 0.0001 & 6670.60 $\pm$ 0.13 & $-1150$ $\pm$ 19 & 14.86 $\pm$ 0.02& 260 $\pm$ 13 \\
   & \ion{C}{iv}& $-$  & 8495.21  & $-$ & $<$13.6  & $-$\\
 5& Ly$\alpha$  & 4.4747 $\pm$ 0.0002 & 6655.45 $\pm$ 0.18 & $-1790$ $\pm$ 10  & 14.76 $\pm$ 0.02 & 220 $\pm$ 15 \\
 6& Ly$\alpha$  & $\sim$4.4648  & $\sim$6643.34 & $\sim -2341$  & $<$14.59 & $\sim$94 \\
 7& Ly$\alpha$  & $\sim$4.4572  & $\sim$6634.10 & $\sim -2754$  & $<$15.67 & $\sim$103 \\
 8& Ly$\alpha$  & $\sim$4.4462  & $\sim$6620.82 & $\sim -3347$  & $<$15.60 & $\sim$60 \\

\hline

\end{tabular}
\tablefoot{The notations used are the same as in Table \ref{tab:absorberfit} (see table notes there). Due to the low S/N of the \ion{C}{iv} spectra extracted from these two regions, the column density results of \ion{C}{iv} absorbers are treated as upper limits with $z$ and $b$ fixed to the fitted \ion{H}{i} values, correspondingly. The exception is the column density of absorber \#1 in the NE, which has a well distributed probability (see Fig \ref{fig:CIVNESWdis}). }
\end{table*}

\end{appendix}